\newcommand{\nc}{\newcommand}
\nc{\beq}{\begin{equation}}   \nc{\eeq}{\end{equation}}
\nc{\bea}{\begin{eqnarray}}   \nc{\eea}{\end{eqnarray}}
\nc{\baa}{\begin{array}}      \nc{\eaa}{\end{array}}
\nc{\bit}{\begin{itemize}}    \nc{\eit}{\end{itemize}}
\nc{\ben}{\begin{enumerate}}  \nc{\een}{\end{enumerate}}
\nc{\bce}{\begin{center}}     \nc{\ece}{\end{center}}
\def\beqa{\begin{eqnarray}}
\def\eeqa{\end{eqnarray}}

\def\h{h}
\def\mh{m_{\h}}

\def\vev#1{\langle #1 \rangle}
\def\etal{{\it et al.}}
\def\half{{1\over 2}}
\def\nsd{N_{SD}}
\def\mtau{m_\tau}
\def\mt{m_t}
\def\mb{m_b}
\def\etc{{\it etc.}}
\def\sw{s_W}
\def\cw{c_W}
\def\tw{t_W}
\def\ntc{N_{TC}}
\def\delm{\Delta m}

\def\mx{M_X}
\def\gam{\gamma}
\def\lam{\lambda}
\def\eps{\epsilon}

\def\br{B}
\def\lsim{\mathrel{\raise.3ex\hbox{$<$\kern-.75em\lower1ex\hbox{$\sim$}}}}
\def\gsim{\mathrel{\raise.3ex\hbox{$>$\kern-.75em\lower1ex\hbox{$\sim$}}}}
\def\sigrts{\sigma_{\tiny\rts}^{}}

\def\mupmum{\mu^+\mu^-}
\def\epem{e^+e^-}
\def\tauptaum{\tau^+\tau^-}

\def\rts{\sqrt s}
\def\ie{{\it i.e.}}
\def\eg{{\it e.g.}}
\def\eps{\epsilon}
\def\anti{\overline}

\def\mw{m_W}
\def\mz{m_Z}
\def\fbi{~{\rm fb}^{-1}}
\def\fb{~{\rm fb}}
\def\pbi{~{\rm pb}^{-1}}
\def\pb{~{\rm pb}}
\def\mev{~{\rm MeV}}
\def\gev{~{\rm GeV}}
\def\tev{~{\rm TeV}}
\def\f{\frac}
\def\pzero{P^0}
\def\mpzero{m_{\pzero}}
\def\gampzero{\Gamma^{\rm tot}_{\pzero}}
\def\pzerop{P^{0\,\prime}}
\def\mpzerop{m_{\pzerop}}

\newcommand{\nn}{\nonumber}
\def\be{\beq}
\def\ee{\eeq}
\def\to{\rightarrow}

\def\PHYSICA #1 #2 #3 {{\sl Physica}~{\bf#1} (#3) #2}
\def\MPL #1 #2 #3 {{\sl Mod.~Phys.~Lett.}~{\bf#1} (#3) #2}
\def\NPB #1 #2 #3 {{\sl Nucl.~Phys.}~{\bf #1} (#3) #2}
\def\NPBPS #1 #2 #3 {{\sl Nucl.~Phys.~B~(Proc. Suppl.)}~{\bf #1} (#3) #2}
\def\PLB #1 #2 #3 {{\sl Phys.~Lett.}~{\bf #1} (#3) #2}
\def\PR #1 #2 #3 {{\sl Phys.~Rep.}~{\bf#1} (#3) #2}
\def\PRD #1 #2 #3 {{\sl Phys.~Rev.}~{\bf #1} (#3) #2}
\def\PRL #1 #2 #3 {{\sl Phys.~Rev.~Lett.}~{\bf#1} (#3) #2}
\def\RMP #1 #2 #3 {{\sl Rev.~Mod.~Phys.}~{\bf#1} (#3) #2}
\def\ZPC #1 #2 #3 {{\sl Z.~Phys.}~{\bf #1} (#3) #2}
\def\IJMP #1 #2 #3 {{\sl Int.~J.~Mod.~Phys.}~{\bf#1} (#3) #2}

\documentstyle[12pt,epsf]{article}
\begin{document}
\newlength{\captsize} \let\captsize=\small 
\newlength{\captwidth}                     

%
\font\fortssbx=cmssbx10 scaled \magstep2
\hbox to \hsize{
%
%
$\vcenter{
\hbox{\fortssbx University of California - Davis}
\hbox{\fortssbx University of Florence}
\hbox{\fortssbx University of Geneva}
}$
\hfill
$\vcenter{
\hbox{\bf UCD-98-13}
\hbox{\bf DFF-316-7-98}
\hbox{\bf UGVA-DPT-07-1011}
\hbox{July, 1998}
}$
}

%
\medskip
\begin{center}
{\Large\bf\boldmath Detecting and Studying the Lightest
Pseudo-Goldstone Boson
at Future $pp$, $e^+e^-$ and $\mu^+\mu^-$ Colliders\\}
\rm
\vskip1pc
{\Large
R. Casalbuoni$^{a,b}$\footnote{On leave from
Dipartimento di Fisica Universit\`a di Firenze,
I-50125 Firenze, Italia}, A. Deandrea$^c$, S. De Curtis$^b$, \\
D. Dominici$^{b,d}$, R. Gatto$^a$ and J. F. Gunion$^e$\\}
\vspace{5mm}
{\it{
$^a$D\'epart. de
Physique Th\'eorique, Universit\'e de Gen\`eve, CH-1211 Gen\`eve
4, Suisse\\
$^b$I.N.F.N., Sezione di Firenze,
I-50125 Firenze, Italia\\
$^c$Centre de Physique Th\'eorique,
CNRS, Luminy F-13288 Marseille, France \\
$^d$Dipartimento di Fisica Universit\`a di Firenze,
I-50125 Firenze, Italia\\
$^e$ Department of Physics, University of California,
Davis, CA 95616, USA}}
\end{center}
\vskip .2in
\begin{abstract}
\noindent For an attractive class of dynamical symmetry breaking
(technicolor) models, the lightest neutral pseudo-Nambu-Goldstone
boson ($\pzero$) contains only down-type techniquarks and
charged technileptons.
We discuss the prospects for discovering and studying the
$\pzero$ of such models at the Tevatron and the LHC and at future
$\epem$ and $\mupmum$ colliders. Depending
upon the number of technicolors, $\ntc$,
discovery of the $\pzero$ at the Tevatron and
the LHC in the $gg\to\pzero\to\gam\gam$
mode could be possible over a wide range of mass. For $\ntc=4$, we estimate
that RunII Tevatron data can be used to exclude or discover a $\pzero$
in the $50-200\gev$ mass range, and
at the LHC, a very precise measurement of $\Gamma(\pzero\to
gg)\br(\pzero\to\gam\gam)$ will be possible.
For $\ntc=4$, discovery of the $\pzero$ at an $\epem$ collider
via the reaction $\epem\to\gam \pzero$ should
be possible for an integrated luminosity of $L=100\fbi$ at $\rts=500\gev$
as long as $\mpzero$ is not near $\mz$.
However, measuring the branching
fractions and couplings of the $\pzero$ with precision would
require much more luminosity.
In the $\gam\gam$ collider mode of operation at an $\epem$
collider, the $\gam\gam\to\pzero\to b\anti b$ signal should
be very robust and could be measured with high statistical
accuracy for a broad range of $\mpzero$ if $\ntc=4$.
For the minimal $\ntc=1$ case, detection of the $\pzero$
at the Tevatron and
in $\epem$ collisions will be very difficult, and the precision
of measurements at the LHC and the $\gam\gam$ collider decline markedly.
Only a $\mupmum$ collider yields a $\pzero$ production rate that
does not depend markedly upon $\ntc$. At a $\mupmum$ collider,
depending upon the luminosity available at low energies,
discovery of the $\pzero$ as an $s$-channel resonance
(given its predicted coupling to $\mupmum$) should prove possible
via scanning, even if it has not already been detected elsewhere.
Once the precise mass of the $\pzero$  is known, operation of the $\mupmum$
collider as a $\pzero$ factory will typically allow precision measurements
of enough observables to determine the number of technicolors
of the theory and (up to a
discrete set of ambiguities) the fundamental parameters
of the low-energy effective Lagrangian describing
the Yukawa couplings of the $\pzero$.
\end{abstract}

\section{Introduction}

Theories of the electroweak interactions based on dynamical
symmetry breaking (DSB) avoid the introduction of
fundamental scalar fields but generally predict
many pseudo-Nambu-Goldstone bosons (PNGB's)
due to the breaking of a large initial global symmetry group $G$.
While observation of any of the states predicted by DSB
would be exciting, it is the PNGB's that are most directly related
to the DSB mechanism. Further, since they do not
acquire mass from the technicolor interactions,
the PNGB's are almost certainly the lightest
of the new states in the physical spectrum predicted by DSB.
Among the PNGB's, the colorless neutral states are unique in that they remain
massless even after the interactions of the color and electroweak gauge
bosons are turned on.  In technicolor models,
their masses derive entirely from
effective four technifermion operators involving two technileptons
and two techniquarks.  Such operators arise from two sources:
the one-loop effective potential generated from
the low-energy effective
Lagrangian that describes the PNGB's and their interactions
with quarks and leptons; and explicit extended-technicolor gauge
boson (technileptoquark) exchanges that change a
techniquark into a technilepton.
The one-loop contributions to the mass-squared matrix
for the PNGB's exhibit an underlying $SU(2)_L\times SU(2)_R$ symmetry.
The technileptoquark gauge boson exchange contributions
automatically preserve this symmetry.
When this symmetry is present, the lightest PNGB, denoted $\pzero$, will
contain only down-type techniquarks (and charged technileptons)
and, in particular,  no up-type techniquarks. As a result,
its mass scale is most naturally determined by the $b$-quark mass
and not the $t$-quark mass (which will set the mass of the next lightest
state). Consequently, it is likely to be
much lighter than all the other PNGB's.
It is the phenomenology of such a $\pzero$ that we shall focus on.

Remarkably, direct observation of a PNGB would not have
been possible at any existing accelerator,
however light the PNGB's are, unless
the number of technicolors, denoted $\ntc$, is very large.
Further, indirect constraints,
\eg\ from precision electroweak data, are model-dependent
and not particularly robust when the number of technicolors is not large.
Thus, it is important to understand, as
a function of $\ntc$, the extent to which future
accelerators will allow us to detect and study the PNGB's
of the DSB model. It is particularly interesting to assess
prospects for detecting and studying the $\pzero$, not just
because it is much the lightest state, but also
because a determination of its couplings
to Standard Model particles would reveal a great deal
about the technicolor model even if none of the heavier PNGB's
and other states are seen. In order to understand how these
prospects depend upon $\ntc$, we will present results for
both the moderate value of $\ntc=4$ and the minimal,
although rather unphysical, reference case of $\ntc=1$.
This comparison will show that sensitivity to the $\pzero$
at the Tevatron, the next linear $\epem$ collider
and a $\gam\gam$ collider declines
significantly as $\ntc$ decreases, whereas a $\mupmum$ collider
will be able to explore $\pzero$ physics more or less equally
well for any value of $\ntc$ in the small to moderate range.

Detection of the PNGB's at the Tevatron and LHC colliders,
has been extensively considered in
Refs.~\cite{production,drk,ehlq,dometal94,chivu,eichtenlane,lane1,elw}.
A convenient review is that of Ref.~\cite{chivu}. For the production modes
studied in the above references, the net result is that PNGB
discovery would not be certain, and, even if a PNGB were
discovered at the Tevatron or LHC,
a detailed study of its properties would be very challenging.
However, inclusive $gg$ fusion production of a neutral PNGB,
followed by its decay to $\gam\gam$, was not given detailed consideration
in the earlier work. Here, the
production occurs via an anomalous $gg\pzero$
coupling with strength proportional to $\ntc$, yielding
a production rate proportional to $\ntc^2$. We find that, for
the modest value of $\ntc=4$, this mode holds considerable
promise for $\pzero$ discovery at the Tevatron and will almost
certainly allow detection of the $\pzero$ at the LHC. At the LHC,
a precise measurement of the product of the $\pzero$'s $gg$
partial width times its branching fraction to $\gam\gam$
would be possible for $\ntc=4$. This measurement
would play a pivotal role in a direct experimental determination
of key parameters of the DSB model. For the minimal reference
value of $\ntc=1$,
Tevatron prospects become poor, but discovery of the $\pzero$
at the LHC in the $\gam\gam$ mode would still be possible
over a large mass range.

Detection of PNGB's at lepton colliders has also been considered.
A useful treatment, in the context of a multi-scale
walking technicolor model, is that of Ref.~\cite{ls}.
Here, we thoroughly survey the role that $\epem$ and $\mupmum$ colliders
can be expected to play in the discovery and study of the $\pzero$.
Indeed, in the type of DSB model we consider
the $\pzero$ might be the only
state that is light enough to be produced at a $\rts\leq 500\gev$
first generation $\epem$ or $\mupmum$ collider.
The most important production process for the $\pzero$ at
an $\epem$ collider is $\epem\to\gam\pzero$. In the $\gam\gam$
collider mode of operation, one searches for $\gam\gam\to \pzero$.
The $\gam\gam\pzero$ coupling required in these two cases
arises from an anomalous vertex graph and is proportional to $\ntc$,
yielding production rates proportional to $\ntc^2$.
For $\ntc=4$, we find that discovery of the $\pzero$ in $\epem\to\gam\pzero$
will be possible for at least a limited
range of masses and that the $\gam\gam$ collider will
provide very robust $\pzero$ signals allowing for fairly precise
measurements of rates in a variety of channels. However,
prospects decline at smaller $\ntc$.
At a $\mupmum$ collider, one has the unique ability to $s$-channel produce
the $\pzero$ ($\mupmum\to\pzero$)
since it has a Yukawa-like coupling to muons. Since this Yukawa coupling is
independent of $\ntc$,
discovery and precision studies of important channel rates will
certainly be possible at a $\mupmum$ collider, regardless
of the value of $\ntc$. The relative rates allow a determination
of the relative strengths of different couplings of the $\pzero$.
Finally,  a measurement
of the $\pzero$'s total width would be possible (by scanning
in the $s$-channel) at a $\mupmum$ collider.
The total width is the crucial ingredient needed to
convert measured production rates into absolute magnitudes for the
parameters of the effective Lagrangian.

It is useful to expand somewhat upon the theoretical context for
our study of PNGB's. For recent reviews
see Refs.~\cite{chivunew,bhateichten}.
The idea of dynamical breaking of electroweak
symmetry based on a new strong interaction goes back to Weinberg  and to
Susskind \cite{techni}.
The original idea of a scaled-up QCD (technicolor)
was extensively developed (see for instance the review by Farhi and
Susskind \cite{technir}). These and many related ideas
usually face substantial difficulties when attempting
to construct a realistic scenario.
First of all, quark and lepton masses demand additional interactions (such
as extended technicolor).
Apart from finding a simple solution for the heaviness of
the top quark (see, for instance,
the recent proposal of top-color assisted technicolor \cite{hill}),
the difficulty in solving the problem of flavor changing neutral
currents (FCNC) blocked concrete progress for many years.

These difficulties eventually led to the realization that the $W$ and $Z$
masses depend on different regions of momenta than do
the quark and lepton masses. As a result, it is possible to construct
theories which yield enhanced contributions in the high
momentum region, without giving rise to strong effects at low momenta.
In this way, it is possible to simultaneously explain the
magnitudes of the intermediate
vector boson masses and of the quark and lepton masses without encountering
severe FCNC problems. In this context, it is important to note that
PNGB masses are expected to arise
from the momentum region important for quark and lepton masses.
These ideas  have
been developed into schemes such as walking technicolor
\cite{Holdom,Yamawaki,Appelquist}.

In this paper, the DSB model will be largely specified
by a low-energy effective theory characterized by a
large initial chiral symmetry group, its subgroup after
spontaneous breaking, and the known local electroweak and color group.
The effective low-energy Lagrangian of the theory
contains a Yukawa coupling component that plays two crucial roles.
First, it determines the most general form of
the couplings of all the PNGB's,
in particular those of the $\pzero$, to SM fermions.
Second, the one-loop potential computed from
the low-energy effective Yukawa couplings gives contributions to the
mass-squared matrix of the PNGB's.
The relative size of these one-loop contributions to the mass-squared
matrix as compared to the contributions from technileptoquark
gauge boson exchange diagrams is model-dependent. However, should
the one-loop contributions be dominant, the $\pzero$ mass
would then be mostly determined by the same mechanism that is responsible
for the quark and lepton masses.

Another possible difficulty for technicolor models is the $S$
parameter of electroweak precision measurements
\cite{peskintakeuchi}. We recall that the calculation of this
parameter is based on dispersion relations saturated with
low-lying resonances assuming that TC dynamics is described by a
simple scaling of QCD. In particular, use is made of the Weinberg
sum rules \cite{Weinberg} and of the KSRF relation
\cite{KSRF} (see refs. \cite{peskintakeuchitwo} and \cite{cahn}). In
the same papers, there are also attempts to model the
current spectral functions with nonresonant contributions.
Already at this level, a great uncertainty is obtained. In fact,
according to the parameterization used, the result can vary by
a factor 2. Furthermore, these results depend crucially on the
assumptions about the dynamics. In particular, the contributions
of vector and axial-vector resonances tend to cancel against
one another in the dispersive representation of $S$
\cite{selfenergies}. It is only the  assumption that TC dynamics
should scale as  QCD that gives a particular relation among the
masses and couplings of these resonances making the cancellation
incomplete. Therefore, relaxing this hypothesis and allowing the
masses and couplings to vary, it is possible to
change completely the previous picture and obtain big
cancellations, and even  a negative value for $S$. For
instance, in Ref.~\cite{cancellations}, in the effective
Lagrangian approach of the present (BESS) model, the only
physical degrees of freedom beyond those of the Standard Model
are (in the unitary gauge) new vector, axial-vector and
pseudo-Goldstone bosons. One finds that there is a substantial
portion of parameter space for which the full set of loop
diagrams, including those containing the new vector, axial-vector
and/or pseudo-Goldstone bosons, can yield a small value for $S$
(as well as values for the $T$ and $U$ parameters consistent with
precision electroweak measurements). If one insists
that technifermions appear as physical degrees of freedom
at sufficiently large energies, the associated
contribution to $S$ 
[$S=1/2\pi(N_{TF}/2)(N_{TC}/3)$]
might or might not be a problem. The additional dynamics
associated with the resonance states might be such as to
generate (negative) cancelling contributions.
Also in dynamical schemes such
as walking technicolor it is  possible to decrease the
value of $S$, see refs. \cite{negatives} and \cite{postmodern}.
In view  of the previous considerations and of the many
uncertainties involved in a sensible evaluation of the $S$ parameter
 in these theories, we will not consider further the
constraints coming from precision electroweak measurements.
Certainly, we cannot guarantee that an attractive
technicolor or other dynamical symmetry breaking model, fully
consistent with precision electroweak
and other current data, will eventually be constructed. However, 
if such a model is nature's choice and if its underlying dynamics gives
rise to
PNGB's, the phenomenology of the lightest $P^0$ PNGB will be
that we delineate in this paper. The present theoretical uncertainties
associated with DSB models, if anything, increase the importance
of searching for a light $P^0$ using the approaches we shall discuss.
Discovery of the $P^0$, and a precision study of its properties, would
be the 
first steps in revealing and unravelling the underlying DSB theory.

In the context of the low-energy
effective theory there are two important scales:
the effective cut-off scale $\Lambda$ and the scale $v$
related to electroweak symmetry breaking. The
overall magnitude of the one-loop contributions to PNGB masses is determined
by $\Lambda m_f/v$ and fermionic couplings
are proportional to $m_f/v$ (where $f$ denotes the relevant fermion type).
Typically, discovery of the $\pzero$ (and other PNGB's) is easier
for small $v$ so long as the mass does not become so large
as to push the discovery channels close
to their kinematic limits. The largest possible value
of $v$ is the standard $v=246\gev$.
This choice is that appropriate in the simplest
technicolor models.  In multi-scale walking technicolor models,
the effective value of $v$ for some of the PNGB's is typically 1/2 or
1/4 as large. For fixed PNGB masses,
the smaller $v$'s would imply production rates that are as much
as an order of magnitude larger than for $v=246\gev$; detection
of the PNGB's becomes much easier. For instance, the optimistic
conclusions of Ref.~\cite{ls} depend crucially on the
presence of such enhancements in the multi-scale model they employ.
We will adopt the most conservative
choice of $v=246\gev$ in our studies.  Nonetheless, we find
that discovery and detailed study of the $\pzero$ should generally be
possible at future colliders.

In large measure, these good prospects
are due to the down-type techniquark (and technilepton) nature
of the $\pzero$, in the class of DSB model we consider,
which has three important implications: (1) it
guarantees that $\mpzero({\rm one-loop}) \sim  \Lambda m_b/v$ will be
small (for $\Lambda\lsim 10\tev$); (2) it makes full strength
for the $\mupmum\pzero$ coupling likely; and (3) it leads
to large numerical values for both the $gg\pzero$ and $\gam\gam\pzero$
anomalous couplings as compared to other eigenstate possibilities.
As alluded to earlier, the importance of the
anomalous couplings is that the $gg\pzero$ coupling-squared
determines the $\pzero$ production rate at hadron colliders and
the $\gam\gam\pzero$ coupling-squared
controls $\pzero$ production rates at $\epem$ and $\gam\gam$ colliders.
The $\mupmum\pzero$ coupling-squared determines the $\pzero$
production rate at a muon collider. Of course, as noted earlier,
since the $gg\pzero$ and $\gam\gam\pzero$
couplings-squared are proportional to $\ntc^2$,
prospects for discovering and studying the $\pzero$ at hadron, $\epem$
and $\gam\gam$ colliders decline significantly for the minimal $\ntc=1$
value as compared to our canonical $\ntc=4$ choice. Thus,
since the $\mupmum\pzero$ coupling is $\ntc$-independent, a muon
collider is the best option at low $\ntc$.

The outline of the paper is as follows.  In Section 2,
we present details of the low-energy effective Lagrangian
describing PNGB's, with gradually increasing focus on the light $\pzero$.
We also discuss technileptoquark contributions to $\mpzero$.
In Section 3, we present results for the branching fractions
and total width of the $\pzero$ for a representative choice
of model parameters. In Section 4, prospects for $\pzero$
detection at the Tevatron and LHC are discussed, with particular
emphasis on the $gg\to\pzero\to\gam\gam$ mode.
In Section 5, $\pzero$ detection at an $\epem$ collider (primarily in
the $\epem\to\gam\pzero$ mode) is studied.
In Section 6, $\pzero$ detection at a $\gam\gam$ collider is discussed.
Section 7 gives detailed results for discovery
and study of the $\pzero$ via $s$-channel production
at a $\mupmum$ collider. In Section 8, we give the procedure
for combining the Tevatron/LHC measurement of $\Gamma(\pzero\to
gg)\br(\pzero\to\gam\gam)$ with the $\gam\gam$-collider
measurement of $\Gamma(\pzero\to\gam\gam)\br(\pzero\to b\anti b)$
and the muon-collider measurements of $\Gamma(\pzero\to
\mupmum)\br(\pzero\to F)$ ($F=\tauptaum,b\anti b,gg$) and $\gampzero$
so as to determine: (a) the number of technicolors;
and (up to a discrete set of ambiguities) (b) the parameters
describing the Yukawa couplings of the $\pzero$ and (c) a tentative
value for the effective cut-off scale of the low-energy theory.
A discussion of expected errors for these determinations is also
presented. Our conclusions are presented in Section 9.

\section{Model details and resulting PNGB mass spectrum and couplings}

We begin by briefly summarizing the framework we employ for computing the
one-loop contributions to the PNGB mass spectrum coming from the interactions
which are responsible for the masses of the
ordinary fermions \cite{mass}.  We will avoid the introduction
of a detailed specific fundamental theory by working
with the low-energy effective theory as
characterized by its chiral symmetry group
$G$,  which is broken down to a  subgroup $H$, and by the gauge group $G_{W}=
SU(3)\otimes SU(2)_L\otimes U(1)_Y$.
The first step is to write the allowed effective Yukawa
couplings between the ordinary fermions and the PNGB's.
These effective Yukawa couplings can be employed independently of the
underlying fundamental mechanism which gives rise to them.

The allowed couplings are those which
are invariant under the gauge symmetry $G_{W}$ that survives in the low-energy
sector. According to the general prescription of Ref. \cite{ccw},
these invariant
couplings are specific functions of the fermion fields and of the
PNGB bound-state fields that emerge in the effective theory
and transform non-linearly under $G$.
The lowest order term in the expansion of these
couplings in inverse powers of the PNGB decay constant
$v$ contains the mass terms for the ordinary fermions.
This allows us to obtain a set of relations among the Yukawa
couplings and the ordinary fermion masses.

In general, the number of independent Yukawa
couplings exceeds the number of  quark and lepton masses, so that some
free parameters remain. Thus, measurements of such observables
as branching fractions and production rates
at high-energy accelerators will be important for
determining other combinations of the Yukawa couplings.
The possibilities for and typical errors in the parameter
determinations will be the main focus of later sections of the paper.
For now, we will discuss how the masses and fermionic
couplings of the PNGB's depend upon the Yukawa coupling parameters.

Given a specific form for the Yukawa couplings,
the natural tool for analyzing
their influence on the vacuum orientation is the effective potential.
The one-loop effective potential, including the effects of both ordinary
gauge interactions and Yukawa couplings, is a major ingredient
in determining the PNGB mass spectrum.
The other possible contributions to the PNGB mass-squared
matrix are those arising at tree-level from technileptoquark exchanges.

We will specialize to the case of a  $SU(8)\times SU(8)$ chiral symmetry,
spontaneously broken to the diagonal $SU(8)$ subgroup. This pattern of global
symmetries is realized, for instance, in the popular technicolor model
\cite{techni,technir} in which the
technifermions consist of a generation $(U,D,N,E)$ of techniquarks  and
technileptons, with the same quantum numbers as the corresponding ordinary
fermions. The spontaneous breaking of $SU(8)\times SU(8)$ to the diagonal
$SU(8)$ subgroup produces 63 Goldstone bosons.
One has 7 colorless goldstones $\pi_{a}$, ${\tilde \pi}_{a}$ and
$\pi_{D}$ associated to the generators $T^{a}$, ${\tilde T}^{a}$ and $T_{D}$
($a=1,2,3$). The states $\pi_{a}$ are those absorbed by the $W$ and $Z$
via the Higgs mechanism. The states $\tilde\pi_3$ and $\pi_D$
are the colorless, neutral states upon which we focus.
In addition, there are 32 color octet goldstones:
$\pi^{\alpha}_{8}$ and $\pi_{8}^{a\alpha}$ ($\alpha =1, \ldots 8$). They are
related to the generators $T_{8}^{\alpha}$ and $T_{8}^{a\alpha}$. Finally,
there are 24 triplet Goldstone bosons: $P_{3}^{\mu i}$ and ${\bar
P}_{3}^{\mu i}$ ($\mu$=0,1,2,3 and $i$=1,2,3). These are linear combinations
of the fields $\pi_3^{\mu i}$ and ${\bar \pi}_3^{\mu i}$ associated with the
generators $T_{3}^{\mu i}$ and ${\bar T}_{3}^{\mu i}$ (for the list of the
$SU(8)$ generators see Ref. \cite{mass}).

In the single-scale technicolor model based on
$SU(8)\times SU(8)$, the low-energy gauge
interactions of the Goldstone bosons are described by the
Lagrangian \cite{chadha}
\be
{\cal L}_{g}= {v^{2} \over 16} Tr \left( {\cal D}_{\mu} U^{\dagger}
 {\cal D}_{\mu} U\right)
\label{Lg}
\ee
where
\be
U=\exp \left({\displaystyle{{2iT^{s} \pi^{s}}\over v}}\right)
\ee
with $v=246~GeV$.
The generators $T^{s}$ ($s=1, \ldots, 63$) are normalized according to
$Tr(T^{r} T^{s})=2 \delta^{rs}$. The covariant derivative ${\cal D}_{\mu} U$ is
given by
\be
{\cal D}_{\mu} U=\partial_{\mu} U +{\cal A}_{\mu} U-U {\cal B}_{\mu}
\ee
where
\bea
{\cal A}_{\mu}=W_{\mu}+{\hat B}'_{\mu}+G_{\mu}~~& &~~
{\cal B}_{\mu}={\hat B}_{\mu}+{\hat B}'_{\mu}+G_{\mu} \cr
W_{\mu}=ig T^{a} W_{\mu}^{a} ~~& &~~
{\hat B}_{\mu}=ig' T^{3} B_{\mu}\cr
{\hat B}'_{\mu}=ig' {T_{D}\over\sqrt{3}} B_{\mu}~~& &~~
G_{\mu}=i{g_{s}\over\sqrt{2}}T_{8}^{\alpha}G_{\mu}^{\alpha}
\label{gauge}
\eea
In Eq. (\ref{gauge}), $W_{\mu}^{a}$ $(a=1,2,3)$, $B_{\mu}$ and
$G_{\mu}^{\alpha}$ $(\alpha=1,\cdots,8)$,
are the gauge
vector fields related to the gauge group $G_W$.

In a multi-scale technicolor model, the above would be generalized
by writing ${\cal L}_g$ as a sum of terms involving different $v_i$'s,
where each term would contain a $U_i$ that depends on a subset
of the $\pi^s$ fields. (Such a structure
would amount to an explicit breaking of the $SU(8)\times SU(8)$
symmetry by ${\cal L}_g$.) By requiring $\sum_i v_i^2=v^2=(246\gev)^2$,
correct values for the $W$ and $Z$ masses are obtained, and,
if each $U_i$ has $1/v_i$ in the exponent, the kinetic energy
terms for the $\pi$'s that are in the exponent will be
correctly normalized. The term of relevance for our purposes would
be that which contains the neutral color-singlet
$\pi_D$ and $\tilde\pi_3$ isosinglet and isotriplet fields.
As regards these fields,
the discussion to be presented would then be unaltered except that
the value of $v$ would be replaced by a smaller value (typically
$v/2$ or $v/4$).

In order to write the Yukawa couplings between the Goldstone bosons and
the ordinary fermions we decompose the matrix $U$ according to
\be
U=\left(\matrix{
U_{uu}^{ij} & U_{ud}^{ij} & U_{u\nu}^{k} & U_{ue}^{k}\cr
U_{du}^{ij} & U_{dd}^{ij} & U_{d\nu}^{k} & U_{de}^{k}\cr
U_{\nu u}^{l} & U_{\nu d}^{l} & U_{\nu\nu} & U_{\nu e}\cr
U_{eu}^{l} & U_{ed}^{l} & U_{e\nu} & U_{ee}\cr}\right)
\ee
The indices $i,j,k$ and $l$ are color indices running from 1 to 3.
The most general Yukawa coupling
invariant with respect to $G_{W}$ is given in \cite{mass}.
With the exception of the coupling
to muons (as needed for the $\mupmum$ collider study),
only the couplings involving the third family are phenomenologically
relevant for accelerator physics. (This is because, as usual,
the magnitudes of the
Yukawa couplings are most naturally set by the scale of the corresponding
fermionic masses.) It is only these Yukawa couplings
that we can hope to determine. Thus, we consider
the effective Yukawa coupling Lagrangian
\bea
{\cal L}_Y=
&-&m_{1} \left( {\bar t_{R}^{i}} U^{\dagger ij}_{uu} t_{L}^{j}+
{\bar t_{R}^{i}} U^{\dagger ij}_{du} b_{L}^{j}+ h.c.\right)\cr
&-&m_{2} \left( {\bar b_{R}^{i}} U^{\dagger ij}_{ud} t_{L}^{j}+
{\bar b_{R}^{i}} U^{\dagger ij}_{dd} b_{L}^{j}+ h.c.\right)\cr
&-&m_{4} \left( {\bar \tau}_{R} U^{\dagger}_{\nu e} \nu_{\tau L}+
{\bar \tau}_{R} U^{\dagger}_{ee} \tau_{L}+ h.c.\right)\cr
&-&m_{4}^{(2)} \left( {\bar \mu}_{R} U^{\dagger}_{\nu e} \nu_{\mu L}+
{\bar \mu}_{R} U^{\dagger}_{ee} \mu_{L}+ h.c.\right)\cr
&-&m_{5} \left( {\bar t_{R}^{i}} U^{\dagger jj}_{uu} t_{L}^{i}+
{\bar t_{R}^{i}} U^{\dagger jj}_{du} b_{L}^{i}+ h.c.\right)\cr
&-&m_{6} \left( {\bar b_{R}^{i}} U^{\dagger jj}_{ud} t_{L}^{i}+
{\bar b_{R}^{i}} U^{\dagger jj}_{dd} b_{L}^{i}+ h.c.\right)\cr
&-&m_{7} \left( {\bar t_{R}^{i}} U^{\dagger}_{\nu\nu} t_{L}^{i}+
{\bar t_{R}^{i}} U^{\dagger}_{e\nu} b_{L}^{i}+ h.c.\right)\cr
&-&m_{9} \left( {\bar b_{R}^{i}} U^{\dagger}_{\nu e} t_{L}^{i}+
{\bar b_{R}^{i}} U^{\dagger}_{ee} b_{L}^{i}+ h.c.\right)\cr
&-&m_{10} \left( {\bar \tau}_{R} U^{\dagger jj}_{ud} \nu_{\tau L}+
{\bar \tau}_{R} U^{\dagger jj}_{dd} \tau_{L}+ h.c.\right)\cr
&-&m_{10}^{(2)} \left( {\bar \mu}_{R} U^{\dagger jj}_{ud} \nu_{\mu L}+
{\bar \mu}_{R} U^{\dagger jj}_{dd} \mu_{L}+ h.c.\right)\cr
&-&m_{11} \left[ \bar t_R\lam^\alpha
\left(\begin{array}{cc} U_{uu}^\dagger &U_{du}^\dagger\end{array}\right)
\left(\begin{array}{cc} \lam^\alpha & 0 \\ 0 & \lam^\alpha \end{array}\right)
\left(\begin{array}{c} t_L \\ b_L \end{array}\right)+h.c.\right]\cr
&-&m_{12} \left[ \bar b_R\lam^\alpha
\left(\begin{array}{cc} U_{ud}^\dagger &U_{dd}^\dagger\end{array}\right)
\left(\begin{array}{cc} \lam^\alpha & 0 \\ 0 & \lam^\alpha \end{array}\right)
\left(\begin{array}{c} t_L \\ b_L \end{array}\right)+h.c.\right]\,.
\label{Ly}
\eea
In the above, we have omitted terms involving only colored PNGB's,
and kept only couplings for the third family, with the exception
of the parameters $m_{4,10}^{(2)}$
which determine the PNGB couplings to muons. In Eq.~(\ref{Ly}),
we have also assumed no generation-mixing terms, thereby
automatically guaranteeing absence of flavor-changing
neutral current interactions. For later reference, it is
useful to note that in the $(U,D,N,E)$ realization of our model,
the $U_{uu}$, $U_{dd}$, $U_{\nu\nu}$ and $U_{ee}$ entries above
come from techni-$U$, $D$, $N$ and $E$ fermion propagator loop corrections,
respectively.

It will be important to understand how the (two) neutral
color-singlet PNGB states enter in the notation of Eq.~(\ref{Ly}).
For this purpose we define the states
\beq
\pzero=\frac{\tilde\pi_3-\pi_D}{\sqrt{2}}\,,~~~~
\pzerop=\frac{\tilde\pi_3+\pi_D}{\sqrt{2}}\,,
\label{eigenstates}
\eeq
which, as discussed shortly, are the mass eigenstates
in the class of models we consider.
Expanding to first order in $1/v$, we have
\bea
U_{uu}^{ij}&\sim& \delta^{ij}\left[1+{i\over v}\sqrt 6{\pzerop\over
3}\right]+\ldots \nn\\
U_{dd}^{ij}&\sim& \delta^{ij}\left[1-{i\over v}\sqrt 6{\pzero\over
3}\right]+\ldots \nn\\
U_{\nu\nu}&\sim& \left[1-{i\over v}\sqrt 6{\pzerop}\right]+\ldots \nn\\
U_{ee}&\sim& \left[1+{i\over v}\sqrt 6{\pzero}\right]+\ldots\,.
\label{specialu}
\eea
Non-neutral entries ($U_{ud}^{ij}$, $U_{\nu e}$ \etc) contain
neither ${\cal O}(1)$ terms nor
neutral $\pzero$ and $\pzerop$ fields (but, rather, contain
charged PNGB's) in their $1/v$ expansion.
We note that this expansion assumes that we begin with the correct vacuum
state, implying that the masses of the PNGB
fields appearing in the $U_{ij}$'s
must all be positive. This will imply constraints on the $m_i$ parameters
of Eq.~(\ref{Ly}).

>From the ${\cal O}(1)$ terms in
the expansion of ${\cal L}_{Y}$ we easily recover the expressions for
the fermion masses
\bea
m_{t}&=&m_{1}^\prime+m_{7} \cr
         m_{b}&=&m_{2}^\prime+m_{9} \cr
         m_{\tau}&=&m_{4}+3 m_{10}  \cr
m_{\mu}&=&m_{4}^{(2)}+3 m_{10}^{(2)}  \cr
         m_{\nu_{\tau}}&=&m_{\nu_{\mu}}=0\,,
\label{mf}
\eea
where
\beq
m_1^\prime\equiv m_1+3m_5+{16\over 3} m_{11}\,,\quad
m_2^\prime\equiv m_2+3m_6+{16\over 3}m_{12}\,.
\label{mprimes}
\eeq
Note that the fermion masses provide 4 constraints on the
independent parameters appearing in ${\cal L}_Y$.

We note that in the multi-scale case, it is precisely the $U_i$ that
contains the color-singlet neutral $\pzero$ and $\pzerop$ that will
be responsible for fermion masses and that would appear in Eq.~(\ref{Ly}).
This means that the $1/v$ that appears in the expansions of
Eq.~(\ref{specialu}) would be replaced by $1/v_i$.  This
$1/v_i$ factor would then emerge in the Yukawa couplings
and mass contributions for the $\pzero$ to be discussed below.

Starting from the interaction Lagrangian ${\cal L}_{g}+{\cal L}_{Y}$,
the next step is to evaluate the one-loop effective potential.
Let us assume for the moment that it is the only contribution to the
PNGB mass-squared matrix. Then, all PNGB masses would be obtained by its
diagonalization as detailed in \cite{mass}.  In particular,
we will focus on the neutral color-singlet PNGB's.
The mass matrix in the $\pzero$--$\pzerop$
basis arises from one-loop diagrams involving the quark
and lepton fields to which they couple, where the momenta of the loops
are integrated up to scale $\Lambda$. Using Eq.~(\ref{Ly}),
one finds that the $\pzero$ and $\pzerop$ are, in fact,
mass eigenstates and that the one-loop contributions
to their masses-squared are given by:
\beq
\mpzero^2({\rm one-loop}) = {4 \Lambda^{2}
\over \pi^{2} v^{2}}\rho_{8}\,,\quad
\mpzerop^{2}({\rm one-loop}) = {4 \Lambda^{2}
\over \pi^{2} v^{2}}\rho_{7}\,,
\label{masses}
\eeq
where
\beq
\rho_{8} = 2 m_{2}^\prime m_{9}
+2 m_{4}m_{10}+2 m_{4}^{(2)}m_{10}^{(2)}\,,\quad
\rho_{7} =  2 m_{1}^\prime m_{7}
\label{rhos}
\eeq
and $\Lambda$ is the $UV$ cut-off, situated in the TeV region.
Notice that, as promised,
$\pzero$ and $\pzerop$ receive no mass contributions
from the gauge interactions in (\ref{Lg}).  Further,
the $\pzero$ mass arises only
from the terms involving $b$, $\tau$ and $\mu$ in ${\cal L}_Y$
while the $\pzerop$ mass arises from terms
involving the $t$. Terms proportional to $m_i^2$
(\eg\ $m_2^{\prime\,2}$, $m_9^2$, $\ldots$  contributions to $\mpzero^2$)
are absent due to the cancellation between the standard
self-energy bubble diagrams resulting
from the `square' of ${\cal O}\left({1/ v}\right)$ terms
in the expansion of the $U$ fields appearing in ${\cal L}_Y$,
as given by using Eq.~(\ref{specialu}), and diagrams arising from
the quadratic couplings that first appear at order
${\cal O}\left({1/ v^2}\right)$ in the expansion
of the $U$ fields in ${\cal L}_Y$.

A useful point of reference for this discussion is
the one-family technicolor scheme of Refs.~\cite{techni,technir}. There,
the $\tilde\pi_3$ and $\pi_D$
states are given in terms of the $(U,D,N,E)$ techniquarks
and technileptons as:
\bea
\tilde \pi_3&=&{1\over\sqrt{24}}(U\anti U-D\anti D-3N\anti N+3E\anti E)
\,,\label{pi3form}\\
\pi_D&=&{1\over \sqrt{24}}(U\anti U+D\anti D-3N\anti N-3E\anti E)
\,,\label{pidform}
\eea
where there is an implicit color summation in the $U\anti U$
and $ D\anti D$ combinations. The $\pzero$ and $\pzerop$
mass eigenstates of Eq.~(\ref{eigenstates}) then take the forms:
\bea
\pzero&=&{1\over \sqrt{12}}(3E\anti E-D\anti D)\,,\label{pzeroform}\\
\pzerop&=&{1\over\sqrt{12}}(U\anti U-3N\anti N)\,,\label{pzeropform}
\eea
which are purely $T_3=-1/2$ and $T_3=+1/2$ isospin states, respectively.
Contributions to the $\pi_D$--$\tilde\pi_3$ mass-squared
matrix come from four-technifermion operators in the low-energy effective
theory that fail to commute with
the currents to which $\pi_D$ and $\tilde\pi_3$ couple.
The only operators that fail to commute with the $\pi_D$
and $\tilde \pi_3$ currents are ones involving two techniquarks and two
technileptons (see, for example, Refs.~\cite{drk} and \cite{technir}).
For instance, the combination $D\anti D E\anti E$ will give mass to
the $\pzero$. This explains the absence of squared masses in
the expressions for $\rho_8$ and $\rho_7$ discussed in
the previous paragraph. Since
Yukawa coupling contributions to PNGB masses must come from
effective four-technifermion operators that contain two techniquark
and two technilepton fields, only $m_fm_{f^\prime}$ cross
terms can contribute. Thus, for example, the $m_2m_9$ term
comes from a $U_{dd}$--$U_{ee}$ cross term that creates via a $b$-quark loop
a $\vev{D\anti D E\anti E}$ effective four-technifermion interaction.

We emphasize that it is the $\pzero$ and $\pzerop$ that are mass eigenstates
of the one-loop effective potential and not
the $\pi_D$ and $\tilde\pi_3$ (isosinglet and isotriplet) states.
This can be understood as being due to the fact that
the underlying $SU(2)_L\times SU(2)_R$ invariance of the techniquark
and technilepton sector of the
theory remains unbroken in the chiral limit (see also \cite{peskinmass}).
This implies that the $m^2$ operator that breaks the chiral symmetry must
commute with the charge operator $T_3\equiv T_{3L}+T_{3R}$
and, therefore, take the form
\beq
m^2=A{\bf 1}+BK_3+CT_3
\label{msqop}
\eeq
where $K_3\equiv T_{3L}-T_{3R}$ (which implicitly contains
a $-\gamma_5$) is the only matrix
other than the unit matrix and $T_3$ itself that
commutes with $T_3$.\footnote{The
generality of this mass matrix is most easily derived in the
$\pzero$--$\pzerop$ basis. From the requirement that $m^2$ and $Q$ commute,
$m^2$ must depend only on $T_{3L}$ and $T_{3R}$. The $\pzero$ and $\pzerop$
belong to the representation (1/2,1/2) of $SU(2)_L\times SU(2)_R$. Therefore,
for them, $T_{3L}^2=T_{3R}^2=-T_{3L}T_{3R}=1/4$ (where, for
the last equality remember that $T_{3L}+T_{3R}=0$ for the neutral states
of interest). Therefore, one needs to keep only linear terms in
$T_{3L}$ and $T_{3R}$ in $m^2$.}
In the $\tilde\pi_3$--$\pi_D$ basis, $m^2$ then necessarily takes the form
$m^2=\left(\matrix{A & B \cr B & A\cr}\right)$,
which is diagonalized in the $\pzero$--$\pzerop$ basis
with eigenvalues $A-B$ and $A+B$, respectively.
As discussed shortly, the $SU(2)_L\times SU(2)_R$ symmetry is preserved
by the other possibly important contributions to the mass-squared matrix,
so that $\pzero$ will, indeed, be a mass eigenstate.
This important result leads to significant
differences in the phenomenology of the $\pzero$ as compared
to what would be the case for $\tilde \pi_3$ and
$\pi_D$ being mass eigenstates.

In the $(U,D,N,E)$ one-family model,
the meaning of $A$ and $B$ of Eq.~(\ref{msqop}) becomes
very apparent.  One finds
\beq
A=m^2_{D,E}+m^2_{U,N}\,,\qquad B=m^2_{U,N}-m^2_{D,E}\,,
\label{decomp}
\eeq
where $m^2_{U,N}$ is the contribution coming from $\vev{U\anti
U}\vev{N\anti N}$ and $m^2_{D,E}$ is the contribution
from $\vev{D\anti D}\vev{E\anti E}$. After diagonalization,
one then finds
\beq
\mpzero^2=2m^2_{D,E}\,,\qquad \mpzerop^2 =2m^2_{U,N}\,.
\label{mpmasses}
\eeq
Since, in this model, $\vev{D\anti D}$ and $\vev{E\anti E}$
determine $m_b$, while $\vev{U\anti U}$ and $\vev{N\anti N}$
determine $m_t$, $\mpzero\ll\mpzerop$, as implicit
in Eqs.~(\ref{mf}), (\ref{masses}) and (\ref{rhos}),
is a very natural expectation.

Let us now go consider the possible four-technifermion operator contributions
to the $\pi_D$--$\tilde\pi_3$ mass-squared matrix in more generality.
In fact, there are only two important sources of such operators.  The first
is the one-loop effective potential, as discussed above, where
the masses-squared for the $\pzero$ and $\pzerop$ were explicitly noted
to have come from operators induced by SM quark and lepton loops
that mixed techniquark and technilepton bilinears.
These derive from the one-loop diagrams constructed from tree-level
exchanges of the extended technicolor (ETC) gauge bosons that mediate
$(U,D,N,E)$ to $(t,b,\nu,\tau)$ transitions.
The second contribution is that from
heavy technileptoquark (LQ) gauge boson exchange
at tree-level, where the LQ gauge bosons are those mediating
$N\to U$ and $E\to D$  transitions in $(U,D,N,E)$ techni-flavor
space.\footnote{These latter can be thought
of as being a subset of Pati-Salam-like $SU(4)$
group gauge bosons operating in the $(U_{1,2,3},N)$
and $(D_{1,2,3},E)$ subspaces (where we have exposed
the color indices $1,2,3$).} These ETC and LQ gauge bosons are
intimately connected by the generator algebra
\beq
[Q^a_L,Q^b_L]=if^{abc}Q^c_L\,,\quad  [Q^a_R,Q^b_R]=if^{abc}Q^c_R\,,\quad
[Q^a_R,Q^b_L]=0\,,
\label{calgebra}
\eeq
and related current algebra, where if, for example, $a$ and $b$
are chosen to give ETC $E\to b$ and $b\to D$ transitions, respectively,
then $f^{abc}$ will be non-zero for the LQ $E\to D$ transition.
This means that if ETC gauge bosons are present then the group
structure demands that the LQ gauge bosons also be present. Further,
they must have chiral structure correlated with that of the ETC gauge bosons.
Using the notation $g^L_{\rm LQ}$ and $g^R_{\rm LQ}$ for the
coefficients\footnote{These $g^{R,L}$ do not include the overall
gauge coupling constant, $g$.} defining the left-handed and right-handed
components of the LQ gauge boson currents, with similar
definitions for the ETC gauge boson currents, Eq.~(\ref{calgebra}) requires
$g^{R,L}_{\rm LQ}=[g^{R,L}_{\rm ETC}]^2$.

Explicit computation of the LQ
extended-technicolor gauge boson exchanges  (following
the techniques of Ref.~\cite{drk}) yields the result that
the LQ contributions to the
$\pi_D$--$\tilde\pi_3$ mass-squared matrix take the
same form as given in Eqs.~(\ref{msqop}), (\ref{decomp}) and (\ref{mpmasses}).
This result obtains
for any relative weight of $g^L_{\rm LQ}$ vs. $g^R_{\rm LQ}$.
Explicitly, the LQ contributions would contribute
(see Refs.~\cite{drk} and \cite{technir})
\beq
 m^2_{D,E}({\rm LQ})\sim g^2{g^L_{\rm LQ}g^R_{\rm LQ}\over m_{\rm LQ}^2}
{\vev{D\anti D}\vev{E\anti E}\over F_T^2}\,,
\label{msqlq}
\eeq
to $\mpzero^2$, with a similar contribution to $\mpzerop^2$
proportional to $\vev{U\anti U}\vev{N\anti N}$.
(Here, $m_{\rm LQ}$ specifies the mass of
the relevant technileptoquark type Pati-Salam gauge bosons and
$F_T$ represents the effective technipion
decay constant.) Thus, even after
including LQ exchanges, the $\pzero$ and $\pzerop$ are the mass eigenstates.
And, once again, the appearance of $\vev{D\anti D}\vev{E\anti E}$
in $m^2_{D,E}$ and of $\vev{U\anti U}\vev{N\anti N}$ in $m^2_{U,N}$
guarantees that the $\mpzero$ and $\mpzerop$
are most naturally related to the $m_b$ and $m_t$
SM-fermion masses, respectively. Because of the possibility
that LQ mass-squared contributions are significant, our
phenomenological studies will not assume a specific relation between the
mass of the $\pzero$ and its Yukawa coupling parameters.
That the mass eigenstate is the $\pzero$ as defined in
Eqs.~(\ref{eigenstates}) and (\ref{pzeroform}) will, however,
be quite important for the phenomenology.

A natural question is whether or not it is possible for the one-loop
mass-squared matrix contributions to dominate over the LQ contributions.
This would produce a very self-contained theory with all masses
as well as couplings determined by the one-loop effective potential.
We need to compare the result of Eq.~(\ref{msqlq})
to the one-loop effective potential terms. In our notation,
the latter are of order $\Lambda^2 m_fm_{f^\prime}/\pi^2 F_T^2$
(where we have used $v\sim 2F_T$, as appropriate
for this model). Using the fact that
$m_f\sim \vev{F\anti F}g^2g^L_{\rm ETC}g^R_{\rm ETC}/m_{\rm ETC}^2$,
where $F$ is the technifermion coupled to $f$ by the ETC gauge boson,
and writing $\Lambda\equiv \lambda m_{\rm ETC}$, the one-loop effective
potential contributes, for example,
\beq
m^2_{D,E}({\rm ETC})\sim {\lambda^2 g^4[g^L_{\rm ETC}g^R_{\rm ETC}]^2\over m_{\rm ETC}^2}
{\vev{D\anti D}\vev{E\anti E}\over \pi^2 F_T^2}\,.
\label{msqetc}
\eeq
In comparing Eq.~(\ref{msqetc}) to Eq.~(\ref{msqlq}), we first recall
that $g^{R,L}_{\rm LQ}=[g^{R,L}_{\rm ETC}]^2$, implying exactly the
same chiral structure for the two contributions. In particular,
one cannot have $m^2(LQ)\to 0$
as a result of $g^L_{\rm LQ}\to 0$ or $g^R_{\rm LQ}\to 0$
without killing the one-loop $m^2({\rm ETC})$ contribution at the same rate.
Further, one cannot take
$m_{\rm LQ}\gg m_{\rm ETC}$ since the ETC gauge bosons cannot
form a subgroup on their own without inclusion of the LQ gauge bosons;
most naturally, $m_{\rm LQ}\sim m_{\rm ETC}$.
Only if the product $\lambda^2 g^2/\pi^2$ is very large could
the one-loop mass contributions dominate over the LQ mass contributions.
Now, $\lambda^2$ implicitly contains the usual
one-loop $C_{\rm ETC}/16$ factor, where $C_{\rm ETC}$ could be computed
for a given ETC group for the two-ETC-gauge-boson exchange
diagram underlying Eq.~(\ref{msqetc}).
Writing $\lambda^2\equiv\kappa^2 C_{\rm ETC}/16$,
$\alpha_{\rm ETC}\equiv g^2/4\pi$,
and taking $C_{\rm ETC}/4\sim 1$ (a not atypical value), we would need to have
$\kappa^2 \alpha_{\rm ETC}/\pi\gg 1$ for one-loop dominance. If
the one-loop ETC contributions to  the mass-squared matrix dominate,
Eq.~(\ref{masses}) could then be used to determine $\Lambda$ from the measured
value of $\mpzero$ once all the parameters in $\rho_8$ of Eq.~(\ref{rhos})
have been determined from experiment. The determination of $\Lambda$
made assuming dominance of the one-loop ETC diagrams can be expected
to at least provide a rough idea (\ie\ to within
a factor of 3 or 4, see below) of the ETC gauge boson scale.

What is a natural expectation for $\lambda$ and thence
$\kappa^2\alpha_{\rm ETC}/\pi$?
In QCD chiral theory, the analogue of $\lambda$ is $4\pi f_\pi/\Lambda_{\rm
QCD}\sim [1.2\gev/300\mev]\sim 4$ \cite{weinberglambda,apich};
the corresponding
$\kappa^2\alpha_s/\pi$ is then of order $\sim [4]^2\times 4\alpha_s/\pi$, which
would, indeed, be much larger than 1.
Another example is provided by the Fermi theory, where the natural
cutoff is about $300\gev$ (the unitarity limit), whereas the mass
of the mediating gauge bosons is about $100\gev$, corresponding to $\lambda\sim
3$. Generally speaking, the cut-off scale at which the
effective theory breaks down is
significantly larger than the typical mass of the underlying
exchanged gauge boson(s). The only exception
arises if there is additional new physics at scales
just above the gauge boson mass scale. There is no requirement
in the context of ETC that this be the case, although it cannot be ruled out.
Thus, we believe it is quite possible that the one-loop mass contributions
could dominate $\mpzero^2$, in which case the physics of the $\pzero$
would be entirely determined by the effective Lagrangian and a determination
of $\Lambda$ would be possible.

Let us now turn to the couplings of the $\pzero$ to SM particles.
By inserting the expansion to  first order in the PNGB fields
given in Eq.~(\ref{specialu}) into
Eq. (\ref{Ly}), we easily obtain the couplings of the $\pzero$ and $\pzerop$
to fermions. Obviously,
the $\pzero$ boson couples to the $T_3=-1/2$ component
of the fermion doublets, while $\pzerop$ couples to the $T_3=+1/2$ component.
It is the $\pzero$ upon which we focus; its Yukawa couplings to
fermions are:
\be
{\cal L}_Y= -i \lambda_b \bar
b\gamma_5 b \pzero-i\lambda_\tau\bar\tau\gamma_5\tau \pzero
-i\lambda_\mu\bar\mu\gamma_5\mu \pzero
\ee
with
\beq
\lambda_b= -{\sqrt{6}\over 3v} \left(m_2^\prime-3m_9\right)\,,\quad
\lambda_\tau={\sqrt{6}\over v}\left(m_4-m_{10}\right)\,,\quad
\lambda_\mu={\sqrt{6}\over v} \left(m_4^{(2)}-m_{10}^{(2)}\right)\,.
\label{yukcoups}
\eeq
It is important to note that $\lambda_b$, $\mpzero$ [Eqs.~(\ref{masses}) and
(\ref{rhos})], and $m_b$ [Eq.~(\ref{mf})] all depend only
on the combination $m_2^\prime$ defined in Eq.~(\ref{mprimes}).
Thus, the complete set of parameters describing the $T_3=-1/2$
sector phenomenology comprises $m_2^\prime$,
$m_4$, $m_4^{(2)}$, $m_9$, $m_{10}$ and $m_{10}^{(2)}$, for a total
of six parameters. Since the above Yukawa couplings involve
different combinations of these six parameters than do
the fermion masses ($\mb$, $m_\tau$ and $m_\mu$), all six mass parameters
can be independently determined using the known fermion mass values
if experimental measurements of $\lambda_b$, $\lam_\mu$ and $\lam_\tau$
are available. Given a determination of the six $m_i$, the value of
$\rho_8$ in Eq.~(\ref{rhos}) can be computed and the measured $\mpzero$
will yield a result for $\Lambda$ through Eq.~(\ref{masses}) if
the LQ contribution to $\mpzero^2$ is small. Clearly, our ability
to measure the Yukawa couplings will be an important focus of our
phenomenological discussion.

As already apparent from Eq.~(\ref{mf}), the parameters $m_i$ are of the
order of the masses of the corresponding fermions. In order to explore
a representative phenomenological case, in Ref.~\cite{mumu}
we made the following choices:
\bea
& &m_1^\prime=m_7=\f{m_t} 2 \cr
& &m_2^\prime=m_9=\f{m_b} 2 \cr
& &m_{10}=-m_4=\f{m_{\tau}} 2 \cr
& &m_{10}^{(2)}=-m_4^{(2)}=\f{m_{\mu}} 2 \,.
\label{choice1}
\eea
This particular set of choices is based on the
assumption of no relevant cancellations.
The corresponding one-loop $\pzero$ and $\pzerop$ masses are
\beq
\mpzero^2({\rm one-loop}) = \frac{2\Lambda^2}{\pi^2v^2} m_b^2\,,\quad
\mpzerop^2({\rm one-loop}) = \frac{2\Lambda^2}{\pi^2v^2} m_t^2\,,
\label{mp0}
\eeq
where we have  neglected contributions to $\mpzero^2$ from the
one-loop potential proportional to $m_\mu^2$ and $m_\tau^2$.
The fermionic couplings of the $\pzero$ corresponding to
Eq.~(\ref{choice1}) are
\beq
\lambda_b = \sqrt{\frac 2 3 }\frac {m_b} v\,,\quad
\lambda_\tau = -\sqrt{6} \frac {m_\tau} v\,,\quad
\lambda_\mu = -\sqrt{6} \frac {m_\mu} v\,.
\label{pcoups}
\eeq
More generally, we would have $\lambda_f=\xi_f m_f/v$ with $\xi_f$ a
number of the order of 1 which depends on the particular choice of the Yukawa
parameters.

Our phenomenological analysis will be performed with
the choice of couplings to fermions given in Eq.~(\ref{pcoups}).
This is a purely representative choice that will allow us
to make a first assessment of the prospects
for determining {\it directly from experiment}
all those $m_i$ parameters in ${\cal L}_Y$
associated with the $\pzero$, as well as other important
parameters of the model.

Also of importance are the couplings of the $\pzero$
to a pair of SM gauge bosons arising
through the ABJ anomaly \cite{production,drk,ehlq,randa,chivu}.
These are model-dependent.  We will employ those obtained in the
standard technicolor theories of Ref.~\cite{production,drk,ehlq}.
The relevant Feynman-rule (which in our notation will include
double Wick contractions when two identical gauge bosons are present)
for such a coupling can be written in the general form:
\beq
g_{P V_1V_2}={\alpha \ntc A_{P V_1V_2}\over \pi v}
\eps_{\lam\mu\nu\rho}p_1^\lam\eps_1^\mu p_2^\nu\eps_2^\rho\,,
\label{pvvcoups}
\eeq
where for $P=\pzero$ we have:
\bea
\label{pgamgamcoup}
A_{\pzero \gam\gam}&=&-{4\over \sqrt 6} \left({4\over 3}\right) \\
\label{pzgamcoup}
A_{\pzero Z\gam}&=&-{4\over 2\sqrt 6} \left( {1-4\sw^2\over 4\sw\cw }-{\tw\over
3}\right) \\
\label{pzzcoup}
A_{\pzero Z Z}&=&-{4\over \sqrt 6}\left({1-2\sw^2\over 2\cw^2}-
{\tw^2\over 3}\right) \\
\label{pggcoup}
A_{\pzero gg}&=&{1\over \sqrt 6}      \,,
\eea
where $\sw=\sin\theta_W$, \etc\
Similar results apply for the $\pzerop$, but are not needed for our
phenomenological studies that will be confined to the lighter $\pzero$.
We note that the couplings above are precisely those obtained
in the one-family $(U,D,N,E)$ technicolor model when
$F_T$ of the model is chosen so as to obtain correct
values for $\mw$ and $\mz$: namely $F_T=v/2$ (with $v=246\gev$).
These ABJ anomaly couplings also provide a nice illustration of
the importance of knowing the mass eigenstate composition.  For example,
the very crucial $\gam\gam$ and $gg$ couplings-squared for
the $\pzero$, $\pi_D$ and $\tilde\pi_3$ are in the ratio
$A_{\pzero\gam\gam}^2:A_{\pi_D\gam\gam}^2:A_{\tilde\pi_3\gam\gam}^2=8:1:9$
and $A_{\pzero gg}^2:A_{\pi_D gg}^2:A_{\tilde\pi_3 gg}^2=1:2:0$,
respectively, implying very different phenomenology were the $\pi_D$
or $\tilde\pi_3$ the lowest mass eigenstate.

We noted earlier that in the multi-scale/walking
technicolor context the value
of $v$ appropriate for determining the $\pzero$ couplings
could be smaller than $v=246\gev$.  From the above explicit formulae,
it is apparent that all couplings of interest are proportional
to $1/v$, implying that a decrease in $v$ could only increase
production rates for the $\pzero$ and, thereby, our
ability to discover and study the $\pzero$ in the various channels
that will be discussed later.

Finally, it is useful to consider the limit in which LQ mass-squared
matrix contributions can be neglected relative
to the one-loop effective potential
contributions of Eq.~(\ref{mp0}).
The magnitude of $\mpzero$ in this limit can
be better appreciated by writing the one-loop contribution 
to $\mpzero$ from Eq.~(\ref{mp0}) in the form
\beq
\mpzero({\rm one-loop})\sim 8\gev \times\Lambda({\rm TeV})\,.
\label{mp02nd}
\eeq
Given that $\Lambda<10\tev$ is most natural in the model being considered,
the $\pzero$ would be likely to have mass below $\mz$. Only if $\Lambda$
is unexpectedly large and/or the LQ contributions
are very substantial is it possible
that $\mpzero$ would be larger than $\sim
200\gev$.\footnote{Walking/multi-scale technicolor models would
have smaller $v$ which would enhance the one-loop contributions
to $\mpzero^2$ and $\mpzerop^2$ and could also lead
to $\mpzero$ values above $200\gev$.} In contrast, because
$\mpzerop/\mpzero$ is most naturally $\propto m_t/m_b$, the
$\pzerop$ would be unlikely to have mass below $\sim 250\gev$, and,
if $\Lambda$ is large and/or LQ contributions substantial,
it would be a good deal heavier.

It is because the mass of the $\pzero$ is most naturally small, in particular
almost certainly much smaller than that of the $\pzerop$, and because
the $\pzerop$ cannot be studied in
the $s$-channel resonance mode at a muon collider (due to the fact
that it does not couple to charged leptons),
that we choose to focus on the phenomenology of the $\pzero$ in this paper.
We will give detailed results for $10\leq \mpzero\leq 200\gev$.

\begin{figure}[p]
\epsfysize=8truecm
\centerline{\epsffile{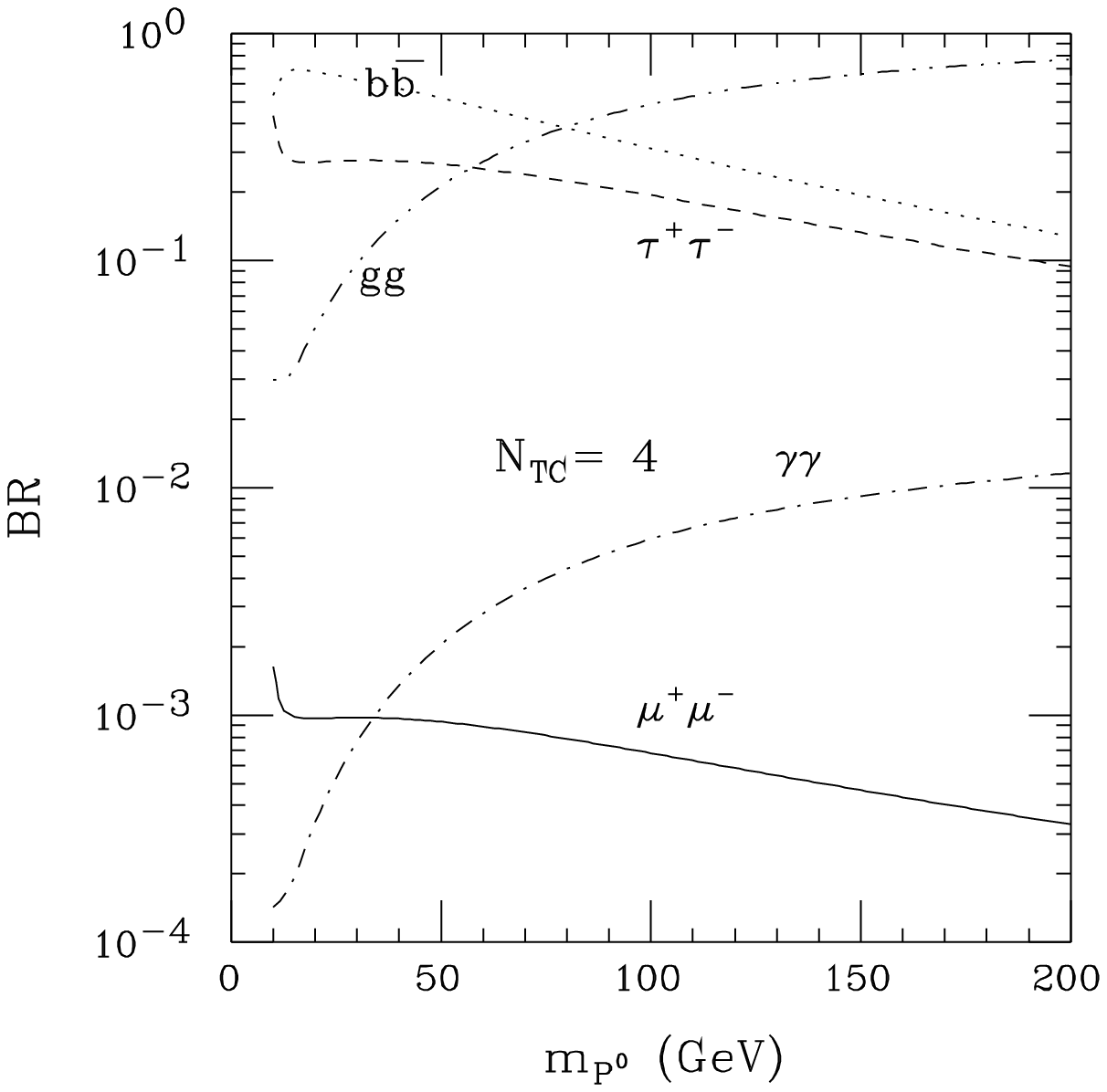}}
\smallskip
\noindent
\caption{Branching fractions for $\pzero$ decay into $\mu^+\mu^-$,
$\tau^+\tau^-$, $b\bar b$, $\gamma\gamma$, and $gg$.
We assume $\ntc=4$ and
employ the couplings of Eqs.~(\ref{pcoups}), (\ref{pgamgamcoup})
and (\ref{pggcoup}).}
\label{figbrs}
\bigskip
\epsfysize=8truecm
\centerline{\epsffile{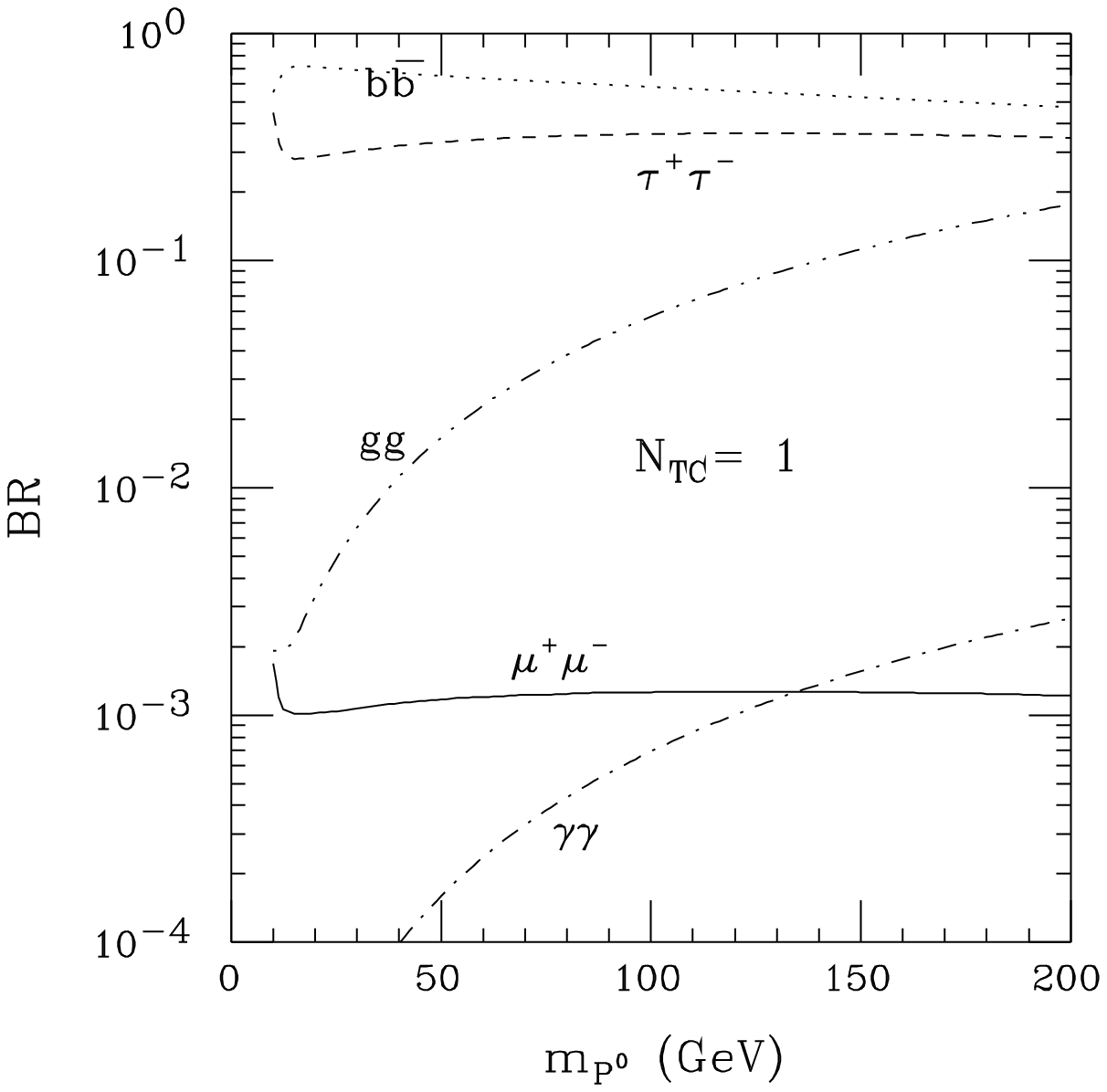}}
\smallskip
\noindent
\caption{As in Fig.~\ref{figbrs}, but for $\ntc=1$.}
\label{figbrsntceq1}
\end{figure}

\begin{figure}[h]
\epsfysize=8truecm
\centerline{\epsffile{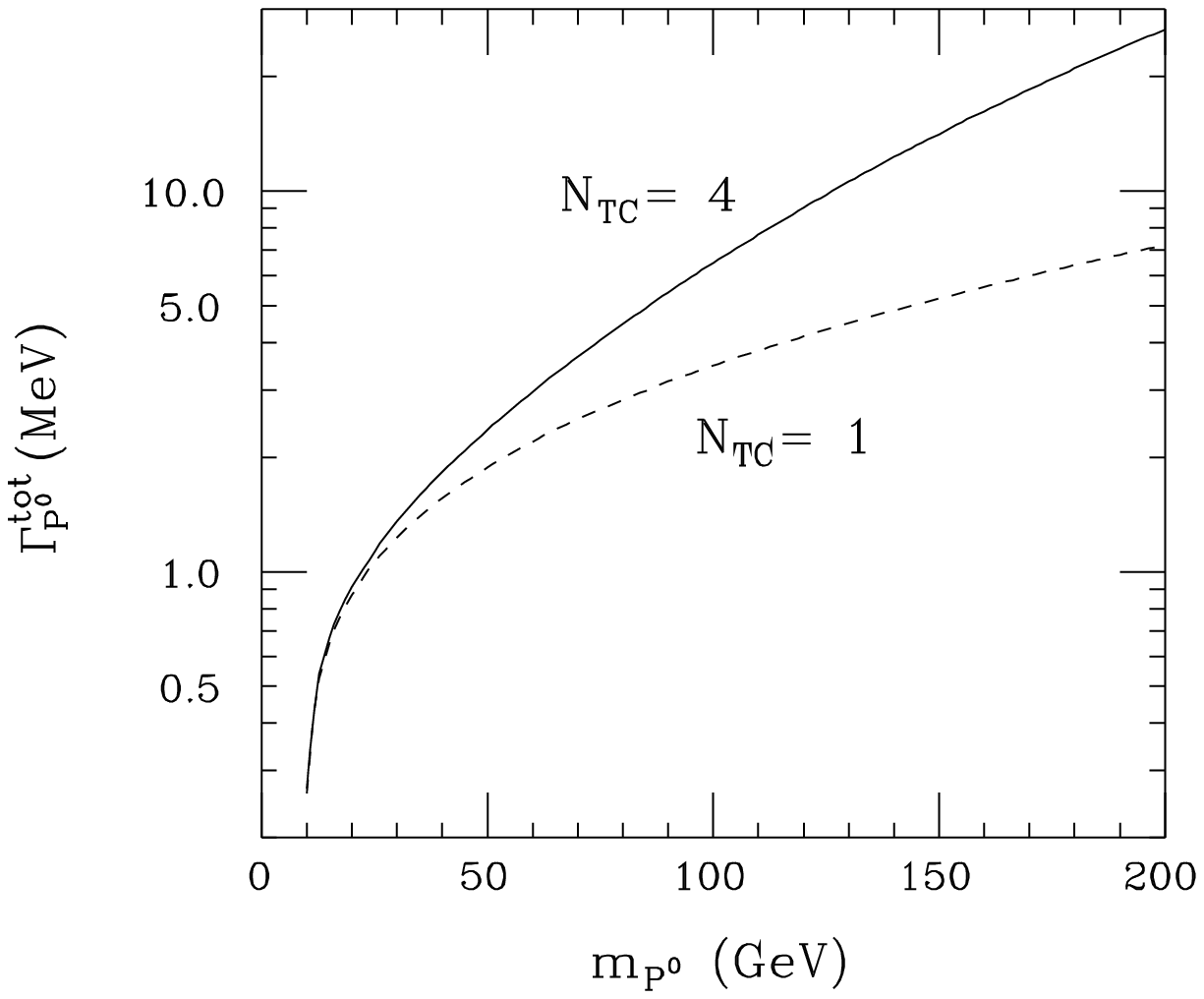}}
\smallskip
\noindent
\caption{$\gampzero$ as a function of $\mpzero$ for $\ntc=4$
and $\ntc=1$. We
employ the couplings of Eqs.~(\ref{pcoups}), (\ref{pgamgamcoup})
(\ref{pggcoup}).}
\label{figgamtot}
\end{figure}

\section{$\pzero$ branching fractions and width}

In this section, we give a brief summary of the branching fractions
and total width of the light $\pzero$.
The $\pzero$ Yukawa couplings to fermions are given in Eq. (\ref{pcoups}).
For $\pzero$ decays, the $\gamma\gamma$ and
gluon-gluon channels are also important; the corresponding
couplings are those summarized in Eqs.~(\ref{pgamgamcoup})
and (\ref{pggcoup}), as generated by the ABJ anomaly.
The corresponding partial widths must be computed
keeping in mind that, for our normalization of
$A_{\pzero\gam\gam}$ and $A_{\pzero gg}$, one must include a factor of 1/2
for identical final state particles:
\beq
\Gamma(\pzero\to VV)=\half C_V {\mpzero^3\over 32\pi}A_{\pzero VV}^2\,,
\label{ptovv}
\eeq
where $C_V=1~(8)$ for $V=\gam$ ($g$).
We list here those partial widths relevant for our analysis:
\bea
\Gamma(\pzero\to \bar f f) &=& C_F \frac {\mpzero}{8\pi} \lambda_f^2
\left(1-\frac
{4 m_f^2} {\mpzero^2} \right)^{1/2}\nn\\
\Gamma (\pzero\to gg)&=& \frac {\alpha_s^2}{48\pi^3 v^2}
\ntc^2\mpzero^3\nn\\
\Gamma (\pzero\to\gamma\gamma)&=& \frac {2\alpha^2}{27\pi^3 v^2}
\ntc^2\mpzero^3\,,
\eea
where $C_F=1(3)$ for leptons (down-type quarks) and $\ntc$ is the number of
technicolors.

\begin{figure}[h]
\epsfysize=8truecm
\centerline{\epsffile{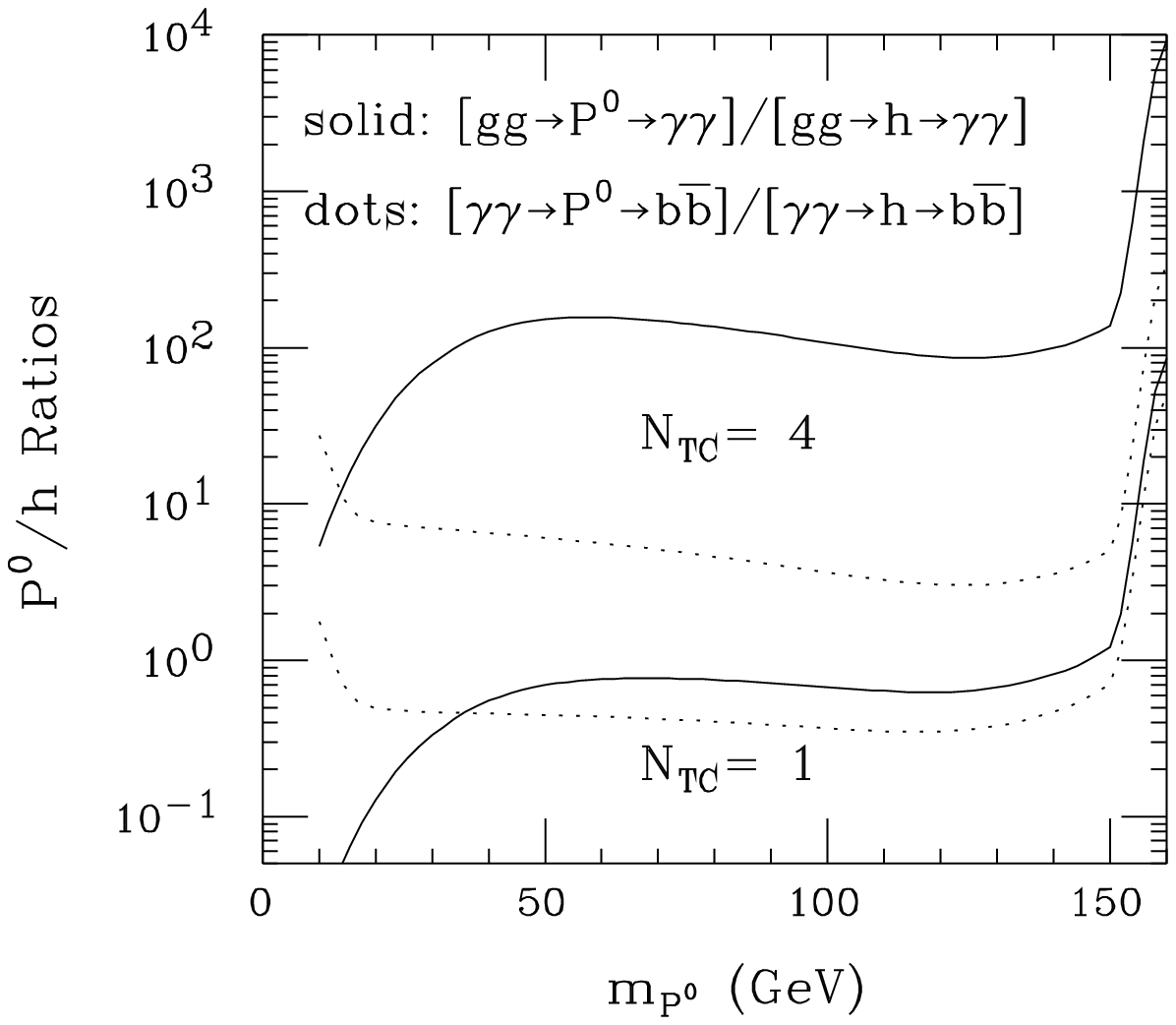}}
\smallskip
\noindent
\caption{The ratios $[\Gamma(\pzero\to gg)\br(\pzero\to \gam\gam)]/
[\Gamma(h\to gg)\br(h\to \gam\gam)]$ (solid curves) and
$[\Gamma(\pzero\to \gam\gam)\br(\pzero\to b\anti b)]/
[\Gamma(h\to \gam\gam)\br(h\to b\anti b)]$ (dotted curves), where $h$
is the SM Higgs boson, are plotted as a function
of $\mpzero$, taking $m_h=\mpzero$. Results are
given for $\ntc=4$ and $\ntc=1$ using
the specific $\pzero$ couplings of Eqs.~(\ref{pcoups}),
(\ref{pgamgamcoup}) and (\ref{pggcoup}).}
\label{fighpgbcomp}
\end{figure}

The resulting branching fractions for $\ntc=4$ and $\ntc=1$ are shown in
Figs.~\ref{figbrs} and \ref{figbrsntceq1}, respectively,
and the corresponding total widths are shown in Fig.~\ref{figgamtot}.
We see that the largest branching fractions are to $b\anti b$, $\tauptaum$
and $gg$. The total width, which is of particular relevance to
muon collider phenomenology, is typically in the few MeV range,
which is similar to that expected for a light SM-like Higgs boson.

There are two important features of these results that
we wish to point out: (a) the ratio of
$\Gamma(\pzero\to gg)$ to $\Gamma(h\to gg)$ is roughly
given by $1.5\ntc^2$; (b) the ratio of
$\br(\pzero\to\gam\gam)$ to $\br(h\to \gam\gam)$ is
of order 4 for $50\leq \mpzero\leq 150\gev$ if $\ntc=4$,
but substantially smaller if $\ntc=1$.
If $\ntc$ and/or $\mpzero$ is large enough that $\pzero\to gg$
is the dominant decay mode (see Fig.~\ref{figbrs}), then
$\br(\pzero\to\gam\gam)$ becomes
independent of $\ntc$ while $\Gamma(\pzero\to gg)$ is proportional
to $\ntc^2$, yielding
\beq
\Gamma(\pzero\to gg)\br(\pzero\to\gam\gam)\to {2\alpha^2\over
27\pi^3}{\ntc^2\mpzero^3\over v^2}\sim 2.4\times 10^{-3}~{\rm MeV}\ntc^2
\left({\mpzero\over 100\gev}\right)^3\,,
\label{gambrlimit}
\eeq
which, for $\ntc=4$,
is typically much larger than the corresponding result for
a SM-like Higgs boson. This will make $\pzero$ discovery
in the $\gam\gam$ final state at a hadron collider
much easier than in the SM Higgs case when $\ntc=4$.
Similarly, for $\ntc=4$, one finds a larger
value of $\Gamma(\pzero\to \gam\gam)\br(\pzero\to b\anti b)$
as compared to the SM $h$ analogue. This implies that
discovery of the $\pzero$ in $\gam\gam$ collisions will be much
easier than for a SM Higgs boson.
The large values of the $\pzero/h$ ratios (when $\ntc=4$)
for these two $\Gamma\br$ products are illustrated
in Fig.~\ref{fighpgbcomp}. Of course, both ratios are smaller
for smaller $\ntc$. In the minimal $\ntc=1$ case, these two ratios
are both of order 0.4 to 0.9 for $30\leq \mpzero\leq 150\gev$, implying
that the ability to detect the $\pzero$ would be about
the same as for the SM Higgs boson over this mass range.

We note that these ratios are quite
sensitive to the composition of the $\pzero$ mass-eigenstate.
In particular, if the $\pi_D$ isosinglet state were
the mass-eigenstate, the phenomenologically relevant products
$\Gamma(\pi_D\to gg)\br(\pi_D\to\gam\gam)$ and
$\Gamma(\pi_D\to\gam\gam)\br(\pi_D\to b\anti b)$ would be
smaller by factors of about 8 and 8/3, respectively,
compared to their $\pzero$ counterparts.  Thus, for $\ntc=4$
(but not for $\ntc=1$), both products would
still be larger than their respective SM Higgs analogues, but
not so dramatically as for the $\pzero$ eigenstate.

It is the possibility of large $\pzero/h$
ratios for the $\Gamma\br$ products
discussed above in combination with
the fact that the $\pzero$ can naturally have a low mass
that makes the $\pzero$ such a unique object.
With this summary of the couplings and basic properties of the $\pzero$,
we are now in a position to consider its phenomenology at various colliders.

\section{$\pzero$ production at the Tevatron and the LHC}

In this section, we discuss the ability of the Tevatron
and LHC to discover the PNGB's. We focus primarily
on the $\pzero$ since, as we have seen, it is the lightest of the PNGB's
and, therefore, possibly the only PNGB that can be produced
and studied at first-generation $\epem$ and $\mupmum$ colliders.

The most important discovery mode for the $\pzero$
at the Tevatron and the LHC is almost certainly production via inclusive
$gg\to\pzero$ fusion followed by $\pzero\to\gam\gam$ decay.
This is the same production/decay mode as
considered for a light Standard Model Higgs boson, $h$.
Indeed, although this mode has hitherto received little attention, it
is likely to be even more robust for the $\pzero$ than for the SM $h$,
and is possibly viable at the Tevatron as well as at the LHC.
This is because the rate
for the $gg\to R\to\gam\gam$ channel ($R=\pzero$ or $h$)
is proportional to $\Gamma(R\to gg)\br(R\to \gam\gam)$
and, as illustrated in Fig.~\ref{fighpgbcomp} in the previous section,
both factors are larger for the $\pzero$ than for the $h$
unless $\ntc=1$. For example, for the modest value of $\ntc=4$,
the $\pzero/h$ ratio is greater than 10 for $\mpzero\gsim 12\gev$ rising
to $>100$ once $\mpzero>30$. For large $\mpzero\gsim 150\gev$, the
ratio becomes very large as a result of the precipitous decline
of $\br(h\to\gam\gam)$ for the SM $h$ due to the opening
up of the $h\to WW,ZZ$ decay modes, whereas for the $\pzero$
the $WW$ mode is absent (due
to the vanishing of the ABJ anomaly for $SU(N)$ type technicolor models)
and the $ZZ$ mode is negligible.
In fact, Figs.~\ref{figbrs} and \ref{figbrsntceq1}
and Eq.~(\ref{gambrlimit}) make
it clear that $\Gamma(\pzero\to
gg)\br(\pzero\to\gam\gam)$ increases at least as
fast as $\mpzero^3$ for large $\mpzero$. As a result,
the signal rate decreases at large $\mpzero$ only as a consequence
of phase space and decreasing gluon distribution functions.

Let us first consider the LHC. We recall that
the LHC ATLAS \cite{ATLAS} and CMS \cite{CMS}
studies claim that the SM Higgs boson
can be discovered using the $gg\to h\to\gam\gam$
mode in the $m_h>70\gev$ mass region, extending
up to about $150\gev$, where the
upper limit derives entirely from the steep decline in $\br(h\to\gam\gam)$
that occurs when the $h\to WW,ZZ$ decay modes open up. From
Fig.~\ref{fighpgbcomp} we feel it is safe to conclude
that, for $\ntc\geq 4$,
the $\pzero$ can be detected in the $gg\to\pzero\to\gam\gam$ mode
for at least $50<\mpzero<200\gev$. The background studies
by the detector collaborations required to determine if the discovery region
is even larger are not available. We expect that
discovery will be possible for a substantial range above $200\gev$ because
of the rapid decline in the background rate as $M_{\gam\gam}$
increases. Extension of the discovery range to lower $\mpzero<50\gev$
will be limited by the rapid rise in the irreducible $\gam\gam$ background
at small $M_{\gam\gam}$
and the increased difficulty in rejecting the reducible $\gam$-jet
and jet-jet backgrounds that arise when the jets fluctuate so
as to look like a photon. For $\ntc=4$, we estimate that discovery reach will
extend down at least as low as 30 GeV (below which the $\pzero/h$
ratio starts to decline rapidly.)
A detailed study by the ATLAS and CMS collaborations is in order.

Of course, it is also interesting to determine the $\pzero$ discovery
reach if $\ntc$ is small, as would be preferred for
consistency with the precision electroweak $S$ parameter
if the technifermions must be counted as real particles as opposed
to a convenient mnemonic for understanding the structure of the
low-energy effective theory.
We have computed the $\pzero$ signal rate for $\ntc=1$
over the $70\gev$ to $150\gev$ range and find that $S/\sqrt B$ 
rises from about 5.8 to 8.1 over this range. Further, since there 
is no precipitous
decline in $\br(\pzero\to\gam\gam)$ for $\mpzero$ values above $150\gev$
(unlike the Higgs case where $\br(h\to\gam\gam)$ declines precipitously
as the $WW$ and $ZZ$ modes open up), the signal rate $S$ remains roughly
constant (for example, $S=80$ for the CMS
detector, using the inputs described later for $L=160\fbi$) up
to at least $\mpzero=200\gev$. In addition, the $\gam\gam$
background rate declines significantly as the mass increases.
Thus, it is clear that $S/\sqrt B$ will continue to rise as
$\mpzero$ increases above $150\gev$ and
we conclude that $\pzero$ detection will be possible
at the LHC for $70\gev\leq\mpzero\leq 200\gev$, and probably higher,
even for the minimal reference value of $\ntc=1$.

\begin{figure}[h]
\epsfysize=8truecm
\centerline{\epsffile{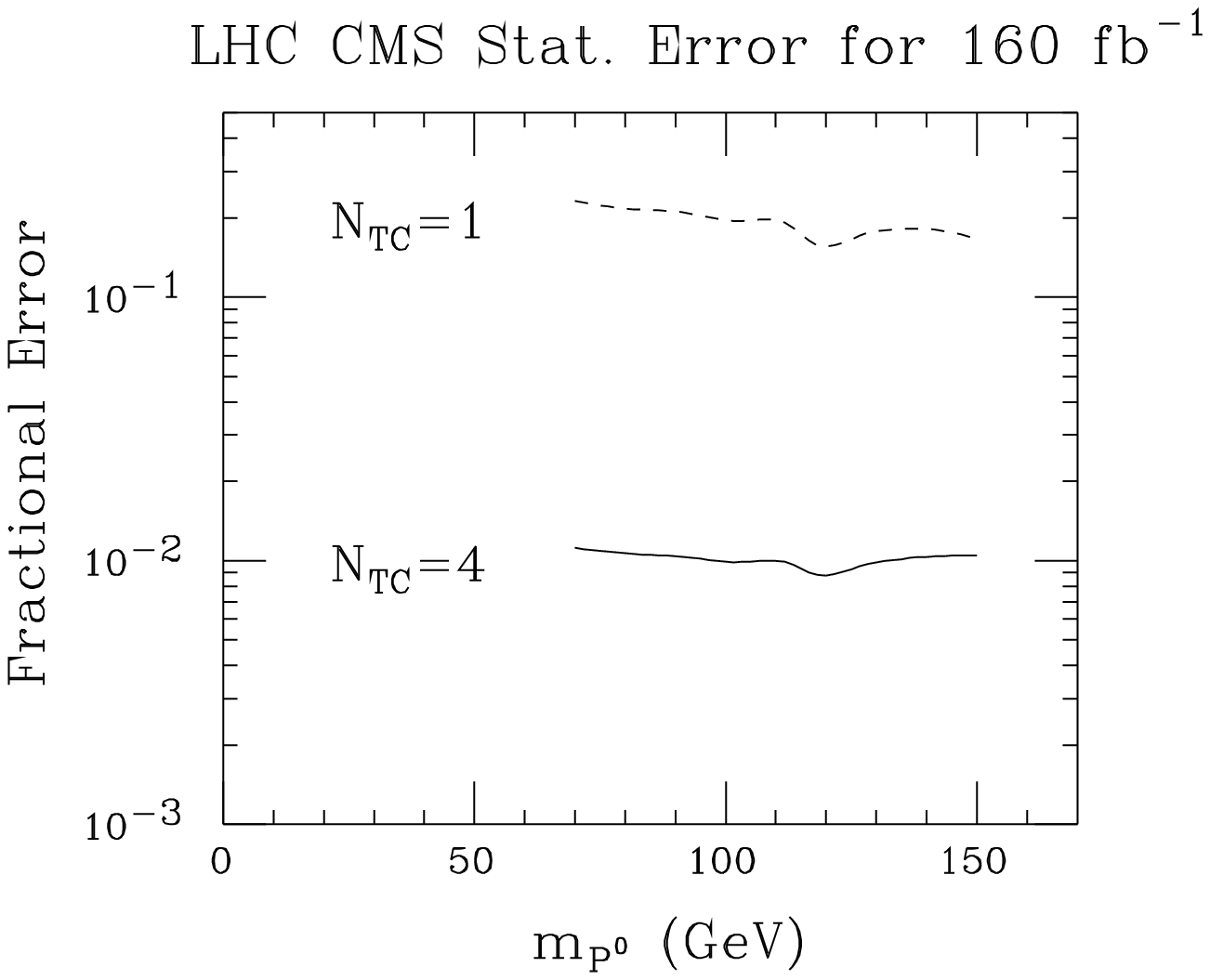}}
\smallskip
\noindent
\caption{The fractional statistical error $(S+B)^{1/2}/S$
for the $gg\to\pzero\to\gam\gam$ rate, for $\ntc=4$
and $\ntc=1$ assuming $L=160\fbi$
of accumulated luminosity at the LHC. Results are based on the rates
quoted by CMS for the SM Higgs boson
when only events containing at least two high $E_T$ (central)
or high $E$ (forward) jets are accepted. The $gg\to\pzero\to \gam\gam$
signal rate $S$ is computed from the SM Higgs rate by multiplying
by the appropriate factor at each $\mpzero$, as plotted in
Fig.~\ref{fighpgbcomp}.}
\label{figcmserrors}
\end{figure}

As we shall see in later sections,
the $gg\to\pzero\to\gam\gam$ LHC mode will be an important ingredient
in determining the parameters of the technicolor model.
Thus, we wish to estimate the statistical errors for the measurement
of the associated rate. We consider
$\pzero$ masses for which the appropriate LHC background and $M_{\gam\gam}$
resolution results are available. Since the CMS $M_{\gam\gam}$
resolution is typically better
than that which will be achieved by ATLAS, we will employ the
results of the CMS technical proposal \cite{CMS} from
their Table~12.2, which covers the mass range $70-150\gev$
and assumes an integrated luminosity of $L=160\fbi$.
The signal and background rates quoted in Table~12.2
are those for the preferred method in which
one accepts only $\gam\gam$ events that contain at least
two jets with either high $E_T$ (for central jets)
or high $E$ (for forward jets). Detailed requirements
may be found in Ref.~\cite{CMS}.
At each mass, we use the $S$ value for the SM 
Higgs boson from the CMS Table~12.2 and the value of $[\Gamma(\pzero\to
gg)\br(\pzero\to\gam\gam)]/[\Gamma(h\to gg)\br(h\to \gam\gam)]$
(see Fig.~\ref{fighpgbcomp})
to compute the expected $S$ for the $\pzero$.  We then
use this modified $S$ and the $B$ value from the CMS Table~12.2
to compute $(S+B)^{1/2}/S$.
The resulting fractional statistical error is plotted as a function
of $\mpzero$ in Fig.~\ref{figcmserrors} for both $\ntc=4$ and $\ntc=1$.
For $\ntc=4$, we find a statistical error for measurement
of the $gg\to\pzero\to\gam\gam$ rate that is essentially constant
at $\sim 1\%$. For $\ntc=1$, the error is of order $20\%$.
It seems likely that these errors will apply
beyond the mass range for which specific CMS SM Higgs boson
studies have been performed, with deterioration expected at small
$\mpzero$.  Overall, we can anticipate
that the $gg\to\pzero\to\gam\gam$ rate will be measured with
substantial statistical accuracy for almost the entire mass range
for which discovery is possible.
Indeed, for $\ntc=4$, systematic uncertainty associated with normalizing
the absolute rate (after cuts) will almost certainly dominate
for all but small $\mpzero$.

\begin{figure}[h]
\epsfysize=7truecm
\centerline{\epsffile{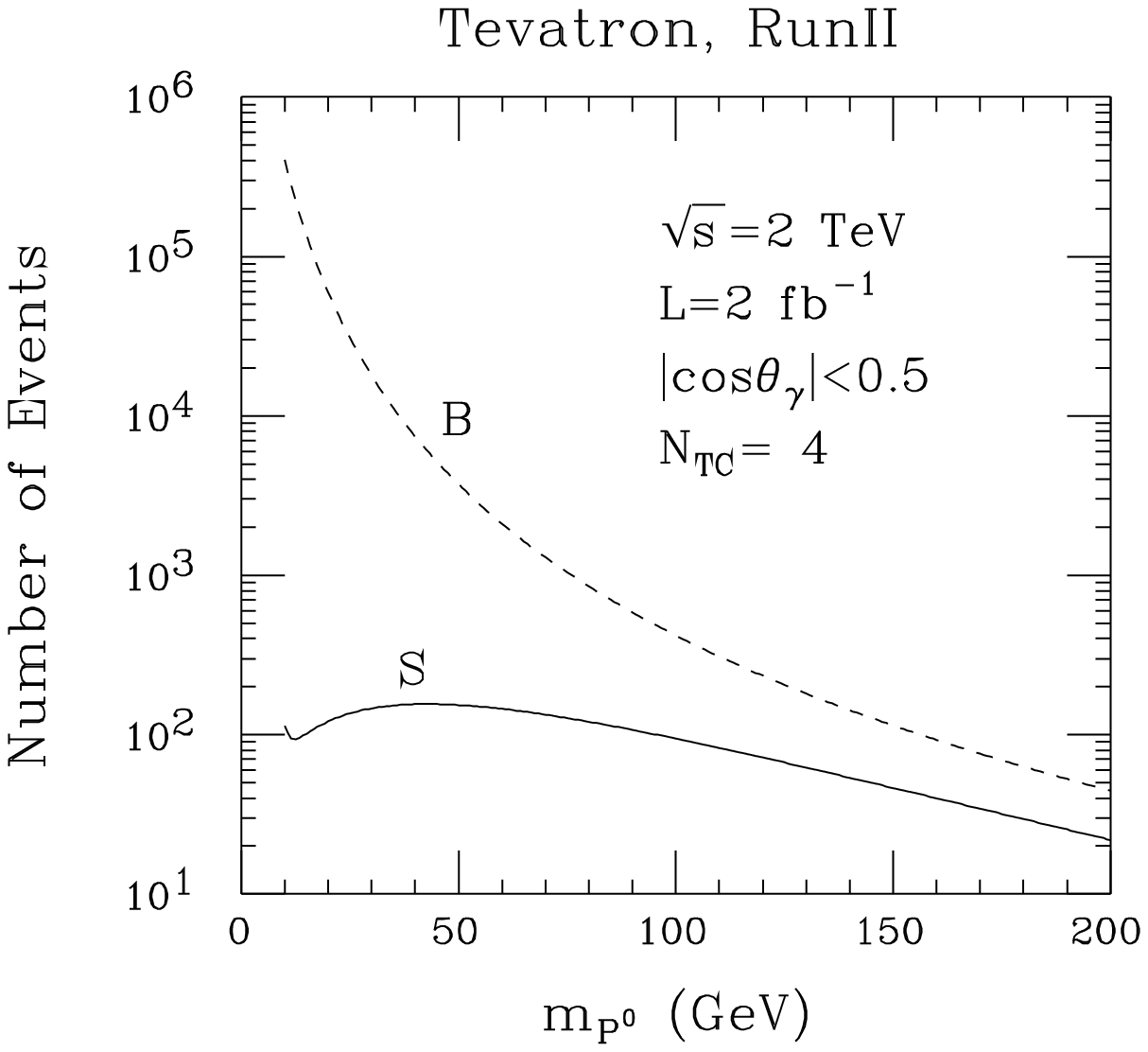}}
\smallskip
\noindent
\caption{The $gg\to\pzero\to\gam\gam$ signal rate $S$
is plotted as a function of $\mpzero$
for $\ntc=4$ in the case of $\protect\rts=2\tev$ and $L=2\fbi$ at the
Tevatron. Also shown is the irreducible background
rate, $B$, obtained by multiplying
$d\sigma(p\anti p\to q\anti q\to\gam\gam)/dM_{\gam\gam}$
by a factor of 1.5 to include the $gg\to \gam\gam$ loop contribution
and by a mass interval size of
$2\Delta M_{\gam\gam}$, where $\Delta M_{\gam\gam}$
is computed for $\Delta E_\gam/E_\gam=0.135/\protect\sqrt{E_\gam}\oplus 0.015$.
In computing both $S$ and $B$,
we required the final state photons to have $|\cos\theta_\gam|<0.5$.}
\label{figtevatrongamgamrates}
\end{figure}

If $\ntc=4$, given the substantial enhancement of
the $gg\to\pzero\to\gam\gam$ rate relative to the case of the SM $h$,
it is also possible that this mode is viable at the Tevatron,
even though for the SM $h$ (and for $\ntc=1$) it is not.
In order to assess the prospects, we have computed signal and irreducible
background rates for RunII parameters, requiring that both of the final-state
photons have $|\cos\theta_\gam|<0.5$.
The $\pzero$ event rate for $\ntc=4$
is plotted in Fig.~\ref{figtevatrongamgamrates}; the rate is
sizeable for the assumed $L=2\fbi$. (In contrast, for $\ntc=1$, $S<1$ for
$L=2\fbi$ for all $\mpzero$.) The background rate shown is that
computed using $B=1.5[d\sigma(p\anti p\to q\anti
q\to\gam\gam)/dM_{\gam\gam}][2\Delta M_{\gam\gam}]$,
where we use a $\gam\gam$ invariant mass window of
$2\Delta M_{\gam\gam}$ with $\Delta M_{\gam\gam}/M_{\gam\gam}
={1\over\sqrt 2}\Delta E_\gam/E_\gam$. Here, we employ
$\Delta E_\gam/E_\gam=0.135/\sqrt{E_\gam}\oplus 0.015$ \cite{frisch}
with $E_\gam=\mpzero/2$. The factor of 1.5 in $B$ above
is included in order to approximately account
for the well-known \cite{ggloop} $p\anti p\to
gg\to\gam\gam$ one-loop contribution
to the inclusive $\gam\gam$ background.

In addition to the irreducible $\gam\gam$ background,
there may be reducible backgrounds 
deriving from $\gam j$ and $jj$ final states in which the $j$ fragments in
such a way that it is identified as a $\gam$ (\eg\ $j\to$ fast $\pi^0$)
\cite{gamjet}. These must be eliminated/reduced by using very tight
cuts and requirements to define a (prompt) isolated photon.
Roughly, it is necessary that
the cuts discriminate at a level of one part
in $10^4$ against a
jet which fluctuates in such a way as to look like a photon
as opposed to a directly produced photon.
ALEPH and RunI Tevatron results \cite{tevgamjet}
suggest that the per-jet discrimination factor
of $\sim 10^{-4}$ can be achieved
for relatively energetic photons.  However, at low $\mpzero$
the photon energies are not large and the jet-$\gam$
discrimination factor worsens.
A detailed detector simulation is required in order to assess
the magnitude of such reducible backgrounds as a function
of $\mpzero$. Here, we note only that both $S$
and the irreducible background $B$ will be reduced
compared to Fig.~\ref{figtevatrongamgamrates} by
the efficiency of the cuts/requirements imposed in order to eliminate
these reducible $j\to\gam$ fluctuation backgrounds.

\begin{figure}[h]
\epsfysize=7truecm
\centerline{\epsffile{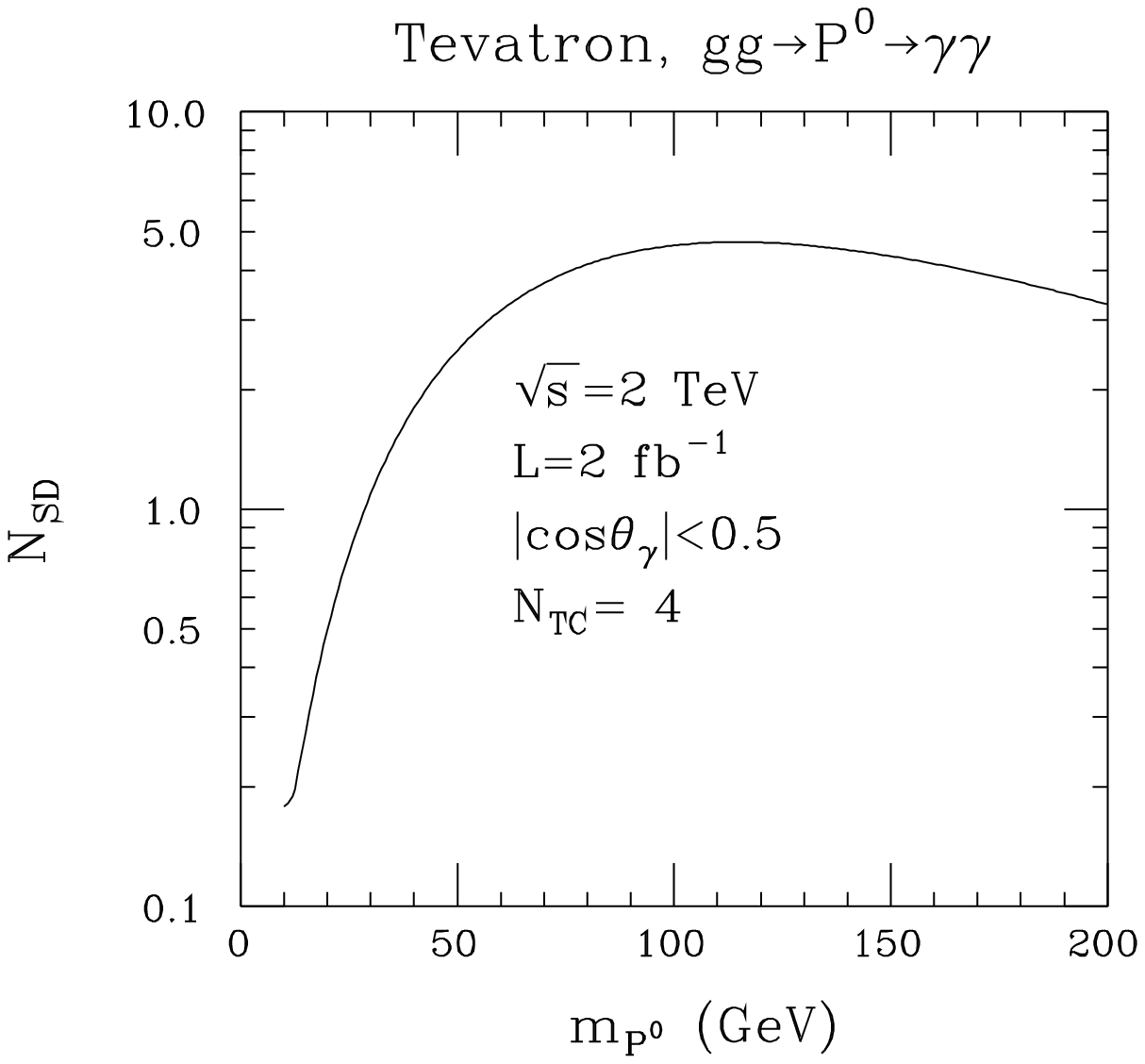}}
\smallskip
\noindent
\caption{The $gg\to\pzero\to\gam\gam$ signal significance,
$\nsd=S/\protect\sqrt B$, is plotted as a function of $\mpzero$
for $\protect\rts=2\tev$ and $L=2\fbi$ at the Tevatron using $S$
and $B$ as plotted in Fig.~\protect\ref{figtevatrongamgamrates}.}
\label{figtevatrongamgamnsd}
\end{figure}

As a very first indication of whether $\pzero$ discovery
during RunII at the
Tevatron is possible for $\ntc=4$ in the $\gam\gam$ mode, we present
in Fig.~\ref{figtevatrongamgamnsd} results for
$\nsd=S/\sqrt{B}$, computed
using the signal and background rates $S$ and $B$
plotted in Fig.~\ref{figtevatrongamgamrates} for $L=2\fbi$.
We see that $\nsd\gsim 3$ is achieved for $\mpzero\gsim 60\gev$.
For $\mpzero\sim 60,100,150,200\gev$, the statistical fractional error
for $\Gamma(\pzero\to gg)\br(\pzero\to \gam\gam)$
(computed as $\sqrt{S+B}/S$) is $0.33,0.24,0.27,0.37$, respectively.

For TeV33 luminosity of $L=30\fbi$, it may prove possible
to probe lower $\mpzero$, pending the
outcome of detailed studies of mass resolution and jet-$\gam$
discrimination at lower $M_{\gam\gam}$. At TeV33,
discovery of the $\pzero$ would certainly be possible
for $\mpzero\gsim 60\gev$, and
the factor of 15 increase in luminosity would allow reasonable precision
($\pm 5\%-\pm 10\%$) for
the determination of $\Gamma(\pzero\to gg)\br(\pzero\to \gam\gam)$.

\begin{figure}[h]
\epsfysize=7truecm
\centerline{\epsffile{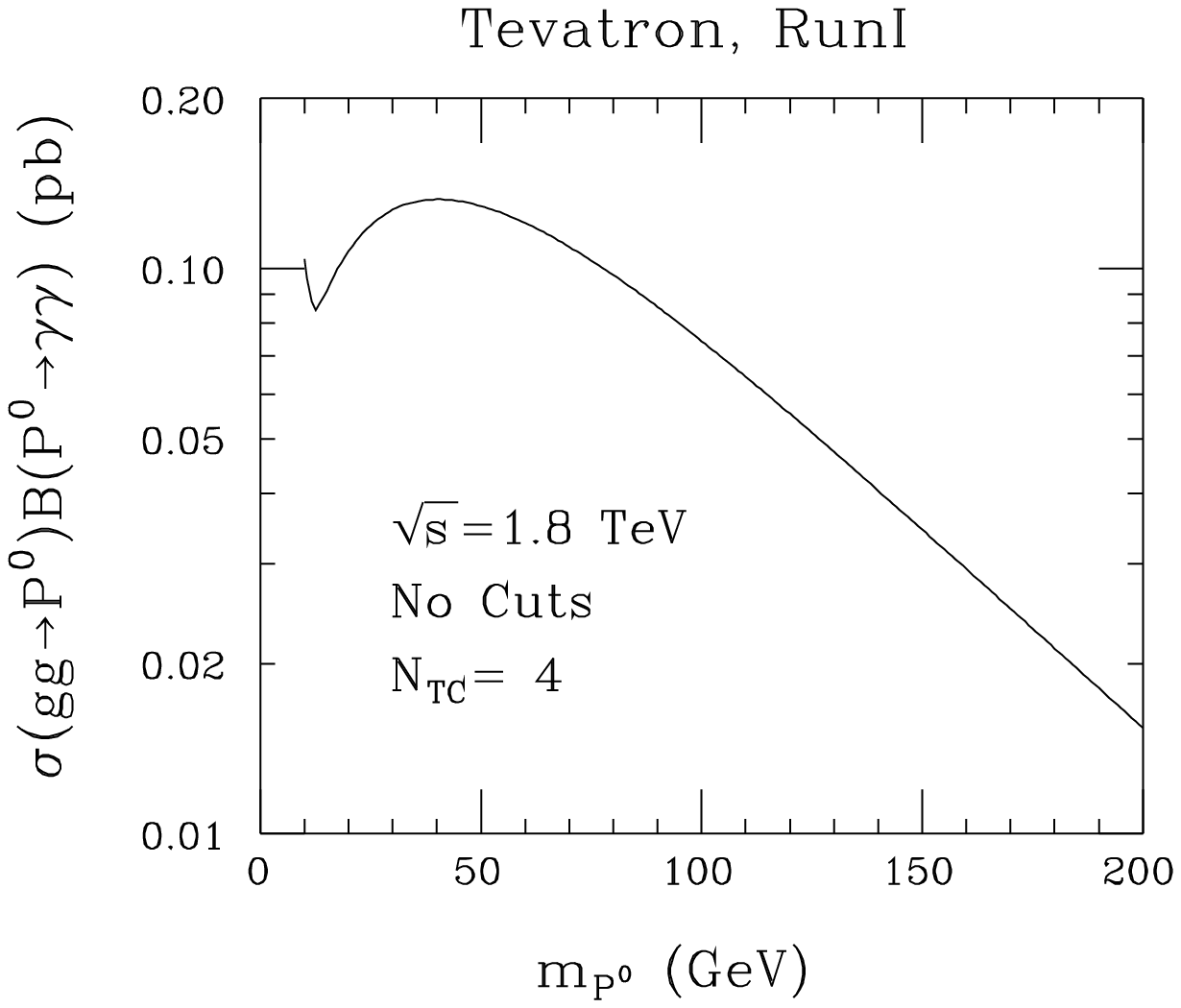}}
\smallskip
\noindent
\caption{$\sigma(p\anti p\to gg\to\pzero)\br(\pzero\to\gam\gam)$
is plotted as a function of $\mpzero$
for $\protect\rts=1.8\tev$ at the Tevatron. $\ntc=4$ is assumed.
No cuts are placed on the $\gam$'s.}
\label{figrts1pt8tev}
\end{figure}

Of course, it is very interesting to know if RunI
results could place any constraints on the $\pzero$.
To assess this, we present in Fig.~\ref{figrts1pt8tev}
a plot of $\sigma(p\anti p\to gg\to\pzero)\br(\pzero\to\gam\gam)$
(taking $\ntc=4$) at $\rts=1.8\tev$; no efficiency reductions
or cuts are included.
We see that with $L=100\pbi$ the raw number of events will
range from $\sim 10$ at $\mpzero\sim 40\gev$ down to $\sim 2$
at $\mpzero\sim 200\gev$.
These numbers will be further reduced by efficiency factors associated
with requiring central rapidity for the photons
and removing the reducible backgrounds from $\gam j$ and $jj$ events.
Consequently, we anticipate that obtaining useful constraints
on the $\pzero$ from RunI data will be very difficult
for moderate $\ntc$ values.

We can roughly verify the above expectation by using
analyzes of the inclusive $\gam\gam$ mass spectrum from RunI
that have already appeared. For example, Ref.~\cite{pjw} gives (Fig.~1)
results for $dN(p\anti p\to\gam\gam)/dM_{\gam\gam}$ for $40\leq
M_{\gam\gam}\leq 350\gev$ and (Fig.~3) resulting
limits on $d\sigma(p\anti p\to\gam\gam)/dM_{\gam\gam}$ (where
the experimental results have been corrected for cuts and efficiencies
to obtain the raw cross section) in the mass
range from $100-350\gev$. We consider first the Fig.~3 limits.
After integrating over the mass bins shown in Fig.~3 of Ref.~\cite{pjw},
which range in size from 15 GeV for $M_{\gam\gam}\sim 120\gev$ to
$30\gev$ at $M_{\gam\gam}\sim 200\gev$,
one obtains $\sigma(p\anti p\to\gam\gam)$ limits of order $0.7\pb$ and
$0.3\pb$ for these respective mass bins. From Fig.~\ref{figrts1pt8tev}, 
we see that sensitivity
would have to be increased by about a factor of 10 in order to place
limits on the $\pzero$ for $\ntc=4$ using this approach.
Possibly of greater utility is the smaller binning
of their Fig.~1. We consider the situation
for a few representative values of $M_{\gam\gam}$.
\bit
\item
At $M_{\gam\gam}=100\gev$, $dN(p\anti p\to\gam\gam)/dM_{\gam\gam}\sim
(0.25+0.15-0.25)/{\rm GeV}$. Multiplying the upper limit by the bin size,
$\sim 4\gev$, gives $N\leq 1.6$ events in this bin (at the
$1\sigma$ level). The expected
QCD background (including fakes) in this bin is, from Fig.~2 of \cite{pjw},
$N\sim 2.4$. The expected ($\ntc=4$) $\pzero$ signal in this bin
is $N=\eps L \sigma\sim 0.8$,
using $\eps\sim 0.105$ \cite{pjw}, $L\sim 100\pbi$ and
$\sigma(p\anti p\to \pzero\to\gam\gam)\sim 0.074\pb$ 
(Fig.~\ref{figrts1pt8tev}).
\item
At $M_{\gam\gam}=200\gev$, $dN(p\anti p\to\gam\gam)/dM_{\gam\gam}\lsim
0.01/{\rm GeV}$. Multiplying by the bin size,
$\sim 8\gev$, gives $N\leq 0.08$ events in this bin. The expected
QCD background in this bin is
$N\sim 0.24$. The expected $\pzero$ signal in this bin is $N\sim 0.2$,
as obtained using $\eps\sim 0.125$ \cite{pjw} and
$\sigma(p\anti p\to \pzero\to\gam\gam)\sim 0.015\pb$.
\item
At $M_{\gam\gam}=50\gev$, a bin size of $\sim 2.5\gev$ gives
$N\sim 25$ events, with an expected background of $N\sim 25$.
Extrapolating linearly
the efficiencies given in Ref.~\cite{pjw} for $100-300\gev$
down to $50\gev$ gives $\eps\sim 0.095$ and an expected signal rate
of $N\sim 1.2$.
\eit
These estimates make it clear
that, for $\ntc=4$, a $\pzero$ is still allowed
for any mass below $200\gev$. However, it is also apparent that
the factor 20 increase in luminosity expected for RunII will
allow either exclusion or discovery of an $\ntc=4$
$\pzero$ in the $100-200\gev$ mass range, with extension to
masses as low as $50\gev$ requiring improved procedures for removing fakes.

Of course, since the signal rate
scales as $\ntc^2$ [see Eq.~(\ref{gambrlimit})] (or faster at $\mpzero$
values low enough that $\br(\pzero\to\gam\gam)$ has not yet saturated),
the RunI limits do allow us to exclude the presence of a $\pzero$
if $\ntc$ is large. For example,
if $\ntc=12$, the number of signal events is at least 9
times larger than quoted above for $\ntc=4$. For the most favorable
mass, $\mpzero=100\gev$, $\ntc=12$ would yield $\sim 7$ events
where $\leq 1.6$ are seen and $\leq 2.4$ are expected from background.
$\ntc=16$ or so would be required in order that a $\pzero$
with $\mpzero=50$ or $200\gev$ have low probability.
A detailed determination of $\mpzero$ mass ranges that can
be excluded (using RunI data)
as a function of $\ntc$ should be performed by the CDF and D0 detector groups.

Our overall conclusion is that there is
significant potential for discovering and studying the
$\pzero$ at the Tevatron using the $gg\to\pzero\to\gam\gam$ production/decay
mode. Our cursory analysis suggests that RunI data can be used
to exclude a $\pzero$ in the $50-200\gev$ mass range for $\ntc\geq 12-16$
(depending upon mass). Given that production rates and branching
ratios depend significantly on $\ntc$,
a detailed analysis by the detector groups of the maximum
$\ntc$ allowed by the RunI data as a function of $\mpzero$ would
provide very important guidance for future searches.
Note that were the isosinglet $\pi_D$ the mass
eigenstate, the $gg\to\pi_D\to\gam\gam$ rate would be much less enhanced
and prospects at the Tevatron would be much worse.

The second decay mode
that can be considered for discovering the $\pzero$
in inclusive gluon-fusion production is $\tauptaum$. However,
the early claims \cite{ehlq} that the $\tauptaum$ decay mode would
easily allow discovery have not been substantiated;
backgrounds and inefficiencies associated with $\tau$ pair
reconstruction are the main difficulties.

One can also consider searching for the $\pzero$
in the $\gam\pzero$ and $Z\pzero$ channels. (Since the
$WW$ coupling to the $\pzero$ is zero,  the $W\pzero$ channel is not
present.) For moderate $\ntc$, the much smaller $ZZ$
coupling of the $\pzero$ as compared to a SM-like Higgs implies that
the $Z\pzero$ cross section will be greatly reduced relative to the Higgs case.
Consequently,
the RunI limits for the $Z\gam\gam$ mode given in Ref.~\cite{pjw} will
only constrain the $\pzero$ when $\ntc$ is quite large.
An analysis as a function of $\ntc$ is in order.
The cleanliness of the $\gam\gam\gam$ channel (see Ref.~\cite{cdfgamgam},
for example) suggests that it too deserves
a detailed examination.
For low $\mpzero$, the $\gam\tauptaum$ and $Z\tauptaum$ channels
might prove useful at large $\ntc$.
We defer an examination of all these channels to a later work.

Most studies of PNGB detection at hadron colliders have focused
on the production of pairs of PNGB's.
Aside from simple Drell-Yan pair production
via $\gam$ and $Z$ exchange, there is the possibility of
resonant production $pp\to V^\pm X \to P^\pm \pzero X$
or $pp\to V^0 X \to P^\pm P^\mp X$
\cite{dometal94}, where $V$ is a vector resonance
such as the technirho (see the introduction).
For the type of model we consider here,
the technirho has mass in the TeV range. Further,
it will not be particularly narrow
and it will have many competing decay modes. In addition,
one will need to find the two PNGB's in channels
such as $P^+\to c\anti b$ and $\pzero\to b\anti b$ or $gg$ that have large
backgrounds and poor mass resolution. To date, there is no
study showing that PNGB pairs can be detected when the technirho
is heavy. More detailed studies will be needed
to fully clarify the prospects.
Thus, Ref.~\cite{lane1} concluded that the Drell-Yan production of pairs
of color-singlet PNGB's is unobservably small compared to backgrounds {\it
unless} there is one or more
fairly strong color-singlet technirho and/or techniomega resonances not
far above threshold. If the technirho is not far above threshold, then,
for example, $\rho_T^\pm\to W^\pm\pzero$ can be the dominant decay.
The $\rho_T^\pm$
is then sufficiently narrow that cuts on the data can be made which
bring out the signal, as found in Ref.~\cite{elw}.
Similarly, for a light techniomega, $pp\to \omega_T X\to\gam \pzero X$
can be the dominant channel and \cite{elw} finds that a viable
signal is possible after suitable cuts.

One could also consider the
process $pp\to gg\to \pzero\pzero$, mediated by the anomalous $gg\pzero$
vertex. The main kinematic trick would be to look for
equal mass pairs. However, backgrounds in the four jet channel
will be very large.

Thus, unless the technirho
(or techniomega) is much lighter than expected in the
model we consider here, the best
mode for $\pzero$ discovery at the Tevatron and LHC
is $gg\to\pzero\to\gam\gam$. In the case of $\ntc=4$, this mode is clearly
viable for $\mpzero\gsim 50\gev$ (up to masses somewhat
above $200\gev$). For this same $\ntc=4$ choice,
viability will probably extend to substantially
lower $\mpzero$ at the LHC and possibly to somewhat lower $\mpzero$
for TeV33 luminosity at the Tevatron --- detailed detector simulations
of backgrounds and $M_{\gam\gam}$ resolution at low $\mpzero$
will be required to confirm these expectations.
If $\ntc$ is minimal, \ie\ $\ntc=1$, prospects for $\pzero$
discovery at the Tevatron are not good, but detection
of $gg\to\pzero\to\gam\gam$ at the LHC will be possible
for (roughly) $70\leq\mpzero\leq 200\gev$ (and possibly higher).
It is in this context that we consider the role of electron
and muon colliders for the discovery and eventual study of the $\pzero$.

\section{$\pzero$ production at an electron collider}

In this section, we give a brief overview of the role that
an $\epem$ collider might play in the study and discovery
of the $\pzero$. As we have already
noted, the partner $\pzerop$ state
with one-loop mass proportional to the $t$-quark mass will, most naturally,
be much heavier and, thus, would have much smaller
cross section. For an ultra-violet cut-off scale, $\Lambda$,
of order $2\tev$ or larger (the most natural range for a technicolor model),
the $\pzerop$ would be too heavy to be produced even at a $\rts\lsim 500\gev$
collider. Although we shall employ the specific $\pzero$ properties,
in particular branching fractions and couplings, as predicted
by the scheme reviewed earlier, strategies for probing the
lightest PNGB in other models of a strongly interacting electroweak sector
would generally not be all that different. In $\epem$ collisions,
the greatest differences would arise from two sources.
First, if the eigenstate composition of the lightest PNGB
were not that of the $\pzero$, then, as noted earlier, the (very crucial)
$\gam\gam$ coupling-squared could be very different, \eg\ a factor
of 8 smaller if the lightest PNGB were the $\pi_D$. This would
reduce the two most relevant cross sections by a factor of 8.
Second, in models such as walking technicolor results would be altered due to:
(a) the possibility that $\mpzero$ is substantially larger
for the same $\Lambda$; and (b) the larger $\pzero\gam\gam$
and $\pzero Z\gam$ couplings as a result of a smaller effective
scale in place of $v$ in Eq.~(\ref{pvvcoups}).

\begin{figure}[h]
\epsfysize=8truecm
\centerline{\epsffile{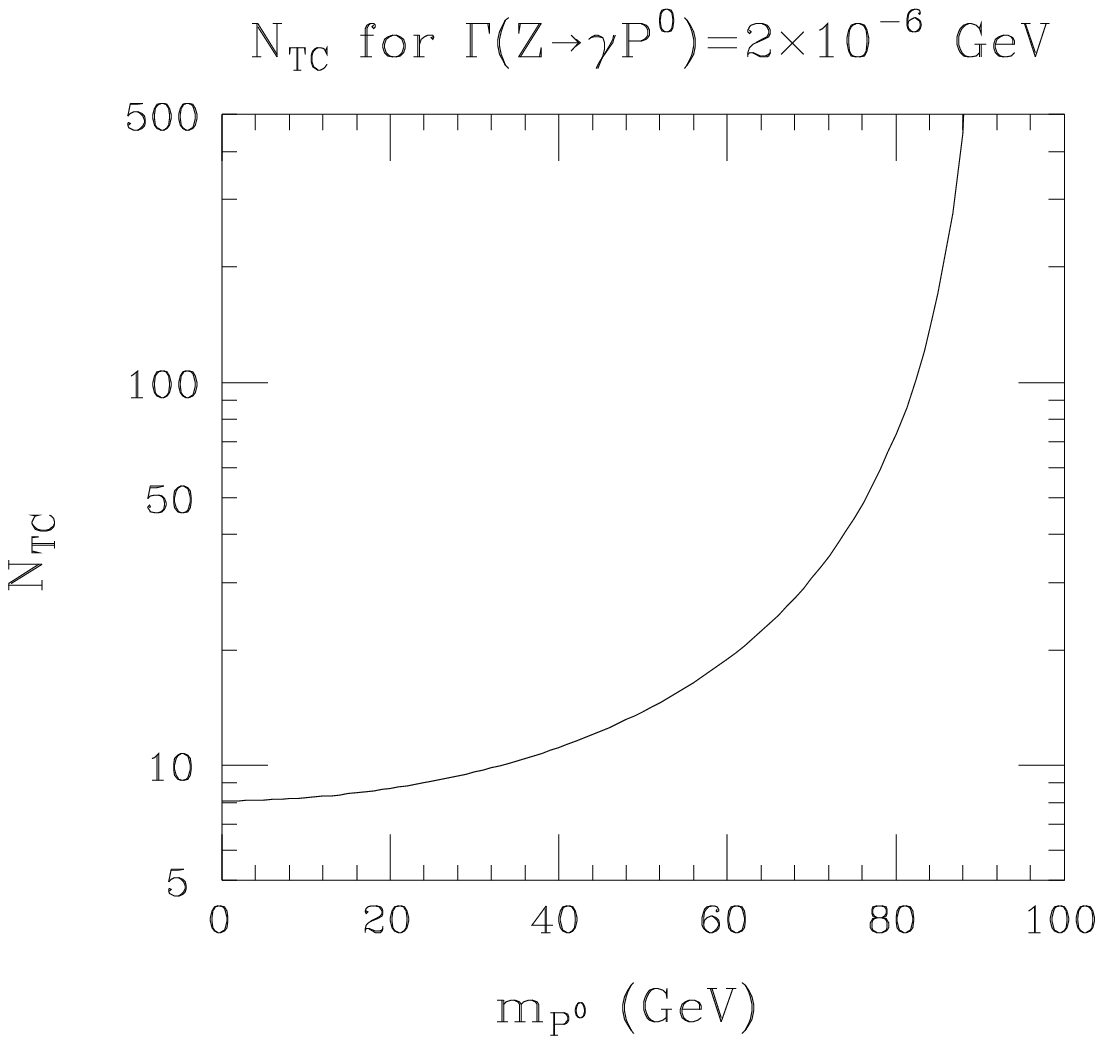}}
\smallskip
\caption{
The value of $\ntc$ required for
$\Gamma(Z\to \gam\pzero)>2\times 10^{-6}\gev$ is plotted as a
function of $\mpzero$.}
\label{ztogampfig}
\end{figure}

First, let us consider whether LEP places any limits on the $\pzero$.
At LEP the dominant production mode is $Z\to\gamma \pzero$.
The width for this decay is given by
\beq
\Gamma(Z\to\gam\pzero)={\alpha^2\mz^3\over 96\pi^3 v^2}\ntc^2 A_{\pzero
Z\gam}^2\left(1-{\mpzero^2\over\mz^2}\right)^3\,,
\label{ztogampzero}
\eeq
where $A_{\pzero Z\gam}$ appeared in Eq.~(\ref{pzgamcoup}).
Let us follow Ref.~\cite{chivu} and require that the $Z\to\gamma \pzero$
decay width be $>2 \times 10^{-6}\gev$ in order for the $\pzero$
to be visible in a sample of $10^7$ $Z$ bosons. The minimum $\ntc$
value required as a function of $\mpzero$ is plotted in Fig.~\ref{ztogampfig}.
We see that $\ntc\gsim 8$ is required at $\mpzero=0$, rising rapidly
as $\mpzero$ increases.\footnote{In a multi-scale model,
where the effective $v$ could be smaller, these results
would be altered.}

We next consider LEP2.
The general form of the cross section for $PV$ production
(from Ref.~\cite{ls}) is
\bea
\sigma(\epem&\to& PV)=
{\alpha^3\ntc^2\over 24\pi^2 v^2}\lam^{3/2}(1,m_P^2/s,m_V^2/s)
\nonumber\\
&\times& \Bigl[A_{PV\gam}^2+
{A_{PV\gam}A_{PVZ}(1-4s_W^2)\over 2c_ws_W(1-\mz^2/s)}
+{A_{PVZ}^2(1-4s_w^2+8s_W^4)\over 8 c_W^2s_W^2(1-\mz^2/s)^2}\Bigr]\,,
\label{epemtopb}
\eea
where $V=\gam,Z$ and $P$ is the PNGB. In the above, we have
neglected the $Z$ width.
As already stated, the best mode for $\pzero$ production
at an $\epem$ collider (with $\rts>\mz$) is $\epem\to\gam\pzero$.
Because the $\pzero Z\gam$ coupling-squared is much smaller than the
$\pzero\gam\gam$ coupling-squared (by a factor of nearly 400),
the dominant diagram is $\epem\to\gam\to\gam\pzero$, proportional
to $A_{\pzero\gam\gam}^2$. Even when kinematically allowed,
rates in the $\epem\to Z\pzero$ channel are substantially smaller,
as we shall discuss. We will give results for the moderate
value of $\ntc=4$.
For $\rts=200\gev$, we find that,
after imposing an
angular cut of $20^\circ\leq\theta\leq 160^\circ$ on the outgoing
photon (a convenient acceptance cut
that also avoids the forward/backward cross section
singularities but is more than 91\% efficient),
the $\epem\to\gam\pzero$ cross section is below $1\fb$ for $\ntc=4$.
Given that the maximum integrated luminosity anticipated
is of order $L\sim 0.5\fbi$, we conclude that LEP2
will not allow detection of the $\pzero$ unless $\ntc$ is very large.

\begin{figure}[h]
\epsfysize=8truecm
\centerline{\epsffile{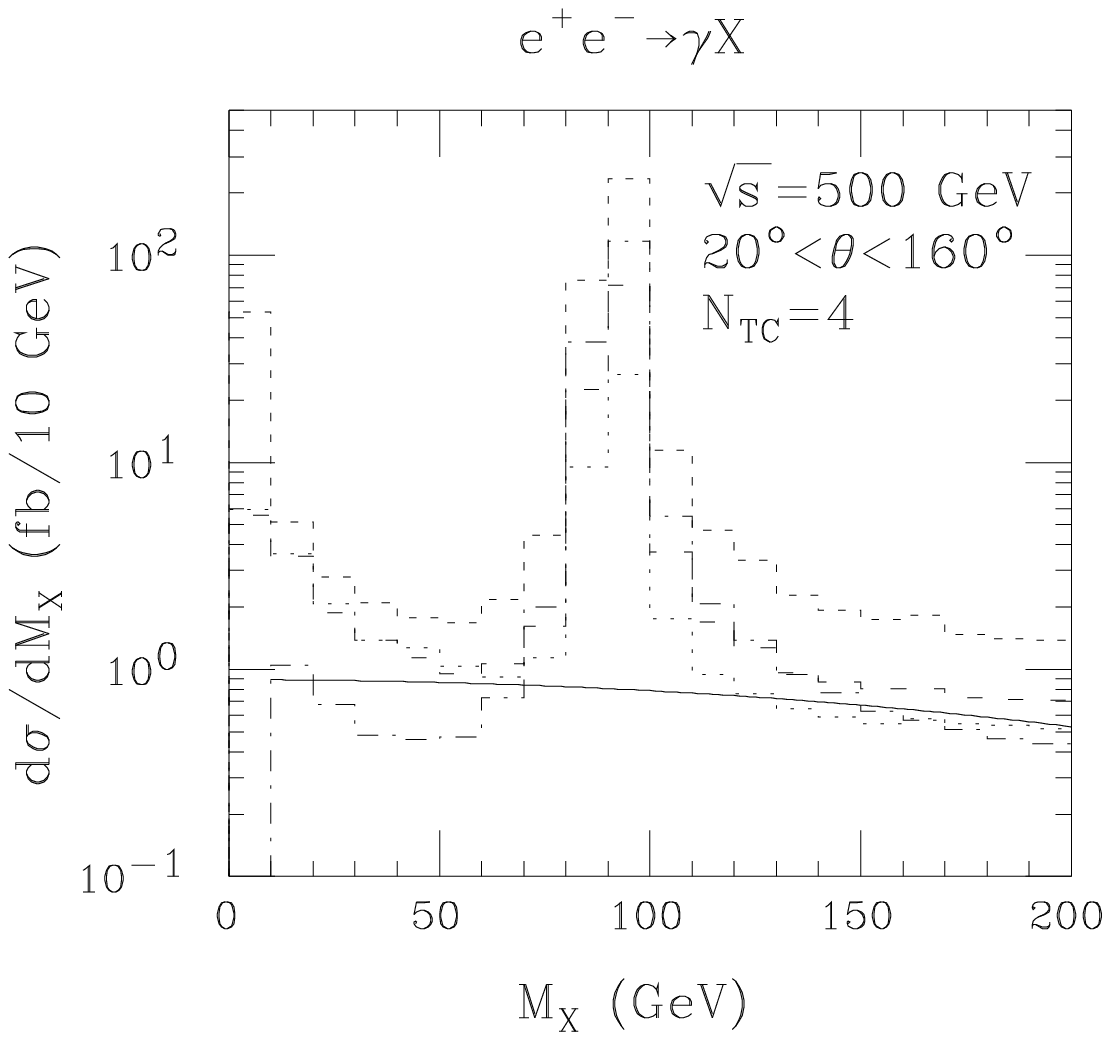}}
\smallskip
\caption{
Taking $\ntc=4$, the cross section (in fb) for
$\epem\to\gam\pzero$ (solid curve) is plotted as a function of $\mpzero$
in comparison to various possible
backgrounds: $\epem\to\gam b\anti b$ (dotdash);
$\epem\to\gam c\anti c$ (dashes);
$\epem\to\gam q\anti q$, $q=u,d,s$ (small dashes); and
$\epem\to\gam\tauptaum$ (dots).
The background cross sections are integrated over a $\Delta M_X=10\gev$
bin width (a possible approximation to the resolution that
can be achieved).  A cut of $20^\circ\leq\theta\leq 160^\circ$ has been
applied to both the signal and the backgrounds. Effects due
to tagging and mis-tagging are discussed in the text.}
\label{figeetogamp}
\end{figure}

The cross section for $\epem\to \gam \pzero$ at $\rts=500\gev$,
after imposing the same angular cut as for LEP2,
is illustrated in Fig.~\ref{figeetogamp} for $\ntc=4$. It ranges from
$0.9\fb$ down to $0.5\fb$ as $\mpzero$ goes from zero up to $\sim 200\gev$.
For $L=50\fbi$, we have at most 45 events with which to discover
and study the $\pzero$.
The $\epem\to Z\pzero$ cross section is even smaller. Without cuts
and without considering any specific $Z$ or $\pzero$ decay modes, it
ranges from $0.014\fb$ down to $0.008\fb$ over the same mass range.
With less than 1 event to work with, there is little point in examining
the $Z\pzero$ mode.  If TESLA is able to achieve
$L=500\fbi$ per year, $\gam\pzero$ production will have a substantial
rate, but the $Z\pzero$ production rate will still not be useful.
Since the $\gam\pzero$ production rate scales as $\ntc^2$, if $\ntc=1$
a $\rts=500\gev$ machine will yield at most 3 (30) events for
$L=50\fbi$ ($500\fbi$), making $\pzero$ detection and study extremely
difficult.  Thus, we will focus our analysis on the $\ntc=4$ case.

In order to assess the $\gam\pzero$ situation more fully,
we must consider backgrounds.
As we have seen, the dominant decay of the $\pzero$ is typically to
$b\anti b$, $\tauptaum$ or $gg$.  For the $b\anti b$ and $gg$
modes, the backgrounds relevant to the $\gam \pzero$
channel are $\gam b\anti b$, $\gam c\anti c$ and $\gam q\anti q$
($q=u,d,s$) production. The cross sections for these processes
obtained after integrating over a 10 GeV bin size in the quark-antiquark
mass (an optimistic estimate of the resolution that could be achieved
using reconstruction of the quark-antiquark or $\tauptaum$ pair) are also given
in Fig.~\ref{figeetogamp}. For $10\lsim\mpzero\lsim 80\gev$
and $\mpzero\geq 100\gev$, the signal to background
ratio is not too much smaller than 1. We will assess the
$\pzero$ discovery potential in
specific channels by assuming that 10 GeV mass resolution can be achieved
for $\mpzero$ in each case.

In order to proceed with a discussion of specific final states, we
must state our assumptions regarding tagging and mis-tagging
efficiencies.  We separate $\tau^+\tau^-$,
$b\anti b$, $c\anti c$ and $q\anti q/ gg$
final states by using topological and $\tau$ tagging with efficiencies
and mis-tagging probabilities as estimated by B. King \cite{bking}
for the muon collider.
These are a bit on the pessimistic side for an $\epem$ collider,
but only slightly so. We take
$\eps_{bb}=0.55$, $\eps_{cc}=0.38$, $\eps_{bc}=0.18$, $\eps_{cb}=0.03$,
$\eps_{qb}=\eps_{gb}=0.03$, $\eps_{qc}=\eps_{gc}=0.32$,
$\eps_{\tau\tau}=0.8$, $\eps_{\tau b}=\eps_{\tau c}=\eps_{\tau q}=0$,
where the notation is that $\eps_{ab}$ is the probability
that a particle/jet of type $a$ is tagged as being of type $b$.
Gluons are treated the same as light quarks. Note that
$\eps_{bq}=1-\eps_{bb}-\eps_{bc}$, $\eps_{cq}=1-\eps_{cc}-\eps_{cb}$,
and $\eps_{qq}=1-\eps_{qb}-\eps_{qc}$.

\begin{figure}[h]
\epsfysize=8truecm
\centerline{\epsffile{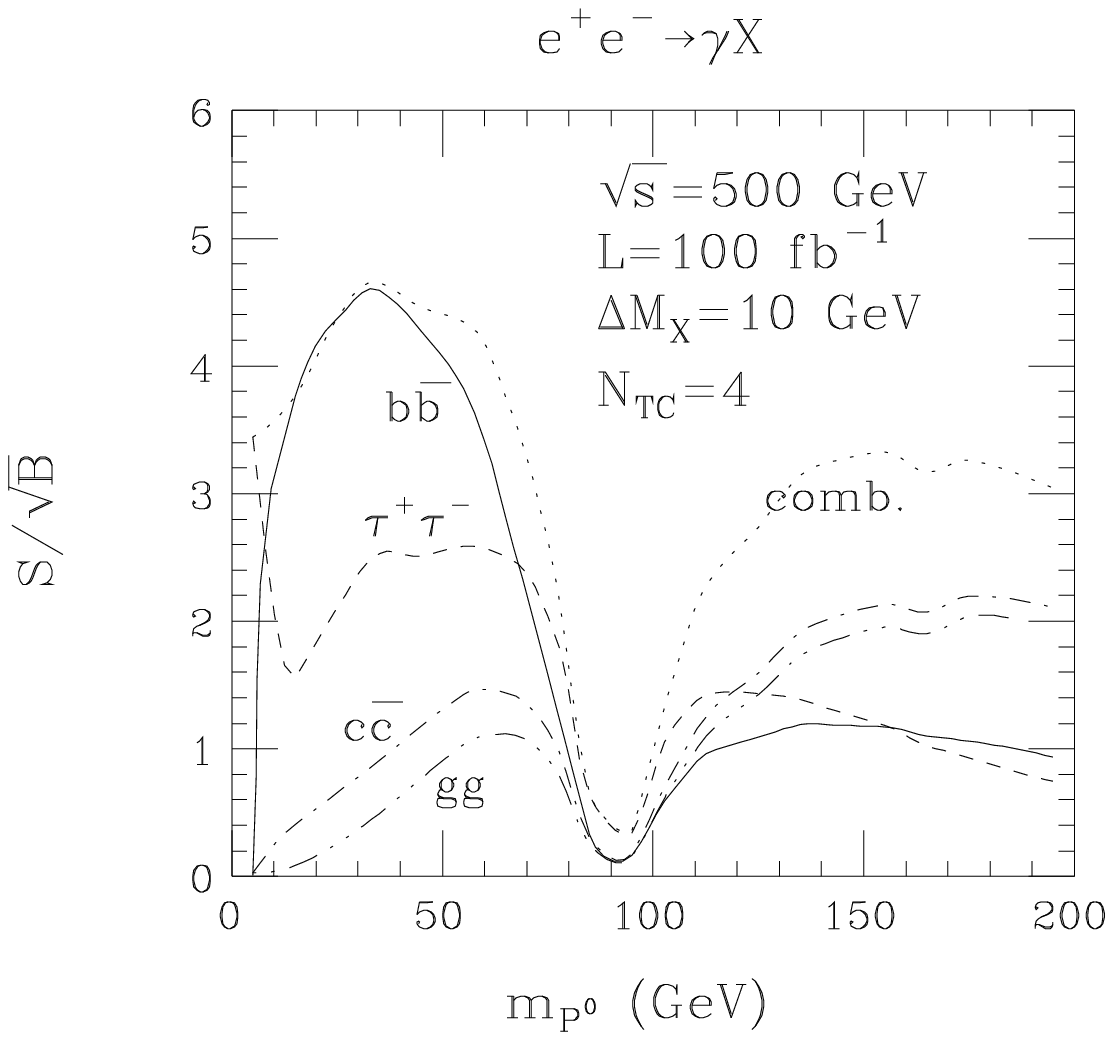}}
\smallskip
\caption{We consider $\epem\to \gam X$ taking $\ntc=4$
and, for $L=100\fbi$ at $\protect\rts=500\gev$,
plot the statistical significances $S/\protect\sqrt B$ for a $\pzero$ signal
in various `tagged' channels as a function
of $\mpzero$. We assume mass resolution of $\Delta M_X=10\gev$ in
each channel and the channel tagging and mis-tagging probabilities discussed
in the text. A cut of $20^\circ\leq \theta\leq 160^\circ$
has been imposed on both the signal and the backgrounds.
The curve legend is: $gg$ (dot-dot-dash); $c\anti c$ (dot-dash);
$b\anti b$ (solid); $\tauptaum$ (dashes). Also shown (dots) is the largest
$S/\protect\sqrt B$ that can be achieved by considering all possible
combinations of channels.}
\label{figeetogampnsd}
\end{figure}

We then identify final states according to the following scheme.
A final state is identified as:
$b\anti b$ if one or more jets is tagged as a $b$; $c\anti c$
if no jet is tagged as a $b$, but one or more jets is tagged as a $c$;
and $q\anti q/gg$ if neither jet is tagged as a $b$ or a $c$.
Background and signal events are analyzed in exactly the same manner.
Note, in particular, that even though the $\pzero$ does not decay
to $c\anti c$, some of its $b\anti b$ and $gg$ decays will be identified
as $c\anti c$.
These individual jet tagging and mis-tagging efficiencies then lead
to the following final state efficiencies:
$\eps_{bb/bb}=0.8$, $\eps_{bb/cc}=0.13$, $\eps_{bb/qq}\sim 0$,
$\eps_{cc/cc}=0.6$, $\eps_{cc/bb}=0.04$, $\eps_{cc/qq}=0.36$,
$\eps_{qq/qq}=0.42$, $\eps_{qq/bb}=0.06$, $\eps_{qq/cc}=0.52$,
$\eps_{\tau\tau/\tau\tau}=0.96$.
Here, the notation is that $\eps_{aa/bb}$ is the probability
that a final state that is truly $aa$ is tagged as the final state $bb$.
We give some examples of how the above channel efficiencies are computed.
For instance, $\eps_{bb/bb}=1-(1-\eps_{bb})^2$,
$\eps_{bb/cc}=2\eps_{bc}\eps_{bq}+\eps_{bc}^2$,
$\eps_{bb/qq}=(1-\eps_{bb}-\eps_{bc})^2$, with the sum of these
three tagging probabilities being 1. The $c\anti c$
final state is treated similarly.
For the $qq$ (or $gg$) final state we have
$\eps_{qq/qq}=(1-\eps_{qb}-\eps_{qc})^2$,
$\eps_{qq/bb}=2\eps_{qb}(1-\eps_{qb})+\eps_{qb}^2$,
and $\eps_{qq/cc}=2\eps_{qc}(1-\eps_{qc}-\eps_{qb})+\eps_{qc}^2$;
again one can check that these sum to 1.

Results for $S/\sqrt B$, in the various tagged channels, for $\ntc=4$ and
assuming $L=100\fbi$ at $\rts=500\gev$,
are plotted in Fig.~\ref{figeetogampnsd}.
We have assumed a mass window of $\Delta M_X=10\gev$ in evaluating
the backgrounds in the various channels. Also shown
in Fig.~\ref{figeetogampnsd}
is the largest $S/\sqrt B$ that can be achieved by considering
(at each $\mpzero$) all possible combinations of the $gg$, $c\anti c$,
$b\anti b$ and $\tauptaum$ channels.  From the figure, we find
$S/\sqrt B\geq 3$ (our discovery criterion)
for $\mpzero\leq 75\gev$ and $\mpzero\geq 130\gev$, \ie\
outside the $Z$ region.  A strong signal, $S/\sqrt B\sim 4$, is only
possible for $\mpzero\sim 20-60\gev$. As the figure shows, the signal
in any one channel is often too weak for discovery, and it is only
the best channel combination that will reveal a signal.
For the TESLA $L=500\fbi$ luminosity, $S/\sqrt B$ should be multiplied
by $\sim 2.2$ and discovery prospects will be improved.

\begin{figure}[h]
\epsfysize=8truecm
\centerline{\epsffile{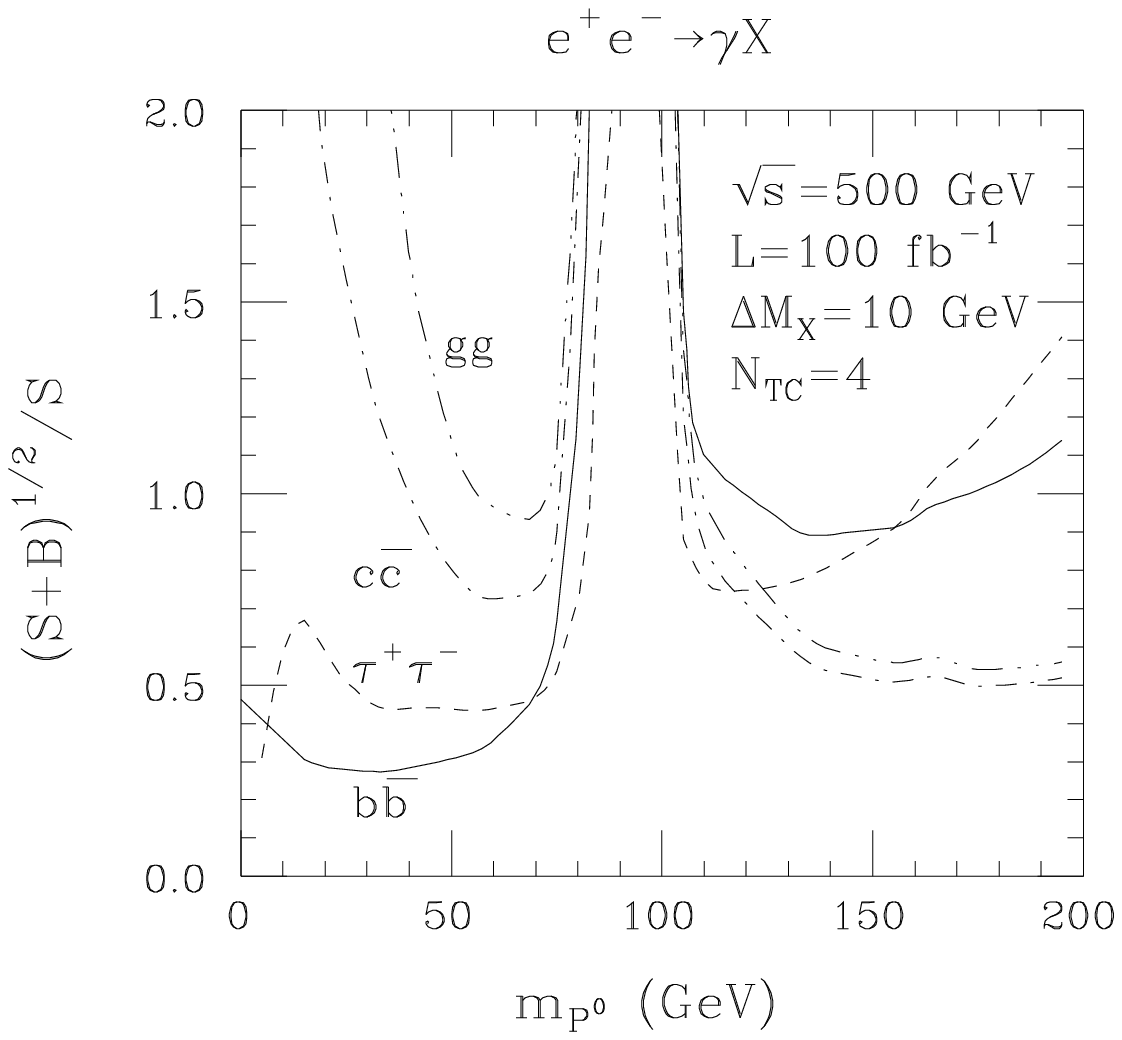}}
\smallskip
\caption{We consider $\epem\to\gam \pzero\to\gam X$ production
for $\ntc=4$, with $L=100\fbi$ at $\protect\rts=500\gev$,
and plot the statistical error $(S+B)^{1/2}/S$ for the
various `tagged' channel rates ($X=\tauptaum$, $b\anti b$,
$c\anti c$, $gg$) as a function
of $\mpzero$. Assumptions and notation as in
Fig.~\protect\ref{figeetogampnsd}.}
\label{figeetogamperrors}
\end{figure}

Once a PNGB is discovered,
one will wish to determine the branching fractions and couplings
as precisely as possible in order to pin down the fundamental parameters
of the model.  In the $\epem\to\gam\pzero$ production mode,
one will begin by extracting ratios of branching fractions
by computing ratios of the rates measured in various final state
channels.\footnote{Note that the reason we focus
on ratios is that the systematic errors
due to uncertainty in the absolute normalization
of the rate in any given channel will cancel out in the ratios.}
As a first indication of how well one can do, we give,
in Fig.~\ref{figeetogamperrors}, the statistical
errors $(S+B)^{1/2}/S$ in each of the tagged channels in the case
of our bench mark example of the $\pzero$. Even if we
decrease the errors of the figure by the $\sqrt 5\sim 2.2$ factor
appropriate for an integrated luminosity of $L=500\fbi$,
the only channel with reasonable error ($\lsim 15\%$) would
be $b\anti b$. Further, in obtaining
results for ratios of $\br(\pzero\to F)$ for $F=gg,b\anti b,\tauptaum$,
one must unfold the mis-tagging (implying
introduction of systematic uncertainties) and combine statistical errors
in the various tagged channels.

The next step, beyond
the extraction of ratios of the $\pzero$ branching fractions,
is the model-independent determination of
the individual $\br(\pzero\to F)$'s
for specific final states $F$ via the ratio of the rate in a
specific final state to the inclusive rate:
\beq
\br(\pzero\to F)={\sigma(\epem\to\gam\pzero)\br(\pzero\to F)\over
\sigma(\epem\to\gam\pzero)}\,.
\label{modind}
\eeq
The crucial issue is then the ability to observe the $\pzero$
inclusively in the $\gam X$ final state
as a peak in the recoil $\mx$ spectrum, and the associated
error in the inclusive cross section $\sigma(\epem\to\gam\pzero)$.
The resolution in $\mx$ is determined by the photon energy resolution.
Using $\Delta E_\gam/E_\gam=0.12/\sqrt{E_\gam}\oplus 0.01$, one
finds $\pm 1\sigma$ mass windows in
$\mpzero$ of $[0,78]$, $[83.5,114]$ and $[193,207]$
(GeV units) for $\mpzero=55$, $100$ and $200\gev$,
respectively. If the resolution could be improved
to $\Delta E_\gam/E_\gam=0.08/\sqrt{E_\gam}\oplus0.005$ \cite{barklow}, then
the mass windows for $\mpzero=55$, $100$ and $200\gev$ become $[36,69]$,
$[91,108]$ and $[196,204]$, respectively. [We note that
$\Delta E_\gam/E_\gam\gsim 0.0125$ ($\gsim 0.0075$)
for $\mpzero\leq 200\gev$ for the first (second) resolution case,
indicating that the constant term is dominant and should be the focus
for improving this particular signal.]

Backgrounds to inclusive $\gam\pzero$ detection will be substantial.
All the backgrounds plotted in Fig.~\ref{figeetogamp}
must be included (integrated over the appropriate mass window),
and others as well. Observation of the $\pzero$ signal in the
recoil $\mx$ spectrum would be difficult, especially for lower
values of $\mpzero$. However, if $\mpzero$ is known ahead of time,
then one can simply employ the appropriate mass window
and estimate the background from $\mx$ bins outside the mass window.
We anticipate that the resulting errors for
$\sigma(\epem\to\gam\pzero)$ will be large, implying that
the corresponding model-independent
determinations of the various $\br(\pzero\to F)$'s from Eq.~(\ref{modind})
will be subject to large statistical uncertainty.
This is an important loss
relative to the usual program for determining the properties
of a Higgs boson in a model-independent manner using the $\epem\to Zh$
signal in the inclusive $\epem\to ZX$ final state.

Overall, one cannot be optimistic
that an $\epem$ collider will provide more than very rough determinations of
the fundamental parameters of the effective Lagrangian described
earlier if $\ntc=4$. If $\ntc=1$,
$L>500\fbi$ would be required in order to even detect
the $\pzero$. We will later show that prospects at a muon collider are
potentially much better, especially if a first signal and approximate
mass determination is provided ahead of time by
the Tevatron and/or LHC or in $\gam\gam$ collisions at an $\epem$ collider.

\section{$\pzero$ production at a $\gam\gam$ collider}

The rate for production and decay of a narrow resonance $R$
in $\gam\gam$ collisions is given by \cite{gunhabgamgam}
\beq
N(\gam\gam\to R\to F)= {8\pi\Gamma(R\to\gam\gam)\br(R\to F)
\over m_R^2E_{\epem}}\tan^{-1}{\Gamma_{\rm exp}\over \Gamma^{\rm tot}_R}
\left(1+\vev{\lam\lam^\prime}\right)G(y_R)L_{\epem}\,,
\label{siggamgam}
\eeq
where $\lam$ and $\lam^\prime$ are the helicities of the colliding photons,
$\Gamma_{\rm exp}$ is the mass interval accepted in the final state
$F$ and $L_{\epem}$ is the integrated luminosity for the colliding
electron and positron beams. In Eq.~(\ref{siggamgam}), $\vev{\lam\lam^\prime}$
and $G(y_R\equiv m_R/E_{\epem})$
depend upon the details of the $\gam\gam$ collision
set-up.  Here, we are interested in exploring the ability of a $\gam\gam$
collider to discover the narrow $\pzero$ resonance and so we choose
laser polarizations $P$ and $P^\prime$ and $\epem$ beam helicities
$\lam_e$ and $\lam_e^\prime$ in the configuration
$2\lam_e P\sim +1$, $2\lam_e^\prime P^\prime\sim+1$,
$PP^\prime\sim +1$ such that $G\gsim 1$ and $\vev{\lam\lam^\prime}\sim 1$
(which suppresses $\gam\gam\to q\anti q$ backgrounds)
over the large range $0.1\leq y_R\leq 0.7$. The $\pzero$ is always
sufficiently narrow that $\tan^{-1}\to \pi/2$. In this limit,
the rate is proportional to $\Gamma(R\to\gam\gam)\br(R\to F)$.
For the $\pzero$, $\Gamma(\pzero\to\gam\gam)$ is large and the total
production rate will be substantial. In this regard,
the importance of the eigenstate composition of the $\pzero$ has already
been noted; \eg\ for the same mass, if the $\pi_D$ were the mass
eigenstate it would have only 1/8 the production rate.

In Fig.~\ref{fighpgbcomp}, we plotted $\Gamma(\pzero\to\gam\gam)\br(\pzero\to
b\anti b)$ divided by $\Gamma(h\to\gam\gam)\br(h\to b\anti b)$,
where $h$ denotes the SM Higgs boson, for both $\ntc=4$ and $\ntc=1$.
Over the $\mpzero=\mh$ range from 15 to 150 GeV, the ratio
for $\ntc=4$ varies from
$\sim 8$ down to $\sim 3$, rising to very large values at masses above
$160\gev$ where the $\h\to WW,ZZ$ decay modes open up.
For $\ntc=1$, this same ratio is order 0.4 to 0.5 over the 15 to 150 GeV
mass range, again rising dramatically at higher masses. Since it is
well-established \cite{gunhabgamgam,baueretal,japanese} that the SM $h$
can be discovered in this decay mode for $40\lsim m_h\lsim 2\mw$, it is clear
that $\pzero$ discovery in the $b\anti b$
final state will be possible up to at least $200\gev$, down to
$\sim 0.1 \rts\sim 50\gev$ (at $\rts\sim 500\gev$), below which $G(y)$
starts to get small. Discovery at lower values of $\mpzero$ would
require lowering the $\rts$ of the machine.

\begin{figure}[h]
\epsfysize=8truecm
\centerline{\epsffile{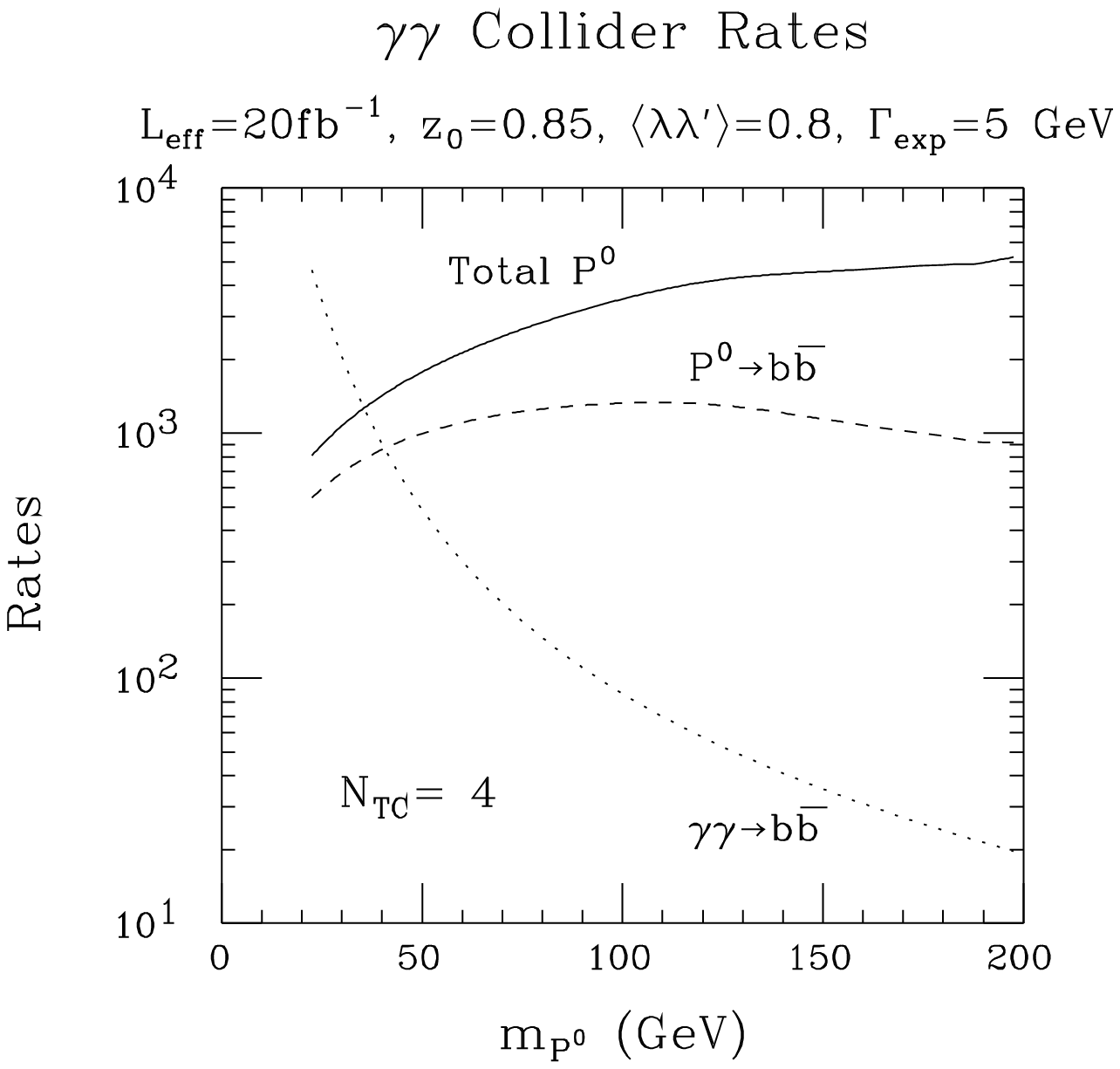}}
\smallskip
\caption{We consider $\gam\gam$ collisions for $L_{\rm eff}=20\fbi$
(assumed independent of $\mpzero$), with an angular cut
of $|\cos\theta|<0.85$ applied to the two-particle final state.
An experimental resolution $\Gamma_{\rm exp}=5\gev$ is assumed
in the final state. We plot, as a function of $\mpzero$:
the total $\gam\gam\to\pzero$ production rate (solid);
the rate for $\gam\gam\to\pzero\to b\anti b$ (dashes);
and the $\gam\gam\to b\anti b$ irreducible background rate (dots).
$\ntc=4$ is assumed.}
\label{gamgamrates}
\end{figure}

\begin{figure}[h]
\epsfysize=8truecm
\centerline{\epsffile{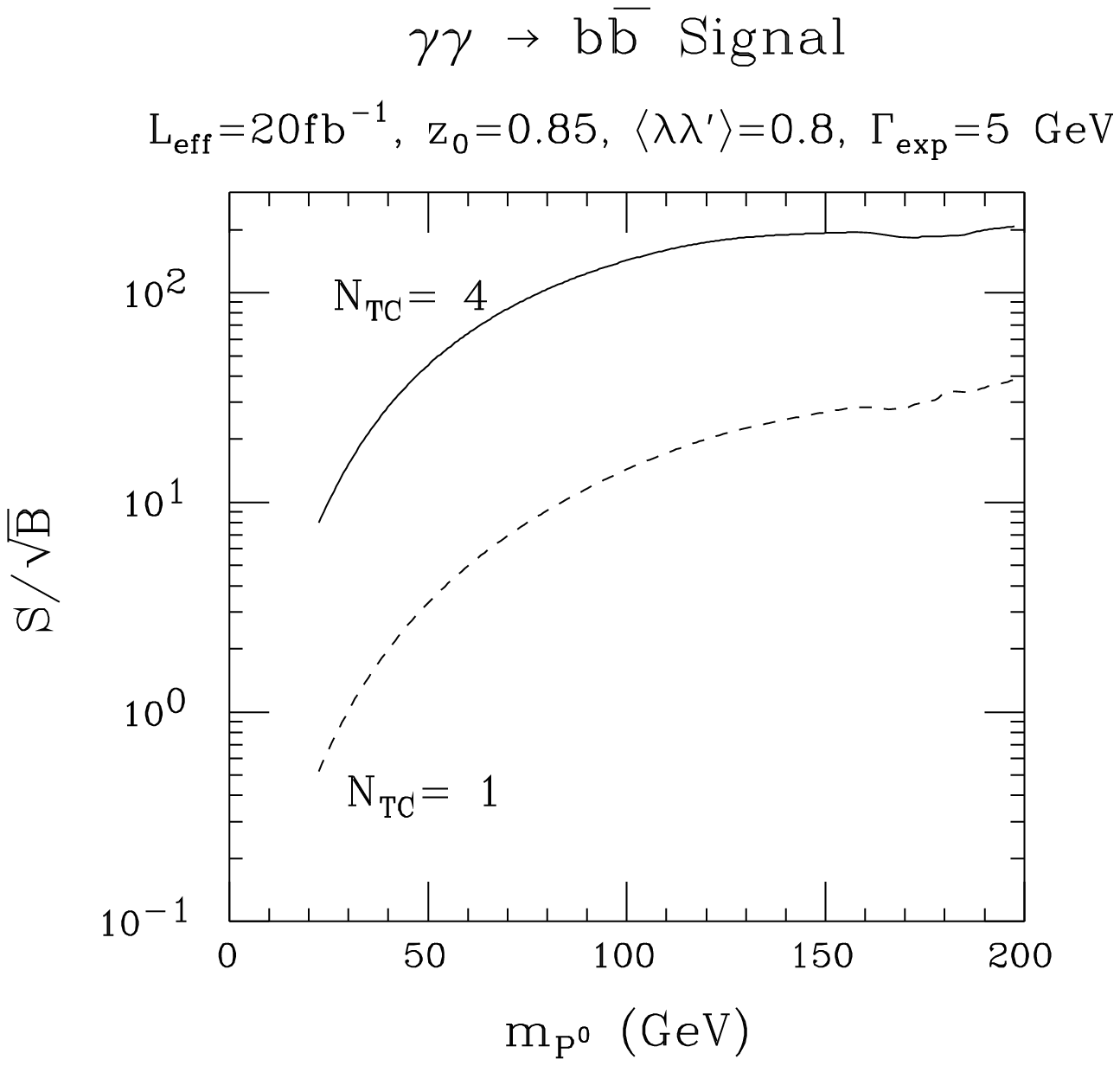}}
\smallskip
\caption{We consider $\gam\gam$ collisions for $L_{\rm eff}=20\fbi$
(assumed independent of $\mpzero$), with an angular cut
of $|\cos\theta|<0.85$ applied to the two-particle final state.
An experimental resolution $\Gamma_{\rm exp}=5\gev$ is assumed
in the final state. We plot, as a function of $\mpzero$,
the statistical significance $S/\protect\sqrt B$ for $\ntc=4$
and $\ntc=1$.}
\label{gamgambbnsd}
\end{figure}

In order to quantify these claims slightly further, we have taken the
results of Ref.~\cite{gunhabgamgam}
(Fig.~2) for the SM Higgs $b\anti b$ signal and
the $b\anti b$ background rate and multiplied the former by the
$\Gamma(\gam\gam)\br(b\anti b)$ ratio plotted in
Fig.~\ref{fighpgbcomp} and by the correction
factor $\tan^{-1}(\Gamma_{\rm exp}/\gampzero)/\tan^{-1}(\Gamma_{\rm
exp}/\Gamma^{\rm tot}_h)$ [see Eq.~(\ref{siggamgam})].  The resulting
signal and background rates are plotted
for $\ntc=4$ in Fig.~\ref{gamgamrates},
assuming that $L_{\rm eff}\equiv
G(y_{\pzero})L_{\epem}=20\fbi$, independent of $\mpzero$.
(As already stated, to achieve $G\gsim 1$ at the lowest masses
would require lowering the machine energy so that $\mpzero/\rts>0.1-0.2$.)
For the $b\anti b$ channel, the signal and background rates plotted
give $S/\sqrt B$ as plotted in Fig.~\ref{gamgambbnsd}.

We have also performed this same study for $\ntc=1$. The signal rate
is, of course, significantly reduced relative to $\ntc=4$
by virtue of the large decrease
in $\Gamma(\pzero\to \gam\gam)$. As Fig.~\ref{fighpgbcomp}
shows, we expect rates similar to those for a SM Higgs;
in the $b\anti b$ final state, $S$ ranges from 40 up to 170 as
$\mpzero$ goes from 20 to 200 GeV. The corresponding
$S/\sqrt B$ values are plotted in Fig.~\ref{gamgambbnsd}.
For $\mpzero>60\gev$, there is an excellent
chance that $\pzero$ detection will be possible
in $\gam\gam$ collisions even for the minimal $\ntc=1$ choice.

Of course, these results are not entirely realistic.
The $\gam\gam\to b\anti b$ background rate plotted
assumes an unrealistically small $b\anti b$
mass resolution of $\Gamma_{\rm exp}=5\gev$.  In addition, 
backgrounds from $\gam\gam \to c\anti c g$ and $\gam\gam\to b\anti b g$ are ignored.
(These are not suppressed by having $\vev{\lam\lam^\prime}\sim 1$.)
However, these three-jet backgrounds can
be largely eliminated by using topological tagging and cuts
designed to isolate the two-jet final state. The resulting
additional efficiency reduction for the $\pzero\to b\anti b$ signal
is typically no smaller than $\gsim 0.5$
(for single-$b$ topological tagging) \cite{japanese}.
Thus, $\pzero$ discovery at a $\gam\gam$ collider
in the $b\anti b$ final state will be very viable over a large mass range.

Once the $\pzero$ has been discovered, either in $\gam\gam$ collisions
or elsewhere, one can configure the $\gam\gam$ collision set-up so
that the luminosity is peaked at $\rts_{\gam\gam}\sim \mpzero$.
A very precise measurement of the $\pzero$ rate in the $b\anti b$
final state will then be possible if $\ntc=4$.
For example, rescaling the SM Higgs `single-tag'
results of Table 1 of Ref.~\cite{japanese} (which assumes
a peaked luminosity distribution with a total of $L=10\fbi$)
for the $106\gev\leq m_{jj}\leq 126\gev$ mass window to the case of
the $\pzero$ using the $[\Gamma(\pzero\to\gam\gam)\br(\pzero\to b\anti b)]/
[\Gamma(h\to\gam\gam)\br(h\to b\anti b)]$ ratio for $\ntc=4$,
plotted in Fig.~\ref{fighpgbcomp},
we obtain $S\sim 5640$ compared to $B\sim 325$,
after angular, topological tagging and jet cuts. This implies a statistical
error for measuring $\Gamma(\pzero\to \gam\gam)\br(\pzero\to b\anti b)$
of $\lsim 1.5\%$. Systematic errors will probably dominate.
Following the same procedure for $\ntc=1$, we find (at this mass) a
statistical error for this measurement of $\lsim 5\%$. Of course,
for lower masses the error will worsen.
For $\ntc=4$, we estimate an error for the $b\anti b$ rate
measurement still below $10\%$ even at a mass as low
as $\mpzero=20\gev$ (assuming the $\rts$ of the machine
is lowered sufficiently to focus on this mass without sacrificing luminosity).
For $\ntc=1$, we estimate an error for the $b\anti b$ rate measurement
of order $15-20\%$ for $\mpzero\sim 60\gev$.

Of course, it would be very interesting to measure rates in other
final state channels as well. The $\ntc=4$ total $\pzero$ rate shown
in Fig.~\ref{gamgamrates} (which can be further increased
once $\mpzero$ is known and the $\gam\gam$ collisions are
configured for a peaked, rather than broad, luminosity spectrum)
indicates that $\gam\gam\to\pzero\to \tauptaum$ and $gg$ will also
have large event rates.  Backgrounds are
probably too large in the $gg$ final state to obtain a robust
$\pzero$ signal. Backgrounds in the $\tauptaum$ channel
are not a large, but there is no sharp mass peak in
this channel. Still, if one configures the machine energy
and $\gam\gam$ collision set-up
so that the $\gam\gam$ luminosity is very peaked at an already known value
of $\mpzero$, a reasonably precise measurement of $\Gamma(\pzero\to
\gam\gam)\br(\pzero\to\tauptaum)$ might prove possible.
Detailed studies of what can be achieved
in the $gg$ and $\tauptaum$ channels should be performed.

For $\ntc=4$, it might even be possible to
detect the $\pzero$ in the $\gam\gam\to \pzero\to\gam\gam$ mode. The
(broad-luminosity-profile) total $\pzero$ production rate plotted in
Fig.~\ref{gamgamrates} is $>3500$ for $\mpzero>100\gev$,
for which masses $\br(\pzero\to \gam\gam)>0.006$ (Fig.~\ref{figbrs}).
The resulting total $\gam\gam\to\pzero\to\gam\gam$ event rate
ranges from a low of $\sim 20$ at $\mpzero\sim 100\gev$ to $\sim 50$
at $\mpzero\sim 200\gev$.  These rates can be
substantially increased if the $\gam\gam$ collision set-up
is optimized for a known value of $\mpzero$. Presumably the (one-loop)
irreducible $\gam\gam\to\gam\gam$ background is quite small.
But, one must worry
about jets that fake photons and possibly about back-scattered
photons that simply pass through into the final state without
interacting. (Presumably a minimum-angle cut could be used
to largely eliminate the latter.)  Once again, a detailed study
will be needed to reliably assess prospects for detection
of the $\pzero\to\gam\gam$ signal. Of course, if $\ntc=1$
the signal rates will be much smaller and $\pzero$
detection in the $\gam\gam$ final state will be quite difficult.

\section{$\pzero$ production at a muon collider}

In this section, we consider $s$-channel production of the
$\pzero$ at a future $\mu^+\mu^-$ collider. Given the mass eigenstate
compositions of Eq.~(\ref{eigenstates}) [see also Eqs.~(\ref{pzeroform})
and (\ref{pzeropform})] it is only the $\pzero$, and not the $\pzerop$,
that has tree-level coupling to $\mupmum$. The $s$-channel production
rate for the $\pzerop$ via its one-loop-induced coupling to $\mupmum$
will generally be much smaller than the tree-level $s$-channel production
rate for the $\pzero$ so long as there is no fine-tuned cancellation
between $m_4^{(2)}$ and $m_{10}^{(2)}$ in the expression for $\lam_\mu$,
Eq.~(\ref{yukcoups}). Our discussion will
expand on the results of Ref.~\cite{mumu}. Related work can
be found in Ref.~\cite{mumualso}. As in
previous sections, we shall employ the specific $\pzero$ properties
as predicted by the couplings of Eq.~(\ref{pcoups}),
(\ref{pgamgamcoup}) and (\ref{pggcoup}). However, the general
features of our results will apply in a broad sense to other
choices for the fermionic couplings and
to other models of a strongly interacting electroweak sector.

Most of our specific numerical results will be presented for $\ntc=4$.
However, we shall also comment on and give some
numerical results for the minimal $\ntc=1$ possibility.
For the moment, we note that $\Gamma(\pzero\to \mupmum)$, which
controls the muon-collider $\pzero$ production rate,
does not depend on $\ntc$. This
is a unique advantage of the muon collider as opposed to all the
other colliders considered; at the other colliders the $\pzero$ production
rate is proportional to the square of the strength of the
$gg$ or $\gam\gam$ coupling of
the $\pzero$, both of which are proportional to $\ntc^2$. Thus, although
discovery of the $\pzero$ for $\ntc=1$ may be possible over
a reasonable mass range at the LHC and at a $\gam\gam$ collider
(see earlier sections), a high precision measurement of
its properties may only be possible at a muon collider. Further, if the
$\pzero$ is very light, it is very possible
that only a muon collider will be able to discover it.

We have already noted that for $\Lambda<10\tev$,
$\mpzero({\rm one-loop})$ will be smaller than $\mz$.
Thus, as in earlier sections, it is important to consider small masses
as well as masses up to the maximum $200\gev$ value considered in this paper.
For much of this mass range a muon collider would provide an excellent probe
of the $\pzero$ via the direct $s$-channel production
process $\mupmum\to\pzero$.  The basic reasons as to why this is so are the
following. First, the $\pzero$ has a sizeable $\mupmum$ coupling
[Eq.~(\ref{pcoups})], and this coupling is independent of $\ntc$. Second,
the muon collider has the (unique) ability to achieve
a very narrow Gaussian spread, $\sigrts$, in $\rts$, as
necessary to obtain a large $\pzero$ cross section given
the very narrow width of the $\pzero$ (as plotted in
Fig.~\ref{figgamtot}).
Indeed, one can achieve $R=0.003\%$ beam energy resolution
with reasonable luminosity at the muon collider, leading to
\begin{equation}
\sigrts\sim 1\mev\left({R\over 0.003\%}\right)\left({\rts\over
50\gev}\right)\,.
\label{sigrtsdef}
\end{equation}
In addition,
the beam energy can be very precisely tuned ($\Delta E_{\rm beam}\sim
10^{-6}$ is achievable \cite{raja}) as is crucial
in scanning for a resonance with a width as small as that predicted for the
$\pzero$.

\begin{figure}[h]
\epsfysize=8truecm
\centerline{\epsffile{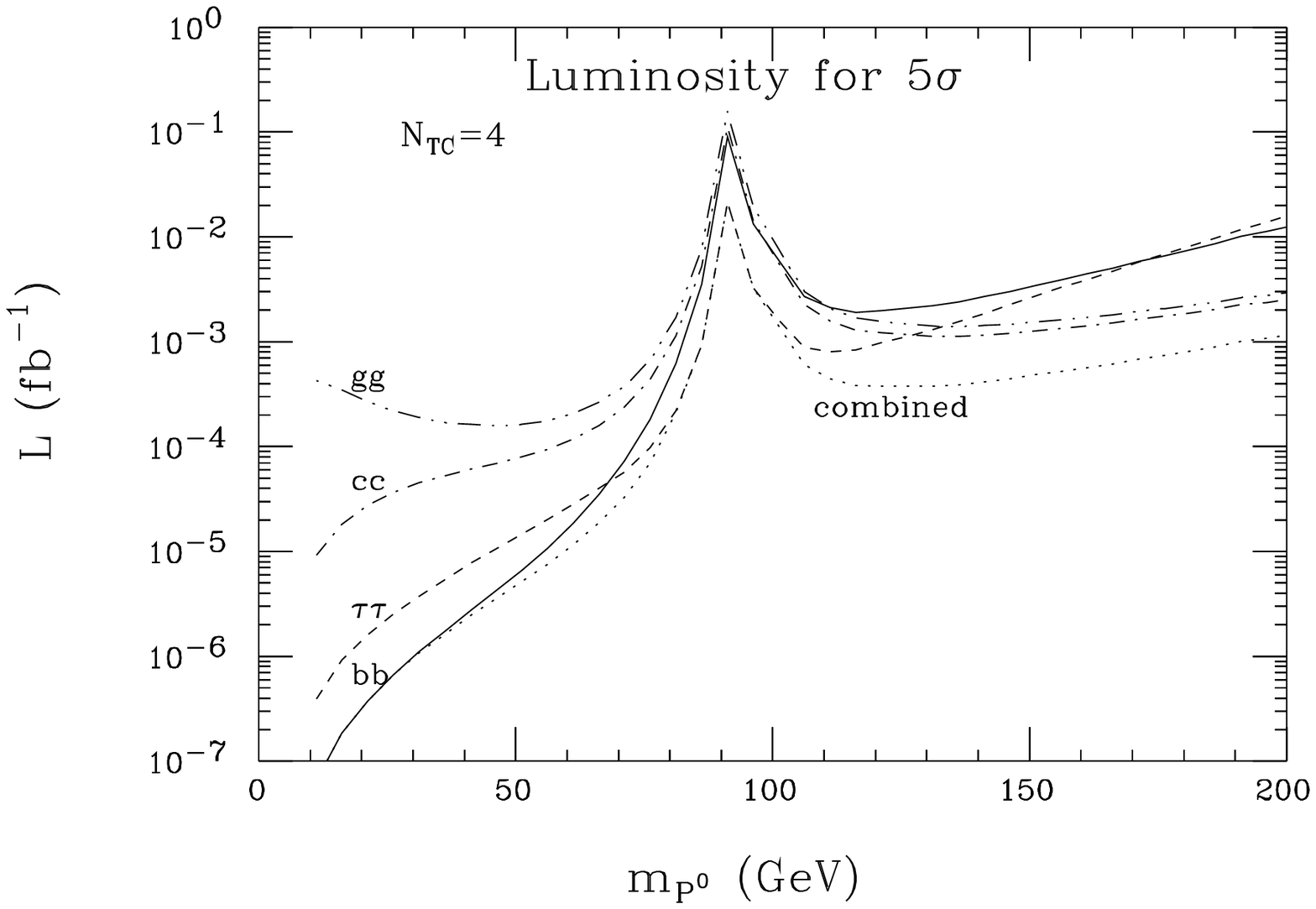}}
\smallskip
\noindent
\caption{Luminosity required for a $S/\protect\sqrt B=5$ signal
for the $\pzero$ in various `tagged' channels
($b\anti b$, $\tauptaum$, $c\anti c$, $gg$) and for the optimal
combination of these four channels. This plot is for $\ntc=4$.}
\label{figlum}
\end{figure}

\begin{figure}[h]
\epsfysize=8truecm
\centerline{\epsffile{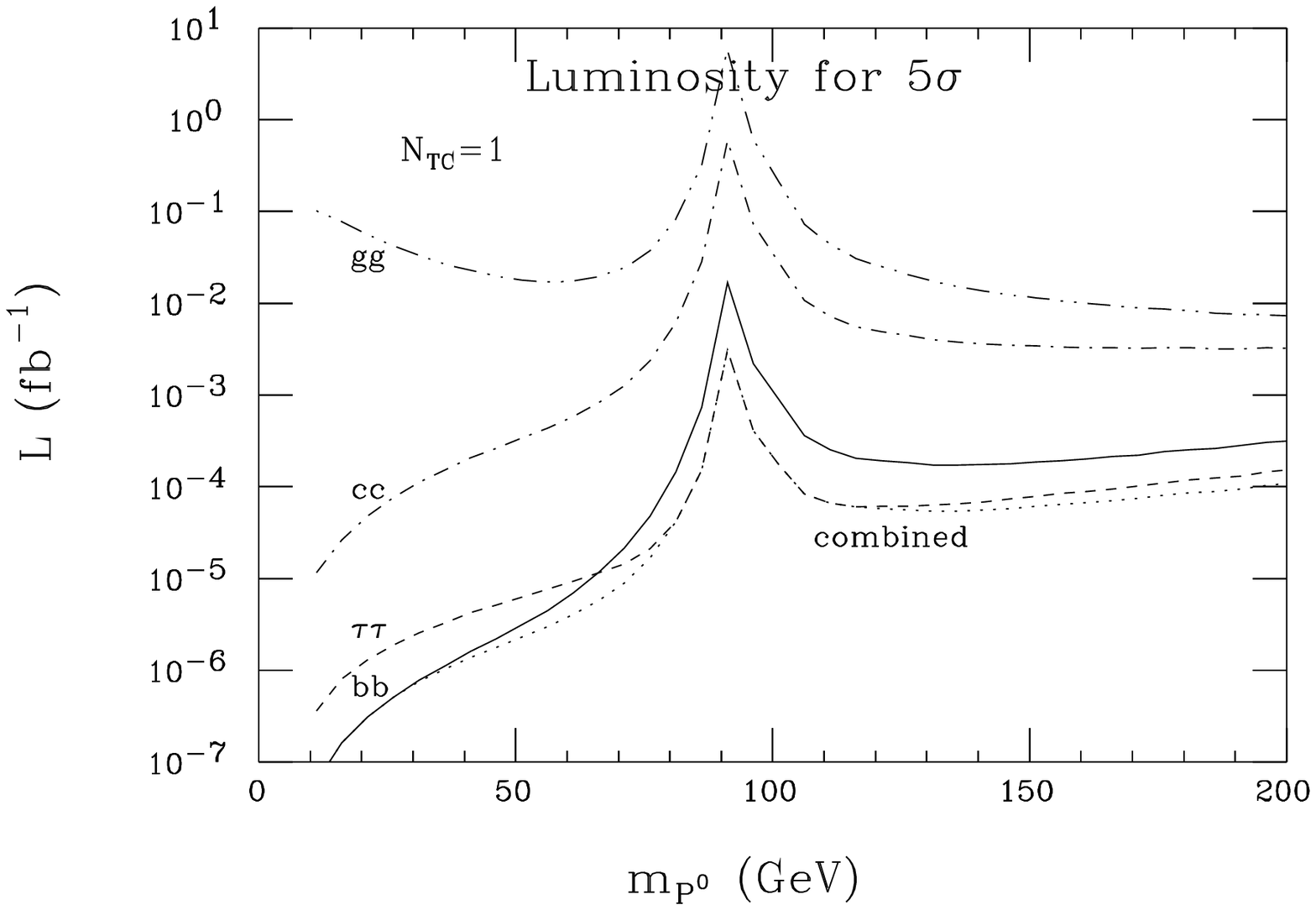}}
\smallskip
\noindent
\caption{As in Fig.~\ref{figlum} but for $\ntc=1$.}
\label{figlumntceq1}
\end{figure}

To quantitatively assess the ability of the muon collider to discover
the $\pzero$ we have proceeded as follows. We compute the $\pzero$
cross section by integrating over the resonance using a $\rts$
distribution given by a Gaussian of width $\sigrts$ (using $R=0.003\%$)
modified by bremsstrahlung photon emission.
(Beamstrahlung is negligible at a muon collider.) See Ref.~\cite{bbgh}
for more details. We separate $b\anti b$, $\tau^+\tau^-$,
$c\anti c$ and $q\anti q/ gg$
final states by using topological and $\tau$ tagging with efficiencies
and mis-tagging probabilities as estimated by B. King \cite{bking}.
These were summarized in the previous section
and used for our $\epem$ collider discussion.
Further, only events in which
the jets or $\tau$'s have $|\cos\theta|<0.94$ (corresponding
to a nose cone of $20^\circ$) are considered.
As in the $\epem$ collider study, a jet final state is deemed
to be: $b\anti b$ if one or more jets is tagged as a $b$; $c\anti c$
if no jet is tagged as a $b$, but one or more jets is tagged as a $c$;
and $q\anti q/gg$ if neither jet is tagged as a $b$ or a $c$.
Background and signal events are analyzed in exactly the same manner.

In Fig.~\ref{figlum}, we plot, for $\ntc=4$,
the integrated luminosity required to
achieve $S_i/\sqrt {B_i}=5$ in a given channel, $i$
(as defined after tagging),
taking $\rts=\mpzero$. We also show the luminosity $L$ needed
for $\sum_k S_k/\sqrt{\sum_k B_k}=5$, where the optimal choice
of channels $k$ is determined for each $\mpzero$. We observe that
very modest $L$ is needed unless $\mpzero\sim\mz$.
Corresponding results are presented for $\ntc=1$ in Fig.~\ref{figlumntceq1}.
For $\ntc=1$, the luminosity required for discovery in the $gg$
final state is,
of course, much higher than for $\ntc=4$
since the $gg$ branching ratio is much smaller.
However, the luminosity required in the $\tauptaum$ and $b\anti b$ channels
is essentially unchanged at low $\mpzero$ and is very substantially smaller
at high $\mpzero$.  Thus, aside from decreased accuracy
for measuring $\Gamma(\mupmum\to \pzero)\br(\pzero\to gg)$ (see later),
small $\ntc$ is actually optimal for $\pzero$ discovery and study
at a muon collider.

\subsection{Scanning and Centering}

Of course, luminosity must be devoted to scanning for the
$\pzero$ if it has not yet been discovered or to centering
on $\rts\simeq\mpzero$ assuming that $\mpzero$
has been measured up to a certain level of uncertainty
at another type of accelerator. As we have discussed,
the $\pzero$ is very likely to have been discovered
at the LHC in the $gg\to\pzero\to\gam\gam$ mode
if $\mpzero>50\gev$ or so, depending on $\ntc$.
If it is observed in this mode,
then its mass will be very accurately measured, almost certainly
to better than $\delm\sim 100\mev$.
We have also seen that for $\mpzero<50\gev$, discovery
or exclusion at the $N_{SD}=3$ level, or better,
will be possible using the $\epem\to\gam\pzero$ mode if $\ntc=4$
(but not if $\ntc=1$). In any case,
the mass uncertainty associated
with $\pzero$ observation in $\epem\to\gam\pzero$ collisions
will be much larger --- $\delm>{\rm few}\gev$ is likely.

\begin{figure}[h]
\epsfysize=8truecm
\centerline{\epsffile{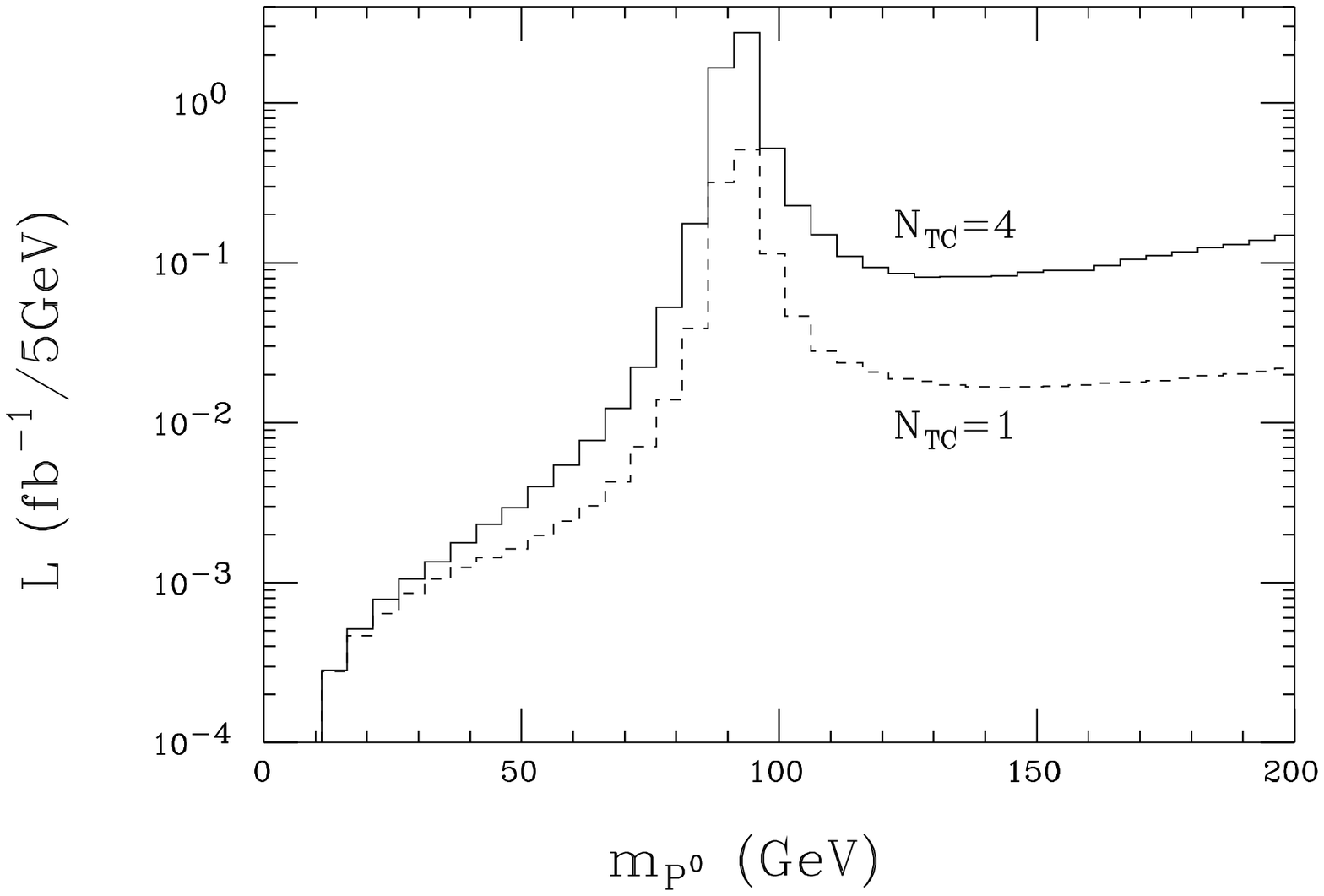}}
\smallskip
\noindent
\caption{Maximum luminosity needed to either discover
or exclude the $\pzero$ at the $S/\protect\sqrt B=3$ level
by scanning in various 5 GeV intervals, based on using the optimal
combination of channels (as defined by the tagging
probabilities given in the text).
Results are given for $\ntc=4$ and $\ntc=1$.}
\label{figscanlum}
\end{figure}

To estimate the luminosity required
at the muon collider for scanning a given interval
so as to either discover or eliminate the $\pzero$, we have adopted
the following approach.  We imagine choosing $\rts$ values separated
by $2\sigrts$. We assume the worst case
scenario in which the resonance sits midway between the two $\rts$
values.  The signal and (separately) background rates for these two $\rts$
values are summed together (for the optimal channel combination)
and the net $N_{SD}\equiv (S_1+S_2)/(B_1+B_2)^{1/2}$ is computed.
We require $N_{SD}=3$ to claim a signal. The luminosity required
for a successful scan of a given interval is computed
assuming that the resonance lies between the  last two scan points.
This, in combination with the fact that $\sigrts$ for $R=0.003\%$ is
typically a factor of two smaller than $\gampzero$
(implying that points further away than $\sigrts$ from the resonance
could be usefully included in establishing a signal) will imply
that the integrated luminosities we compute in this
way are quite conservative. It is important to note that
if a signal is seen for the $\pzero$ in the above-described scanning
procedure, then $\mpzero$ will be determined to within $\lsim \sigrts$;
in addition, we will obtain a first rough measurement of $\gampzero$.

In Fig.~\ref{figscanlum}, we give the integrated luminosity
required to either discover or exclude the
$\pzero$ at the $N_{SD}=3$ level
after scanning the indicated $5\gev$ intervals, assuming
that $\mpzero$ lies within that interval.
As expected, rather modest luminosity is required except
for intervals near the peak in the background
at $\rts\sim \mz$ due to the $\mupmum\to Z$ process.
Note that the scan luminosity required for $N_{SD}=3$ at large $\mpzero$
is substantially smaller for $\ntc=1$ than for $\ntc=4$.

\begin{figure}[h]
\epsfysize=8truecm
\centerline{\epsffile{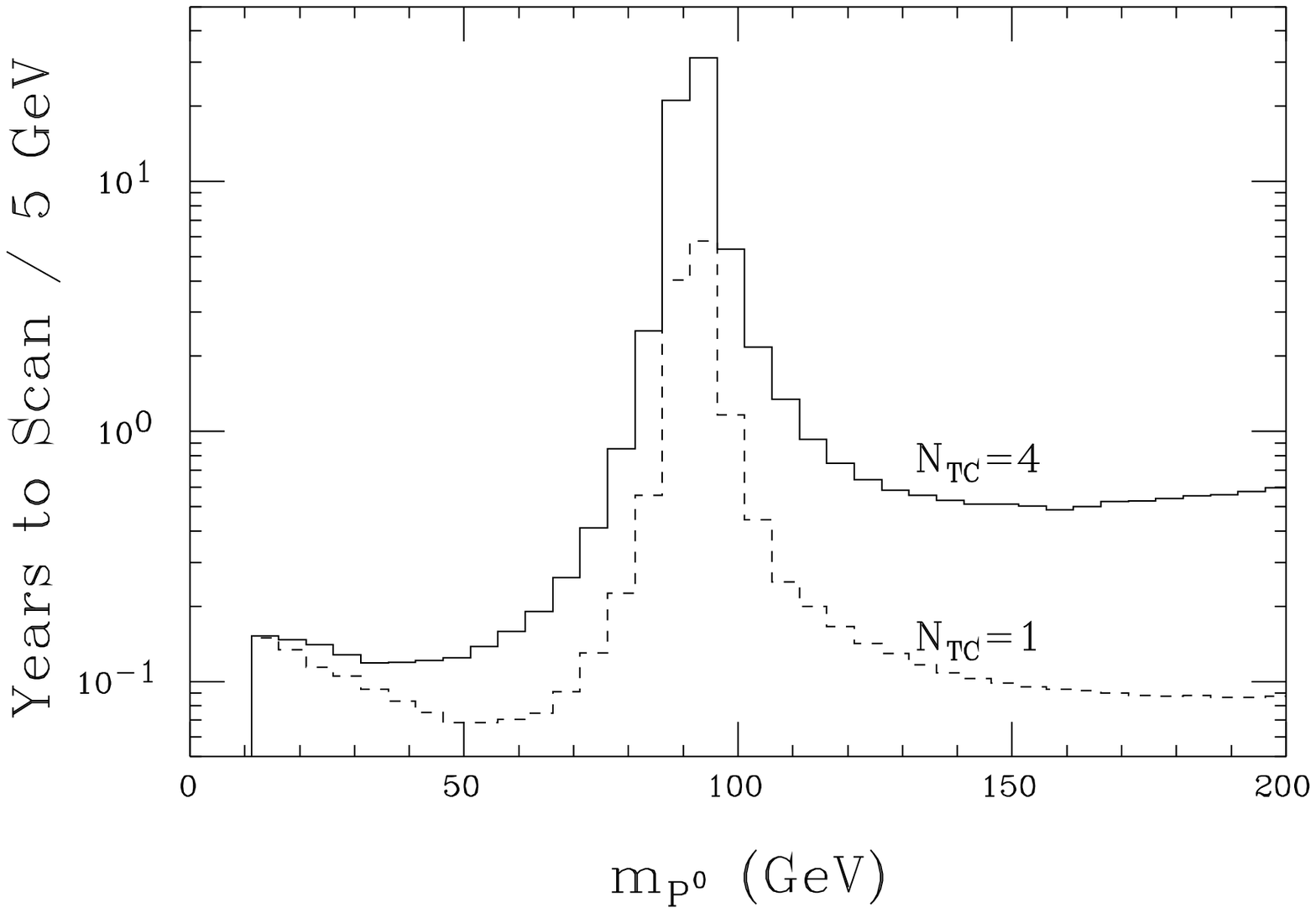}}
\smallskip
\noindent
\caption{The number of years required to either discover
or exclude the $\pzero$ at the $S/\protect\sqrt B=3$ level
by scanning in various 5 GeV intervals, based on
the results of Fig.~\protect\ref{figscanlum} and assuming
luminosity as given in Eq.~(\protect\ref{lumscaling}).
Results are given for $\ntc=4$ and $\ntc=1$.}
\label{figscanyears}
\end{figure}

To assess the implications of Fig.~\ref{figscanlum}, we must
discuss the luminosity that will be available at a $\mupmum$
collider as a function of $\rts$.  We will do so assuming
that the first muon collider is designed for optimum performance
at $\rts\sim 100\gev$, and that for $R=0.003\%$
a yearly integrated luminosity of $L=0.1\fbi$ can be achieved \cite{palmer}.
For energies on either side of
$\rts=100\gev$, the integrated luminosity will vary according to
\cite{palmer}:
\bea
L_{\rm year}&=&0.1\fbi
\left({\rts\over100\gev}\right)^2\,,\quad \rts<100\gev\,,\nn\\
L_{\rm year}&=& 0.1\fbi
\left({\rts\over100\gev}\right)^{4/3}\,,\quad \rts>100\gev\,.
\label{lumscaling}
\eea
Using these expectations, we can determine the maximum (given
the rather pessimistic assumptions employed
for the scan) number of years of operation required to
scan each of the 5 GeV intervals of Fig.~\ref{figscanlum} and
either exclude or discover the $\pzero$ at the $N_{SD}=3$ level.
The results for $\ntc=4$ and $\ntc=1$
are plotted in Fig.~\ref{figscanyears}.
We note that at most $\sim 3$ years would be required to scan
the entire range of $\mpzero< 80\gev$
in which $\mpzero$ is expected to lie if one-loop mass contributions
are dominant and assuming $\Lambda<10\tev$.
However, should $\mpzero$ be near $\mz$, then prior discovery
of the $\pzero$ at another accelerator will be crucial
in order that we be able to center on $\rts\sim\mpzero$
and perform detailed studies.

>From our earlier discussions,
such prior discovery at the LHC appears to be quite
likely if $\mpzero>50\gev$ (possibly lower for $\ntc=4$, pending a detailed
study), and will result in mass uncertainty of less than $\delm\sim 100\mev$.
Prior discovery at an $\epem$ collider is also reasonably likely
if $\mpzero\lsim 50\gev$ and $\ntc=4$ (or larger)
although the mass uncertainty will
be much larger, $\delm\sim {\rm few}\gev$.
The very worst case, and one for which LHC discovery will provide vital
input, is $\mpzero\simeq\mz$. For this case, if $\ntc=4$ ($\ntc=1$)
about $0.016\fbi$ ($0.003\fbi$)
(see Figs.~\ref{figlum} and \ref{figlumntceq1}, optimal channel combination)
would be required to exclude
or detect the $\pzero$ at the $N_{SD}=3$ level at
each of $\sim 100\mev/\sigrts=25$ points separated by $2\sigrts\sim 4\mev$.
At most, this would require about 1/2 (1/10) year of collider operation,
assuming $L\sim 0.1\fbi$ per year at $\rts\sim \mz$. Conversely, without
the LHC input we would be very lucky to
find the $\pzero$ by scanning at the muon collider when
$\mpzero\sim \mz$.

Let us next suppose that $\mpzero<30-50\gev$
and that the LHC does not discover the $\pzero$. However,
let us assume that $\ntc=4$ and
that an $\epem$ collider does discover the $\pzero$
and measures $\mpzero$ to within (to be on the pessimistic
side) $\delm\sim 5\gev$. From Fig.~\ref{figscanyears},
we see that no more than $\sim 0.15$ years is required to scan any
$5\gev$ interval in the $\mpzero<50\gev$ range so as to
discover and center on $\rts\simeq\mpzero$ within $<\sigrts$.
Of course, if $\ntc=1$ and $\mpzero<60\gev$,
it is possible that neither the LHC nor the $\gam\gam$ collider will
be able to discover the $\pzero$. Then, the previously-discussed
muon collider scan, possibly taking up to a year, will be necessary.

Overall, it will almost certainly
be possible to center on $\rts\simeq\mpzero$
at a muon collider in significantly less than a year or two
of operation.  Once centered, the important question is how
well we will be able to determine the 6 $m_i$ parameters of
the effective Yukawa Lagrangian [see Eq.~(\ref{Ly})] associated
with the $T_3=-1/2$ sector
--- $m_2^\prime$, $m_4$, $m_4^{(2)}$, $m_9$,
$m_{10}$ and $m_{10}^{(2)}$ --- and the value of $\ntc$. All of these
go into determining the properties of the $\pzero$ in production
and decay and are potentially measurable if enough accurate
experimental input is available. In order to
assess the situation in this regard, we must first determine
the accuracy with which the basic experimental
observables can be measured.

\subsection{Precision Measurements}

At the muon collider, the quantities that we can hope to
measure with precision are the rates in the various tagged
channels and the total width of the $\pzero$.  This
set of observables is the same as for the SM Higgs boson,
for which the procedures have been studied and discussed
in detail in Ref.~\cite{bbgh}. We will follow very similar
procedures here. In particular,
for the determination of $\gampzero$ we employ the optimized
three-point scan described in Ref.~\cite{bbgh} in which
luminosity of amount $L_1$ is employed at $\rts\simeq\mpzero$
and of amounts $2L_1$ at each of the two points $\rts\simeq\mpzero\pm 2\sigrts$,
implying a total luminosity for the scan of $5L_1$. We will
assume that 4 years of running at $R=0.003\%$
is employed for the three-point scan, corresponding to a
total integrated luminosity of $5L_1$
equal to 4 times the yearly luminosity given in Eq.~(\ref{lumscaling}).
The equivalent luminosity that would have
to be expended at $\rts\simeq\mpzero$ in order
to obtain the same statistics for the various channel
rate measurements is about 3/4 of the $L$ employed for the scan.
(This is a somewhat larger fraction than in the Higgs case
due to the fact that, for the parameters chosen, $\gampzero$
is significantly larger than the Higgs width at the same mass.)
In other words, the statistics for the channel rates will
be approximately the same as obtained if $L$ equal
to 3 times the yearly luminosity of Eq.~(\ref{lumscaling})
is accumulated at $\rts\simeq\mpzero$.

\begin{figure}[h]
\epsfysize=8truecm
\centerline{\epsffile{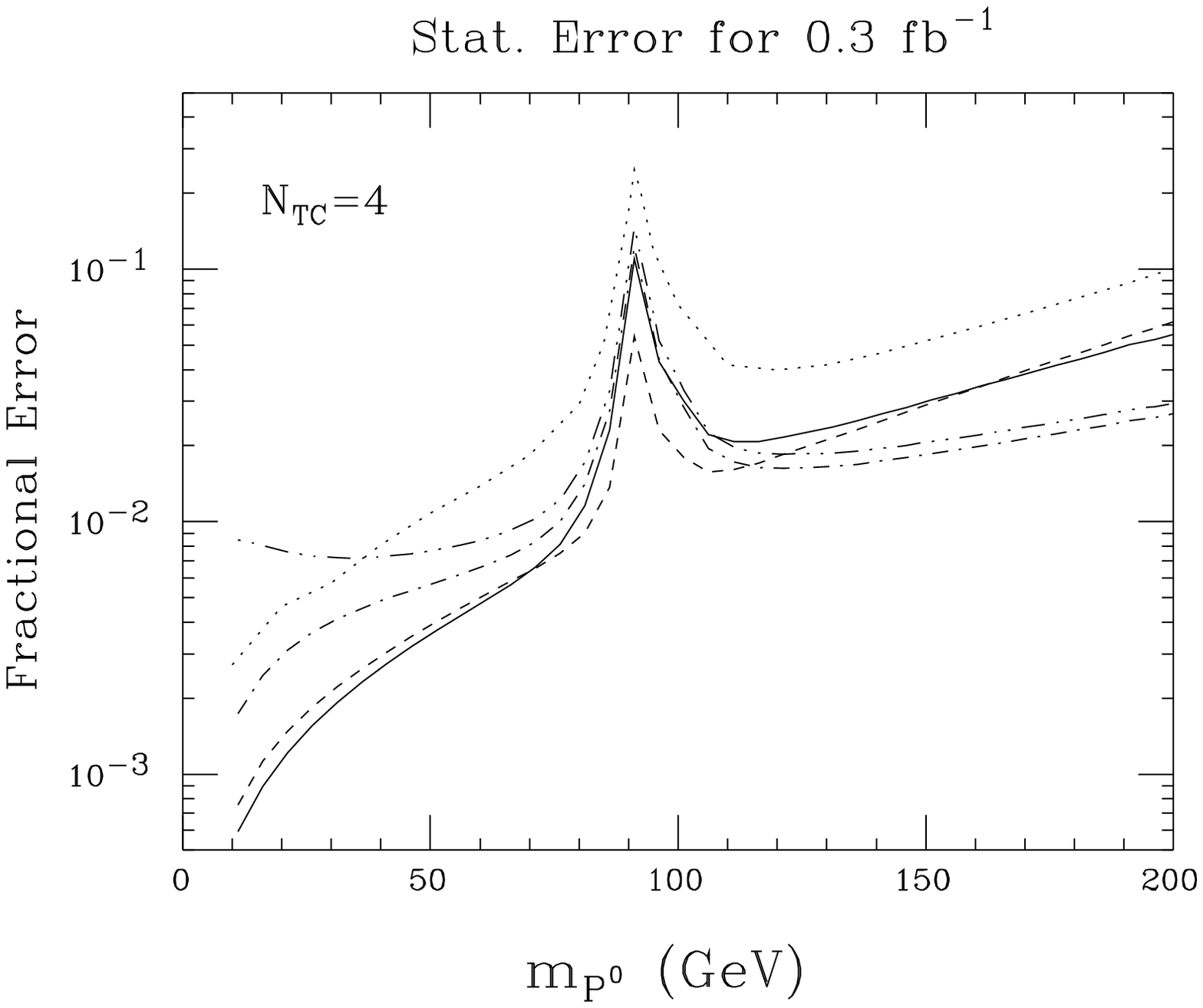}}
\smallskip
\noindent
\caption{The fractional statistical error
for the various tagged channel rates obtained
if $L=0.3\fbi$ is accumulated (with $R=0.003\%$)
at $\protect\rts\simeq\mpzero$.
Also shown is the error for $\gampzero$ obtained
using the optimal three-point scan and $L=0.4\fbi$.
Results are for $\ntc=4$.
Curve notation is: $b\anti b$ (solid); $\tauptaum$ (dashes);
$c\anti c$ (dot-dash); $gg$ (dot-dot-dash); $\gampzero$ (dots).}
\label{figerrors}
\end{figure}

\begin{figure}[h]
\epsfysize=8truecm
\centerline{\epsffile{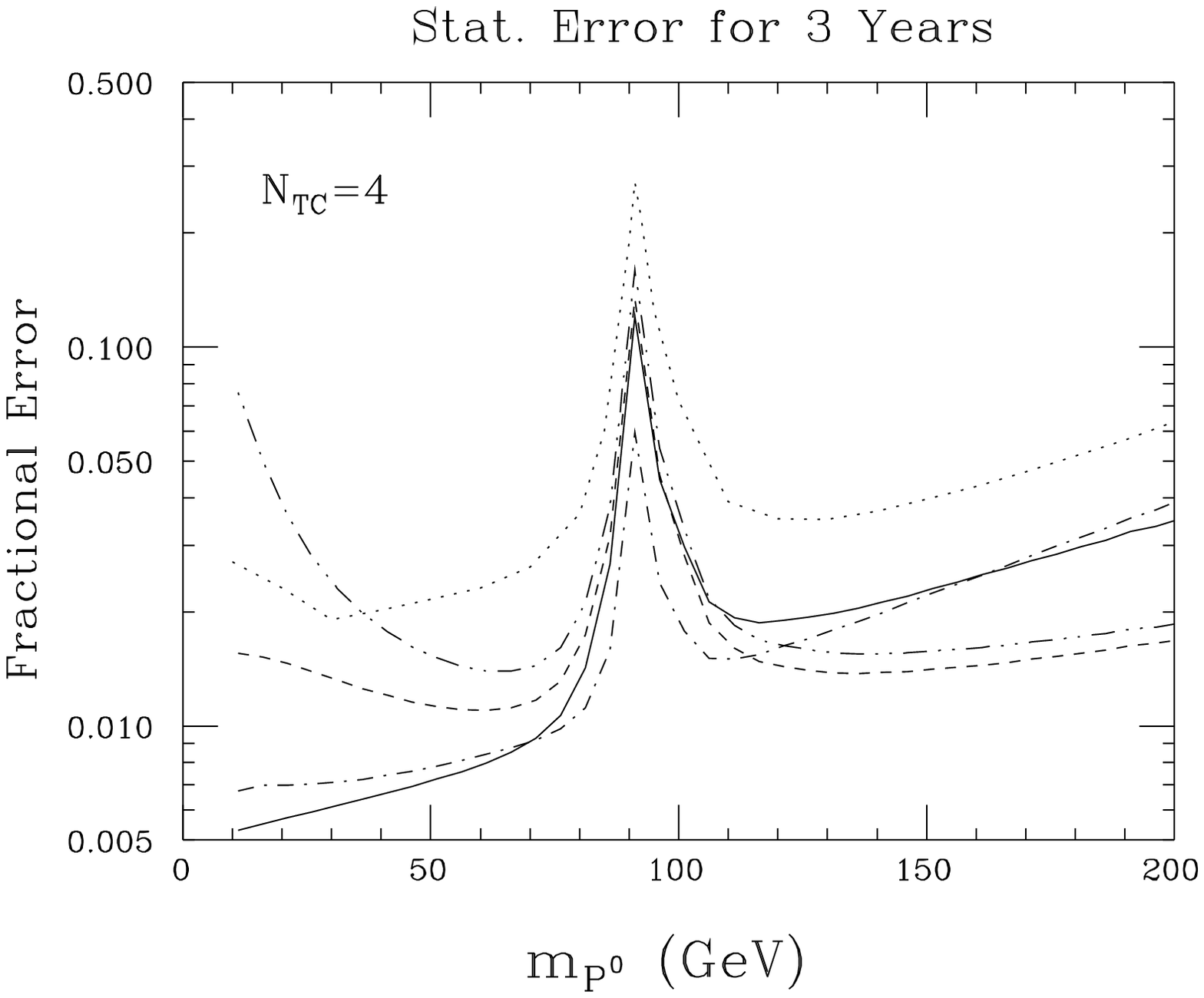}}
\smallskip
\noindent
\caption{The fractional statistical error
for the various tagged channel rates obtained after 3 years
of operation at $\protect\rts\simeq\mpzero$  or, equivalently, after
4 years of running devoted to the optimal three-point scan,
assuming yearly luminosity as given in Eq.~(\protect\ref{lumscaling})
at $R=0.003\%$. Also shown is the fractional error for $\gampzero$ obtained
after devoting 4 years to the optimal three-point scan.
Results are for $\ntc=4$.
Curve notation is: $b\anti b$ (solid); $\tauptaum$ (dashes);
$c\anti c$ (dot-dash); $gg$ (dot-dot-dash); $\gampzero$ (dots).}
\label{figerrorsnyear}
\end{figure}

\begin{figure}[ht]
\epsfysize=8truecm
\centerline{\epsffile{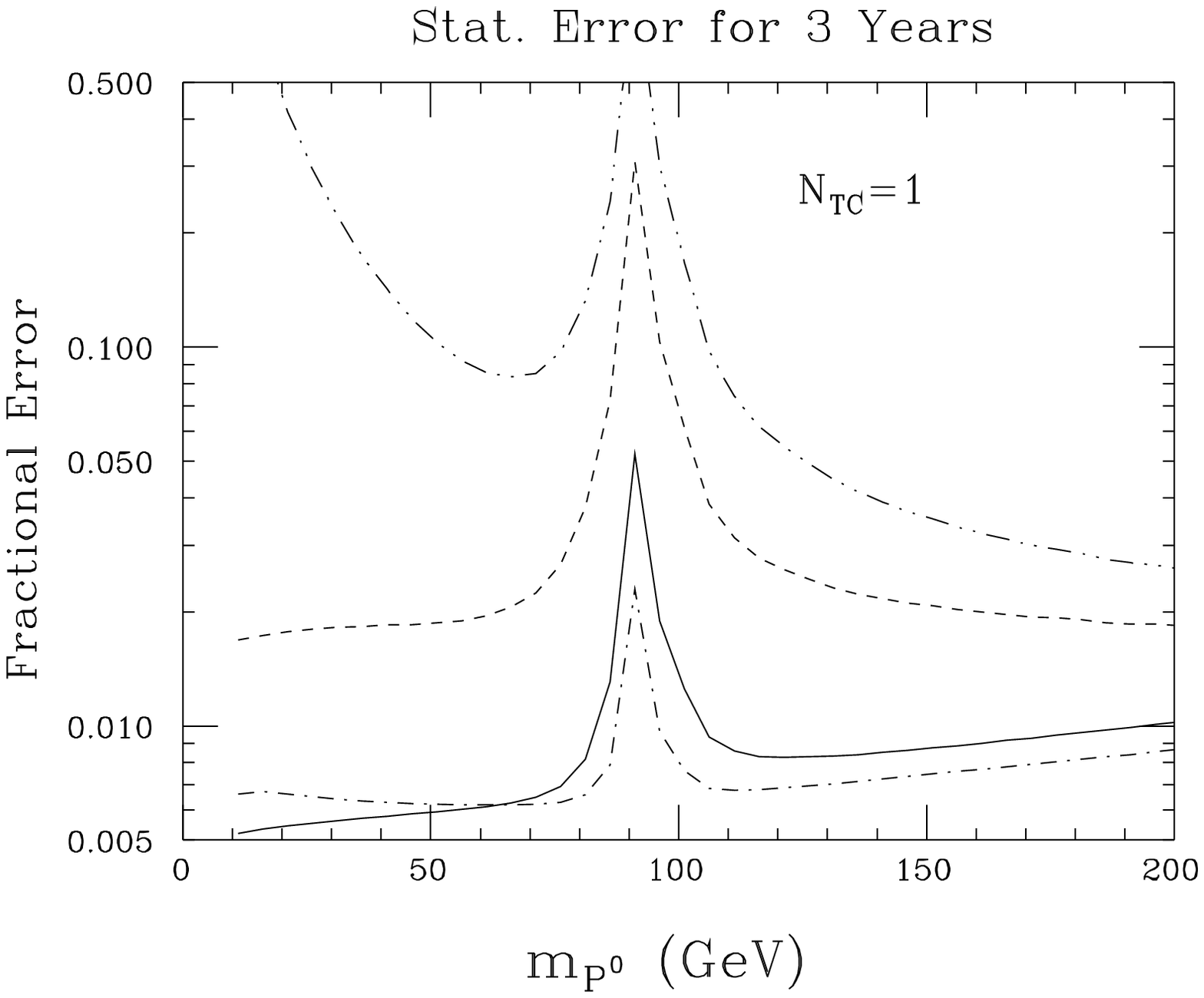}}
\smallskip
\noindent
\caption{The fractional statistical error
for the various tagged channel rates obtained after 3 years
of operation at $\protect\rts\simeq\mpzero$  or, equivalently, after
4 years of running devoted to the optimal three-point scan,
assuming yearly luminosity as given in Eq.~(\protect\ref{lumscaling})
at $R=0.003\%$. Results are for $\ntc=1$.
Curve notation is: $b\anti b$ (solid); $\tauptaum$ (dashes);
$c\anti c$ (dot-dash); $gg$ (dot-dot-dash).}
\label{figerrorsnyearntceq1}
\end{figure}

The only difference in our optimal three-point
scan procedures for determining $\gampzero$ as compared to the Higgs
boson case is the following.  At each $\mpzero$, we compute the
error in $\gampzero$ by using the total signal $S=\sum_k S_k$ and background
$B=\sum_k B_k$ rates
for the combination of channels $k$ which gives the largest net $S/\sqrt B$,
as described earlier.

We begin by plotting in Fig.~\ref{figerrors}
the statistical errors for the $b\anti b$,
$\tauptaum$, $c\anti c$ and $gg$ channel rates using the
channel identification (and mis-identification) efficiencies
quoted earlier, assuming that $L=0.3\fbi$ is accumulated at
$\rts\simeq\mpzero$.
Also shown is the result for the error in the total width
obtained for $L=0.4\fbi$ by employing the optimal three-point
scan procedure. (As noted earlier, the $L=0.4\fbi$ scan
gives approximately the same statistics for the channel
rates as does $L=0.3\fbi$ sitting on-resonance, $\rts\simeq\mpzero$.)
The results in this figure are for $\ntc=4$.

These errors must be adjusted for
the fact that the machine luminosity is a variable function
of $\rts$.  Thus, in Fig.~\ref{figerrorsnyear} we plot
the tagged channel fractional statistical errors assuming
3 years of running at $\rts\simeq\mpzero$
with luminosity given by Eq.~(\ref{lumscaling}).
As described earlier, these errors are approximately the same as
would be achieved after devoting 4 years of machine operation
to the optimal three-point scan of the $\pzero$ resonance.
Also shown is the result for the $\gampzero$ error
if 4 years of running with yearly luminosity as given in Eq.~(\ref{lumscaling})
is devoted to the three-point optimal scan. These results are for
$\ntc=4$. The resulting errors are encouragingly small.

For small values of $\ntc$, the error for the $gg$ channel rate will
be substantially worse, while the errors for the $b\anti b$ and $\tauptaum$
channels will be better.  This is illustrated in
Fig.~\ref{figerrorsnyearntceq1} in the case of $\ntc=1$.
(The error for $\gampzero$ is not very different from the $\ntc=4$ case.)
We see that, with the exception of small $\mpzero$ (and, of course,
$\mpzero\sim m_Z$), the $gg$ channel rate error is nonetheless not particularly
large, only rising above $10\%$ for $\mpzero\lsim 50\gev$.

\section{Determining the Parameters of the Model}

In the previous sections, we have described the experimental
information, and associated statistical errors,
that will probably become available regarding the $\pzero$.
We now consider the strategy for using this experimental
information to extract the underlying parameters of the model.

Let us first consider the limit
in which the various probabilities for
channel identification and mis-identification are precisely known
and, in addition, uncertainty
arising from the absolute normalization of event rates is small.
In this limit, the errors for the various quantities we consider
are purely statistical. In practice,
additional contributions to the errors for the model parameters
will arise from systematic errors associated
with the determination of the channel
tagging and mis-tagging probabilities (by simulation and experimental
studies) and overall event rate normalization.  However,
without a detailed detector scenario and study, we cannot assess
the level of the systematic errors that will arise.

The experimentally well-measured quantities
that could be available to us are:
\bit
\item $\gampzero$ from the muon collider;
\item $\Gamma(\pzero\to gg)\br(\pzero\to \gam\gam)$ from the LHC;
\item $\Gamma(\pzero\to \gam\gam)\br(\pzero\to b\anti b)$ from
the $\gam\gam$ collider; and
\item $\Gamma(\pzero\to\mupmum)\br(\pzero\to F)$
for $F=b\anti b,\tauptaum,gg$ from the muon collider
(after unfolding the channel mis-identification).
\eit
We have discounted the $\epem$ data because of the very large
errors that are anticipated; see Fig.~\ref{figeetogamperrors}
for $\ntc=4$ results --- for $\ntc=1$, discovery
of the $\pzero$ is problematical.
Using the measured value of $\gampzero$, we can convert
the second, third and fourth items of the above list into
results for $\Gamma(\pzero\to gg)\Gamma(\pzero\to\gam\gam)$,
$\Gamma(\pzero\to \gam\gam)\Gamma(\pzero\to b\anti b)$
and $\Gamma(\pzero\to \mupmum)\Gamma(\pzero\to F)$
($F=b\anti b,\tauptaum,gg$), respectively.

If we employ the constraints on the parameters 
deriving from the quark and lepton masses, Eq.~(\ref{mf}), then the experimentally
measured quantities listed have the following reduced parameter dependences:
\bea
\Gamma(\pzero\to gg)\Gamma(\pzero\to \gam\gam)&\propto&
{\ntc^4\over v^4}\,;\label{gggamgam}\\
\Gamma(\pzero\to\gam\gam)\Gamma(\pzero\to b\anti b)&\propto&
{\ntc^2\over v^2} {\left[4m_2^\prime-3\mb\right]^2\over v^2}\,;\label{gamgambb}\\
\Gamma(\pzero\to\mupmum)\Gamma(\pzero\to b\anti b)&\propto&
{\left[{4\over 3}m_4^{(2)}-{1\over 3}m_\mu\right]^2\over v^2}\
{\left[4m_2^\prime-3\mb\right]^2\over v^2}
\,;\label{mumubb}\\
\Gamma(\pzero\to\mupmum)\Gamma(\pzero\to\tauptaum)&\propto&
{\left[{4\over 3}m_4^{(2)}-{1\over 3}m_\mu\right]^2\over v^2}\
{\left[{4\over 3}m_4-{1\over 3}\mtau\right]^2\over v^2}
\,;\label{mumutautau}\\
\Gamma(\pzero\to \mupmum)\Gamma(\pzero\to gg)&\propto&
{\left[{4\over 3}m_4^{(2)}-{1\over 3}m_\mu\right]^2\over v^2}
{\ntc^2\over v^2}
\,.\label{mumugg}
\eea
The following strategy then emerges.
\bit
\item Determine $\ntc/v$ from Eq.~(\ref{gggamgam});
\item Determine $|4m_2^\prime-3m_b|/v$ from $\ntc/v$ and Eq.~(\ref{gamgambb});
\item Determine $|{4\over 3}m_4^{(2)}-{1\over 3}m_\mu|/v$
from $\ntc/v$ and Eq.~(\ref{mumugg});
\item Determine $|{4\over 3}m_4-{1\over 3}m_\tau|/v$
from $|{4\over 3}m_4^{(2)}-{1\over 3}m_\mu|/v$
and Eq.~(\ref{mumutautau});
\item Determine $|4m_2^\prime-3m_b|/v$ from
$|{4\over 3}m_4^{(2)}-{1\over 3}m_\mu|/v$ and Eq.~(\ref{mumubb}).
\eit
Note that the $\gam\gam$ collider measurement allows a consistency
cross check, which amounts to checking
that the formulas for the $gg\pzero$ and $\gam\gam\pzero$
anomalous couplings are both correct.
At this point we can also check if the $\mu$ and $\tau$
Yukawa couplings obey
\beq
{\lam_\mu\over\lam_\tau}={|{4\over 3}m_4^{(2)}-{1\over 3}m_\mu|
\over |{4\over 3}m_4-{1\over 3}m_\tau|}={m_\mu\over m_\tau}
\,.\label{lepuniv}
\eeq
While Eq.~(\ref{lepuniv}) would be quite a natural possibility,
it is not required by the theory. If the relation holds,
then, unless $m_4$ and $m_4^{(2)}$
obey the fine-tuned relation $(m_4/m_\tau+m_4^{(2)}/m_\mu)=1/2$,
we must have $m_4^{(2)}/m_\mu=m_4/m_\tau$, which, in turn,
implies $m_{10}^{(2)}/m_\mu=m_{10}/m_\tau$ and $m_4^{(2)}/m_{10}^{(2)}=
m_4/m_{10}$.

To proceed further we must assume a value for $v$.
In the context of the present model, $v=246\gev$.\footnote{In a more general
walking technicolor model, the effective value of $v$ could, however,
be different.}
We can then use the above information as follows. First, the
(unambiguous) determination
of $\ntc/v$ would yield a result for $\ntc$.
Second, because
only the absolute values of the Yukawa couplings
can be experimentally measured, for the given $v$ we would be left with
two-fold ambiguities for $m_4^{(2)}$,
$m_4$ and $m_2^\prime$.  For each of the eight choices, we would
determine $m_9$, $m_{10}$ and $m_{10}^{(2)}$ from
$m_2^\prime$, $m_4$ and $m_4^{(2)}$ using Eq.~(\ref{mf}).
As noted above, if Eq.~(\ref{lepuniv}) holds, then the only
natural (non-fine-tuned) solutions would have
$m_4^{(2)}/m_{10}^{(2)}=m_4/m_{10}$,
thereby reducing the potential eight-fold ambiguity to a four-fold
ambiguity.

One could also ask how consistent each of the solutions
is with the LQ mass-squared contributions being small in comparison to
one-loop Yukawa contributions.
\bit
\item If $\rho_8<0$, then either LQ contributions must be
assumed to be dominant and positive, or the solution must be discarded.
\item If $\rho_8>0$, then LQ contributions could be small.
Assuming they are, we can compute the value for
$\Lambda$ using $\rho_8$, the chosen value of $v$ and Eq.~(\ref{masses}).
\eit
Note that if it is believed that LQ mass-squared contributions are
small, then the four-fold ambiguity noted above would generally
be further reduced by keeping only $\rho_8>0$ cases.

An example of how this might proceed is provided by the sample
parameter choices of Eq.~(\ref{choice1}),
which yield $\rho_8=\half (m_b^2-m_\tau^2)$ (neglecting
the $-\half m_\mu^2$ contribution).  The alternative $m_i$
choices giving the same values of $|{4\over 3}m_4^{(2)}-{1\over 3}m_\mu|$,
$|{4\over 3}m_4-{1\over 3}m_\tau|$  and $|4m_2^\prime-3m_b|$,
while maintaining consistency with Eq.~(\ref{mf}), are:
\beq
(m_2^\prime=m_b,m_9=0)\,; \quad
(m_4=m_\tau,m_{10}=0)\,;\quad {\rm and/or} \quad
 (m_4^{(2)}=m_\mu, m_{10}^{(2)}=0) \,.
\label{choice2}
\eeq
The alternative values that emerge for $\rho_8$ are:
(1) $\rho_8=0$ for $m_9=m_{10}=0$; (2) $\rho_8=\half m_b^2$
for $m_9=m_b/2$ and $m_{10}=0$;
and (3) $\rho_8=-\half m_\tau^2$ for $m_9=0$ and $m_{10}=m_\tau/2$. Of these,
(1) and (3) are clearly unphysical. Thus,
if LQ contributions to the mass-squared matrix are not substantial,
the only physically consistent value of $(m_2^\prime,m_9)$ would
be $(\half m_b,\half m_b)$, and [maintaining Eq.~(\ref{lepuniv})]
the only ambiguity would be between the choices
$(m_4,m_{10})=(-m_\tau/2,m_\tau/2)$,
$(m_4^{(2)},m_{10}^{(2)})=(-m_\mu/2,m_\mu/2)$
vs. $(m_4,m_{10})=(m_\tau,0)$, $(m_4^{(2)},m_{10}^{(2)})=(m_\mu,0)$.

Whether or not we pursue this last strategy that assumes
one-loop dominance of the mass-squared matrix,
we see that a detailed study of the $\pzero$
would provide a lot of information about the low-energy
effective Lagrangian of the model.

Let us now consider the errors that will arise in the determination
of the parameters through the above equations, given
the statistical errors plotted in Fig.~\ref{figcmserrors}
for $gg\to\pzero\to \gam\gam$,
in Figs.~\ref{figerrorsnyear} and \ref{figerrorsnyearntceq1}
for $\mupmum\to\pzero\to b\anti b,\tauptaum,gg$, and
noted in text for $\gam\gam\to\pzero\to b\anti b$.
For $\ntc=4$, the statistical errors for the CMS measurement of
$\Gamma(\pzero\to gg)\br(\pzero\to\gam\gam)$ and the $\gam\gam$-collider
measurement of $\Gamma(\pzero\to \gam\gam)\br(\pzero\to b\anti b)$
will be at the $1-2\%$ level. For $\ntc=1$, these errors will be
no worse than $10-20\%$.
In considering the statistical errors for the tagged channel
rates at the $\mupmum$ collider, plotted in Figs.~\ref{figerrorsnyear}
and \ref{figerrorsnyearntceq1} for $\ntc=4$ and $\ntc=1$, respectively,
we must recall that the rate for a given tagged channel $F$ is,
to a good approximation,
proportional to $\sum_X\Gamma(\pzero\to \mupmum)\br(\pzero\to X)\eps_{X/F}$,
where the $\eps_{X/F}$ are the probabilities (detailed earlier)
that a channel $X$ is identified as channel $F$. Thus, a certain
amount of unfolding will be required to get to the rates
for the true channels. As discussed earlier, there will be systematic
errors associated with this process. In addition, there will
be systematic errors coming from uncertainties in the overall
absolute rate in the $gg\to\pzero\to\gam\gam$
CMS measurement of $\Gamma(\pzero\to
gg)\br(\pzero\to \gam\gam)$, in the absolute rate for
the $\gam\gam$-collider measurement of
$\Gamma(\pzero\to\gam\gam)\br(\pzero\to b\anti b)$
and in the overall absolute rate
for $\pzero$ production in the sum of all channels at the muon
collider. It is hard to imagine that these systematic errors will
be smaller than $\sim 10\%$.  Given that the statistical errors are,
for the most part (\ie\ except if $\mpzero\sim\mz$)
much smaller than 10\% (the only exception
being the $F=gg$ final state if $\ntc=1$ and
$\mpzero$ is small), there is little
point in doing a detailed error analysis until the
muon-collider systematic
errors can be better estimated on the basis of a detailed
muon-collider detector design.  If the systematic errors are
of order 10\%, the parameters of the model (for any one of
the choices consistent with the
experimental determinations of $|{4\over 3}m_4^{(2)}-{1\over 3}m_\mu|$,
$|{4\over 3}m_4-{1\over 3}m_\tau|$  and $|4m_2^\prime-3m_b|$)
will have systematic
errors that are somewhat larger, perhaps 15\%--20\%.
We also note that if the systematic errors cannot be brought
below the 10\% level, or if $\ntc=1$ so that
the statistical errors for the LHC, $\gam\gam$ collider
and $\mupmum\to \pzero\to gg$
measurements are of order $10-20\%$, then one could devote
less than 4 years of muon collider
running time to the optimal three-point scan without significantly
impacting the overall errors for the model parameters.

\section{Conclusion}

In this paper, we have considered the production and study of the
lightest pseudo-Nambu Goldstone state (denoted $\pzero$)
of a typical technicolor model at future $pp$, $\epem$
and $\mupmum$ colliders.  In the broad class
of models considered, the $\pzero$ is of particular interest
because it contains only down-type techniquarks (and charged technileptons)
and thus will have a mass scale that is most naturally  set by the
mass of the $b$-quark. Other color-singlet PNGB's will have masses
most naturally set by $\mt$, while color non-singlet PNGB's will
generally be even heavier. Vector resonance states such
as the technirho and techniomega are likely to be heavier still.
Thus, it is of great importance to understand what
the prospects for discovering the $\pzero$ are and whether
we can study its properties in enough detail to determine
key features and parameters of the technicolor model.
We have focused our study on the $\mpzero$ mass range that is typically
suggested by technicolor models, $10\gev<\mpzero<200\gev$. The majority
of our numerical results have been presented using
a moderate value for the number of technicolors, $\ntc=4$. However,
we have also given many results for the minimal $\ntc=1$ choice.
This comparison is important since
the $\pzero$ rates at a hadron collider (where the production cross section
is proportional to the square of the $gg\pzero$ coupling)
and at $\epem$ and $\gam\gam$
colliders (where the production cross sections are proportional to the square
of the $\gam\gam\pzero$ coupling) are all proportional to $\ntc^2$
due to the fact that the anomalous couplings are proportional to $\ntc$.
In contrast, the $\pzero$ production rate at a $\mupmum$ collider
depends only on the square of the $\mupmum\pzero$ coupling, which is
independent of $\ntc$.

For $\ntc=4$,
we have found that discovery of the $\pzero$ in the $gg\to\pzero\to\gam\gam$
mode at the LHC will almost certainly be possible unless its
mass is either very small ($\lsim 30\gev$?) or very large ($\gsim
200\gev$?), where the question marks are related to uncertainties in LHC
backgrounds in the inclusive $\gam\gam$ channel.
In addition, over much of this mass range,
the rate in this channel (which is
typically much larger than would arise for a SM Higgs boson)
will be measured with very high statistical accuracy ($\lsim \pm 1\%$,
\ie\ such that systematic uncertainty in the overall rate will dominate).
The $gg\to\pzero\to\gam\gam$ mode is also predicted to be viable, although
over a smaller mass range, at the Tevatron. For the integrated luminosity
of TeV33, a reasonably precise ($\lsim \pm 10\%$)
determination of the signal rate will typically be possible
for $\mpzero$ values such that discovery during RunII is possible.
RunI data can already be used to exclude a $\pzero$ in the $50-200\gev$
mass range for $\ntc> 12-16$. For $\ntc=1$, Tevatron rates become
extremely marginal, but $\pzero$ discovery at the LHC will still be
possible for (roughly) $70\leq\mpzero \leq 200\gev$ and the rate will
be measured with an accuracy of about 20\%.

In contrast, an $\epem$ collider, while able to discover the $\pzero$
via $\epem\to\gam\pzero$, so long as $\mpzero$ is not close to $\mz$
and $\ntc\geq 3$,
is unlikely (unless the TESLA $500\fbi$ per year option
is built or $\ntc$ is very large) to be able to determine the rates for
individual $\gam F$ final states ($F=b\anti b,\tauptaum,gg$
being the dominant $\pzero$ decay modes) with sufficient accuracy as
to yield more than very rough indications regarding the important
parameters of the technicolor model. For $\ntc=1$, discovery of
the $\pzero$ in the $\epem\to\gam\pzero$ mode will be very difficult.

The $\gam\gam$ collider option at an $\epem$ collider is actually a more
robust means for discovering the $\pzero$ than direct operation
in the $\epem$ collision mode.  For $\ntc=4$,
we find that $\gam\gam\to\pzero\to b\anti b$
should yield an easily detectable $\pzero$ signal for
$0.1\lsim{\mpzero\over\rts}\lsim 0.7$ when the $\gam\gam$ collision
set-up is chosen to yield a broad luminosity distribution. Once
$\mpzero$ is known, the $\gam\gam$ collision set-up can be re-configured
to yield a luminosity distribution that is strongly peaked at
$\rts_{\gam\gam}\sim \mpzero$ and,
for much of the mass range of $\mpzero\lsim 200\gev$,
a measurement of $\Gamma(\pzero\to \gam\gam)\br(\pzero\to b\anti b)$
can be made with statistical accuracy in the $\lsim 2\%$ range.
More detailed studies will be required to determine what
is possible in the $\pzero\to gg$, $\tauptaum$ and $\gam\gam$ final states.
For $\ntc=1$, $\pzero$ discovery in the $\gam\gam\to\pzero\to b\anti b$
channel will remain possible for $\mpzero\geq 60\gev$
or so, but the accuracy with
which the rate will eventually be measured worsens to $5-10\%$.

A $\mupmum$ collider would be crucial for detecting a light
$\pzero$ ($\mpzero\lsim 30\gev$ for $\ntc=4$ or $\mpzero\lsim 60\gev$ for
$\ntc=1$) and would play a very special role
with regard to determining key properties of the $\pzero$.
In particular, the $\pzero$, being comprised
of $D\anti D$ and $E\anti E$ techniquarks, will naturally have couplings
to the down-type quarks and charged leptons of the SM.
Thus, $s$-channel production ($\mupmum\to\pzero$) is predicted
to have a substantial rate for $\rts\sim\mpzero$.
Because the $\pzero$ has a very narrow width (not much
larger than that of a SM Higgs boson of the same mass, assuming
a moderate value of $\ntc$), in order to maximize this rate
it is important that one operates the $\mupmum$ collider
so as to have extremely small beam energy spread; $R=0.003\%$
can be achieved and is very crucial to the conclusions stated below.
For this $R$, the resolution in $\rts$ of the muon collider, $\sigrts$,
is of order $\sigrts\sim 1\mev (\rts/50\gev)$, whereas the $\pzero$
width varies from $0.3\mev$ to $26\mev$ as $\mpzero$ ranges
from $10\gev$ up to $200\gev$ (for $\ntc=4$ and the particular choice
of other model parameters that we have employed). Thus, $\sigrts<\gampzero$
is possible and leads to very high $\pzero$ production rates
for typical $\mupmum\to\pzero$ coupling strength.
We, of course, have employed the luminosity that is currently
anticipated to be achievable (as a function of $\rts$) for this $R$.

If $\ntc=4$ and $\mpzero>30\gev$ or $\ntc=1$ and $\mpzero>60\gev$,
it is reasonable to assume
that the $\pzero$ will be discovered at the Tevatron, the LHC
or at an $\epem$ collider operating
in the $\gam\gam$ collider mode. The $\mupmum$
collider could quickly (in less, possibly much less if discovered
at the LHC, than a year) scan the mass
range indicated by the previous discovery (for the expected
uncertainty in the mass determination)
and center on $\rts\simeq\mpzero$ to within
$<\sigrts$. If $\mpzero$ is small ($\lsim 30$ for $\ntc=1$
or $\lsim 60\gev$ for $\ntc=4$),
the $\pzero$ might not be detected even at
the LHC or in $\gam\gam$ collisions,
in which case a $<1$ year scan of the $\lsim 60\gev$ mass
region at the $\mupmum$
collider would be sufficient to either exclude the $\pzero$
or detect it and determine its mass with high accuracy.
Once the $\pzero$ is discovered and its mass accurately determined,
we then imagine devoting a period of several years
to the optimal three-point scan (see main text) of the $\pzero$
resonance in order to measure with high statistical precision
all the $\mupmum\to\pzero\to F$ ($F=b\anti b,\tauptaum,gg$)
channel rates and the
total width $\gampzero$. For $\ntc=4$ and the particular technicolor
model parameters we have chosen, 4 years devoted to the scan
would yield statistical precisions well-below $10\%$
when $\mpzero$ is not close to $\mz$, and of order $10\%$
for $\mpzero\sim \mz$. For $\ntc=1$, the $F=gg$ channel rate
error would be $>10\%$ if $\mpzero<50\gev$ or $\mpzero\sim \mz$,
but the $F=b\anti b,\tauptaum$ channel rates and $\gampzero$
would again have tiny statistical errors unless $\mpzero\sim \mz$.

Finally, we discussed how an accurate $gg\to\pzero\to\gam\gam$
LHC measurement can be combined with the precision
$\mupmum$ collider measurement of $\gampzero$ to determine $\ntc$,
and how the values of $\ntc$ and $\gampzero$ can be combined with
the precision $\mupmum$ measurements of various channel
rates to determine (up to a discrete set of ambiguities)
the six mass parameters of the effective
low-energy Yukawa Lagrangian that determine $T_3=-1/2$
fermion masses and their couplings to the $\pzero$, and how all
the above can, in turn, be used to determine
the cut-off scale $\Lambda$ from the measured value of $\mpzero$
under the assumption that mass-squared contribution coming
from one-loop diagrams involving SM fermions dominate over
technileptoquark mass-squared contributions. For $\ntc=4$,
we found that the accuracies for the measurements would very probably
be systematics dominated.  For systematic errors
(associated with separation of different final states,
\eg\ using $b$-tagging, and with absolute rate normalization)
in the expected $\sim 10\%$ range, all the above parameters could
then be determined within $\sim 15\%$ overall systematic error, with
somewhat smaller statistical error. Finally, we noted
that the accurate $\gam\gam$-collider
measurement of the rate for $\gam\gam\to\pzero\to b\anti b$
would allow us to check the consistency of the formulae
for the $\gam\gam\pzero$ and $gg\pzero$ anomalous couplings.
Although errors in the above procedure would worsen for $\ntc=1$,
due to statistical errors for some of the measurements increasing
to the $10-20\%$ level, a roughly $20-25\%$ determination of the model
parameters would still be possible.
Thus, we conclude that a detailed study of the lightest PNGB
state of the broad class of
technicolor models discussed here is possible and will reveal a great
deal about the low-energy effective Lagrangian for the model.
As described, a muon collider would play an absolutely essential role in
such a study, especially if the number of technicolors is small and/or
if the PNGB state is very light.

\begin{center}
\begin{bf}
Acknowledgements
\end{bf}
\end{center}
The research of RC, SDC, DD, AD and RG has
been carried out within the Program
Human Capital and Mobility: ``Tests of electroweak symmetry
breaking and future European colliders'', CHRXCT94/0579  and No. OFES 950200.
AD acknowledges support of a ``Marie Curie'' TMR research fellowship
of the European Commission under contract No. ERBFMBICT960965.
RC would like to thank Prof. J.P. Eckman for the kind hospitality of the
physics department of the University of Geneva.
JFG is supported by the U.S. Department of Energy
under grant No. DE-FG03-91ER40674
and by the U.C. Davis Institute for High Energy Physics.
JFG would also like to thank the University of Florence and the
INFN for support and hospitality during the completion of this project.
JFG thanks T. Barklow and H. Frisch for providing important
details regarding detector parameters, H. Frisch and P. Wilson
for details regarding the CDF RunI inclusive $\gam\gam$ analysis,
and R.S. Chivukula and C. Hill for valuable theoretical discussions.
RG thanks M. Mangano for information on the CDF $\gam\gam$ study.



\begin{thebibliography}{99}
\bibitem{palmer} These are the current best estimates of R. Palmer, private
communication.
\end{thebibliography}
\end{document}